\documentclass[twocolumn]{aastex62}
\usepackage{multirow}
\usepackage{color}
\usepackage{lineno}
\usepackage[online]{threeparttable}
\usepackage{tablefootnote}

\usepackage{graphicx}
\usepackage{float}
\usepackage{amsmath}
\usepackage{xcolor}

\defcitealias{Jones18}{J18}
\defcitealias{Scolnic18}{S18}

\usepackage{hyperref}
\usepackage{enumitem}
\setlist{parsep=0pt,listparindent=\parindent}

    {\pagebreak[4]\global\pdfpageattr\expandafter{\the\pdfpageattr/Rotate 90}}%
    {\pagebreak[4]\global\pdfpageattr\expandafter{\the\pdfpageattr/Rotate 0}}%

\newcommand{\JHU}{Department of Physics and Astronomy, The Johns Hopkins University, Baltimore, MD 21218.}
\newcommand{\STScI}{Space Telescope Science Institute, Baltimore, MD 21218.}
\newcommand{\Kavli}{University of Chicago, Kavli Institute for Cosmological Physics, Chicago, IL, USA.}

\newcommand{\Harvard}{Harvard-Smithsonian Center for Astrophysics, 60 Garden Street, Cambridge, MA 02138, USA}
\newcommand{\IfA}{Institute for Astronomy, University of Hawaii, 2680 Woodlawn Drive, Honolulu, HI 96822, USA}
\newcommand{\Rutgers}{Department of Physics and Astronomy, Rutgers, The State University of New Jersey, 136 Frelinghuysen Road, Piscataway, NJ 08854, USA}
\newcommand{\Moore}{Gordon and Betty Moore Foundation, 1661 Page Mill Road, Palo Alto, CA 94304, USA}
\newcommand{\Villanova}{Department of Astrophysics and Planetary Science, Villanova University, Villanova, PA, 19085 USA}
\newcommand{\UCSC}{Department of Astronomy and Astrophysics, University of California, Santa Cruz, CA 92064, USA}
\newcommand{\QUB}{Astrophysics Research Centre, School of Mathematics and Physics, Queen's University Belfast, Belfast BT7 1NN, UK}
\newcommand{\Ohio}{Astrophysical Institute, Department of Physics and Astronomy,
    251B Clippinger Lab, Ohio University, Athens, OH 45701, USA}

\def\fp{$Fitprob$}
\def\w{$-0.938 \pm 0.053$}
\def\wspec{$-0.933 \pm 0.061$}
\def\om{$0.353 \pm 0.037$}
\def\wo{$-0.810 \pm 0.144$}
\def\wa{$-0.791 \pm 0.785$}
\def\wreducpercent{7\%}
\def\wshiftpercent{5\%}
\def\lowzdistoff{$0.046 \pm 0.027$}
\def\lowzdistsig{$1.7 \sigma$}
\def\lowzdistsighyphen{$1.7$-$\sigma$}
\def\biascorsyspercent{55\%}

\begin{document}

\title{The Foundation Supernova Survey: Measuring Cosmological Parameters with Supernovae from a Single Telescope}

\author{D. O. Jones}
\affiliation{\UCSC}

\author{D. M. Scolnic}
\affiliation{\Kavli}
\affiliation{Hubble, KICP Fellow}
  
\author{R. J. Foley}
\affiliation{\UCSC}
  
\author{A. Rest}
\affiliation{\STScI}
\affiliation{\JHU}

\author{R. Kessler}
\affiliation{\Kavli}

\author{P. M. Challis}
\affiliation{\Harvard}

\author{K. C. Chambers}
\affiliation{\IfA}

\author{D. A. Coulter}
\affiliation{\UCSC}

\author{K. G. Dettman}
\affiliation{\Rutgers}

\author{M. M. Foley}
\affiliation{\Harvard}

\author{M. E. Huber}
\affiliation{\IfA}

\author{S. W. Jha}
\affiliation{\Rutgers}
  
\author{E. Johnson}
\affiliation{\JHU}

\author{C. D. Kilpatrick}
\affiliation{\UCSC}
  
\author{R. P. Kirshner}
\affiliation{\Harvard}
\affiliation{\Moore}
  
\author{J. Manuel}
\affiliation{\Villanova}

\author{G. Narayan}
\affiliation{\STScI}

\author{Y.-C. Pan}
\affiliation{Division of Theoretical Astronomy, National Astronomical
  Observatory of Japan, 2-21-1 Osawa, Mitaka, Tokyo 181-8588, Japan}
\affiliation{Institute of Astronomy and Astrophysics, Academia Sinica, Taipei 10617, Taiwan}
\affiliation{EACOA Fellow}

\author{A. G. Riess}
\affiliation{\STScI}
\affiliation{\JHU}

\author{A. S. B. Schultz}
\affiliation{\IfA}

\author{M. R. Siebert}
\affiliation{\UCSC}

\author{E. Berger}
\affiliation{\Harvard}

\author{R. Chornock}
\affiliation{\Ohio}

\author{H. Flewelling}
\affiliation{\IfA}

\author{E. A. Magnier}
\affiliation{\IfA}

\author{S. J. Smartt}
\affiliation{\QUB}

\author{K. W. Smith}
\affiliation{\QUB}

\author{R. J. Wainscoat}
\affiliation{\IfA}

\author{C. Waters}
\affiliation{\IfA}

\author{M. Willman}
\affiliation{\IfA}
  
\correspondingauthor{D. O. Jones}
\email{david.jones@ucsc.edu}
  
\begin{abstract}

  Measurements of the dark energy equation-of-state parameter, $w$,
  have been limited by uncertainty in the selection effects and photometric calibration
  of $z<0.1$ Type Ia supernovae (SNe\,Ia).  The Foundation Supernova Survey is designed to lower
  these uncertainties by creating a new sample of $z<0.1$ SNe\,Ia observed on the Pan-STARRS system.
  Here, we combine the Foundation sample with SNe from the Pan-STARRS
  Medium Deep Survey and measure cosmological parameters with 1,338 SNe from a
  single telescope and a single, well-calibrated photometric system.
  For the first time, both the low-$z$ and high-$z$ data are predominantly discovered by surveys that do
  not target pre-selected galaxies, reducing selection bias uncertainties.
  The $z>0.1$ data include 875 SNe without spectroscopic classifications and we
  show that we can robustly marginalize over CC\,SN contamination.
  We measure Foundation Hubble residuals to be fainter than the pre-existing low-$z$ Hubble residuals by
  $0.046 \pm 0.027$~mag (stat+sys).
  By combining the SN\,Ia data with cosmic microwave background constraints, we
  find $w=-0.938 \pm 0.053$, consistent with $\Lambda$CDM.
  With 463 spectroscopically classified SNe\,Ia alone, we measure $w=-0.933\pm0.061$.
  Using the more homogeneous and better-characterized Foundation sample gives a 55\% reduction in the systematic
  uncertainty attributed to SN\,Ia sample selection biases.  Although
  use of just a single photometric system at low and high redshift increases
  the impact of photometric calibration uncertainties in this analysis, previous
  low-$z$ samples may have correlated calibration uncertainties that were neglected
  in past studies.  The full Foundation sample will observe up to 800 SNe to
  anchor the LSST and {\it WFIRST} Hubble diagrams.

\end{abstract}


\section{Introduction}
\label{sec:intro}

Since the discovery of dark energy 20 years ago
\citep{Riess98,Perlmutter99}, measurements of the dark
energy equation-of-state parameter, $w$, have been steadily
improving \citep{Garnavich98,Knop03,Tonry03,Riess04,Astier06,
  Riess07,WoodVasey07,Kowalski08,Kessler09,Sullivan11,Betoule14,Scolnic18}.
In support of a better understanding of dark energy,
recent cosmic microwave background (CMB) experiments have yielded
improved measurements of the cosmic matter density at $z \approx 1090$ \citep{Planck18},
and baryon acoustic oscillations (BAO) have given excellent
constraints on the acoustic scale from $z \approx 0.3$ to $z \approx 2$ \citep{Anderson14,Ross15,Alam17}.
As Type Ia supernova (SN\,Ia) sample sizes have steadily increased,
  their systematic uncertainties have steadily decreased.  Their reduced
  systematic uncertainties are primarily due to
improvements in photometric calibration
and a better understanding of the ways in which SN\,Ia distance
measurements are biased by selection effects.
Systematic and statistical uncertainties
on $w$ have been approximately equal in most recent measurements
(e.g., \citealp{Betoule14}; \citealp{Brout18}; \citealp{Jones18},
hereafter \citetalias{Jones18}; \citealp{Scolnic18}, hereafter \citetalias{Scolnic18}).

Counterintuitively, many of the dominant sources of systematic
uncertainty in dark energy measurements stem from the
nearest SNe\,Ia.  While high-$z$ SN\,Ia samples from the Sloan
Digital Sky Survey (SDSS; \citealp{Kessler09}), the Supernova Legacy Survey (SNLS; \citealp{Conley11,Sullivan11}),
and Pan-STARRS (PS1; \citealp{Rest14}; \citealp{Scolnic14}; \citetalias{Scolnic18}; \citetalias{Jones18}) are observed on
photometric systems with mmag-level systematic uncertainties, the heterogeneous
low-$z$ SN samples are observed on more than 13 different photometric
systems, each with their own systematic uncertainties.  These systematic uncertainties
may be correlated in ways that are difficult to predict.  There could be additional
unknown systematic uncertainties associated with the fact that much of the data were taken
at a time when cosmological analyses were not yet concerned with or
limited by mmag-level systematic uncertainties.  The sample selection
criteria and follow-up criteria for low-$z$ SNe\,Ia are heterogeneous
and sometimes not well documented, leading to systematic uncertainties
in the observational biases and selection effects.

Unlike the high-$z$ SN\,Ia data, samples of
SNe\,Ia at $z < 0.1$ were predominantly found by
surveys that targeted pre-selected sets of galaxies
(e.g., the Lick Observatory Supernova Search; \citealp{Filippenko01}).
There is some evidence that SN\,Ia selected by targeted
surveys have different biases than those from untargeted
surveys \citep{Jones18b}.
The Foundation Supernova Survey \citep{Foley18} aims to create a single, low-$z$ sample
that is more similar in calibration uncertainty and sample selection characteristics to the
high-$z$ data.  Foundation uses the PS1 telescope to follow
SNe\,Ia discovered primarily by untargeted searches such as
ASAS-SN \citep{Holoien17}, ATLAS \citep{Tonry18},
Gaia \citep{Prusti16}, and the Pan-STARRS Survey for Transients
(PSST; \citealp{Huber15}).
Although some Foundation data were discovered by targeted searches,
untargeted surveys would likely have reported many of these
targeted events if targeted surveys with greater depth or
higher cadence (e.g., DLT40; \citealp{Tartaglia18})
had not discovered them first \citep{Foley18}.
Untargeted surveys independently discovered 94\% of
the \citet{Foley18} sample.

Foundation aims to compile a sample of up to 800 SNe\,Ia observed in $griz$
over the next several years.  The Foundation Data Release 1 (DR1) includes
180 SNe\,Ia that pass the \citet{Foley18} sample criteria for inclusion in a cosmological
analysis, approximately equal to the number of published, cosmologically
useful SNe\,Ia from all previous low-$z$ samples combined
(\citealp{Betoule14}; \citetalias{Jones18}; \citetalias{Scolnic18}).  The Dark Energy Survey
cosmological analysis \citep{DES18}, for example, uses
just 122 low-$z$ SNe\,Ia.

Here, we combine Foundation data with high-$z$ data observed
using the same telescope and the same photometric system,
creating for the first time a unified sample with
a significant number of SNe\,Ia ($>$25) at both $z < 0.1$ and $z > 0.1$
observed on the same photometric system.  The data reduction and calibration of
these data are nearly homogeneous, with the caveats
that the use of redder bands at high-redshift and the exposure
time increase for the high-$z$ data introduce modest sample-to-sample differences.
We use both the subset of SNe from the PS1 Medium Deep Survey (MDS) with
spectroscopic classifications, and the full photometrically
classified MDS sample.  The photometrically classified
MDS sample includes $\sim$5\% core-collapse (CC) SN contamination, but
this contamination can be marginalized over in a Bayesian
framework (\citealp{Kunz07,Hlozek12}; \citealp{Jones17}).

We expand the CC\,SN simulations in \citet{Jones17} to gain improved constraints on the effect of 
CC\,SN contamination on our cosmological measurements.
We also measure star-formation rates
for the entire data set using PS1 and SDSS photometry, where available,
to test the effect of star-formation on SN\,Ia shape- and color-corrected
magnitudes, ensuring that uncertainty in the relationship between
SN\,Ia properties and their host galaxies is not biasing the cosmological
parameters.

Because of the limited wavelength coverage of the SALT2.4 model
that is used to measure SN\,Ia distances, the present analysis is
restricted primarily to the Foundation $gr$ photometry (6 Foundation SNe
are at high enough redshift to include $i$ data).
The SALT2 color law is only trained from 2800-7000 \AA\ \citep{Guy07,Guy10,Betoule14}.
Relative to the previous low-$z$ data, this reduces the precision of the measured color of each SN
and limits our ability to verify the SALT2 color law with Foundation data.
However, an extended SALT2 model, re-trained with Foundation data,
would allow us to take advantage of the available $iz$ observations in the future.

In \S\ref{sec:data}, we present the MDS and Foundation data sets.
In \S\ref{sec:cosmoest}, we outline our cosmological parameter
estimation methodology.  Our results are in \S\ref{sec:results},
including a discussion of the consequence of replacing the current low-$z$
sample with the Foundation sample.
We discuss future prospects for cosmology with the Foundation data set
in \S\ref{sec:discussion} and
our conclusions are in \S\ref{sec:conclusions}.

\section{Data}
\label{sec:data}

\begin{figure}
\includegraphics[width=3.5in]{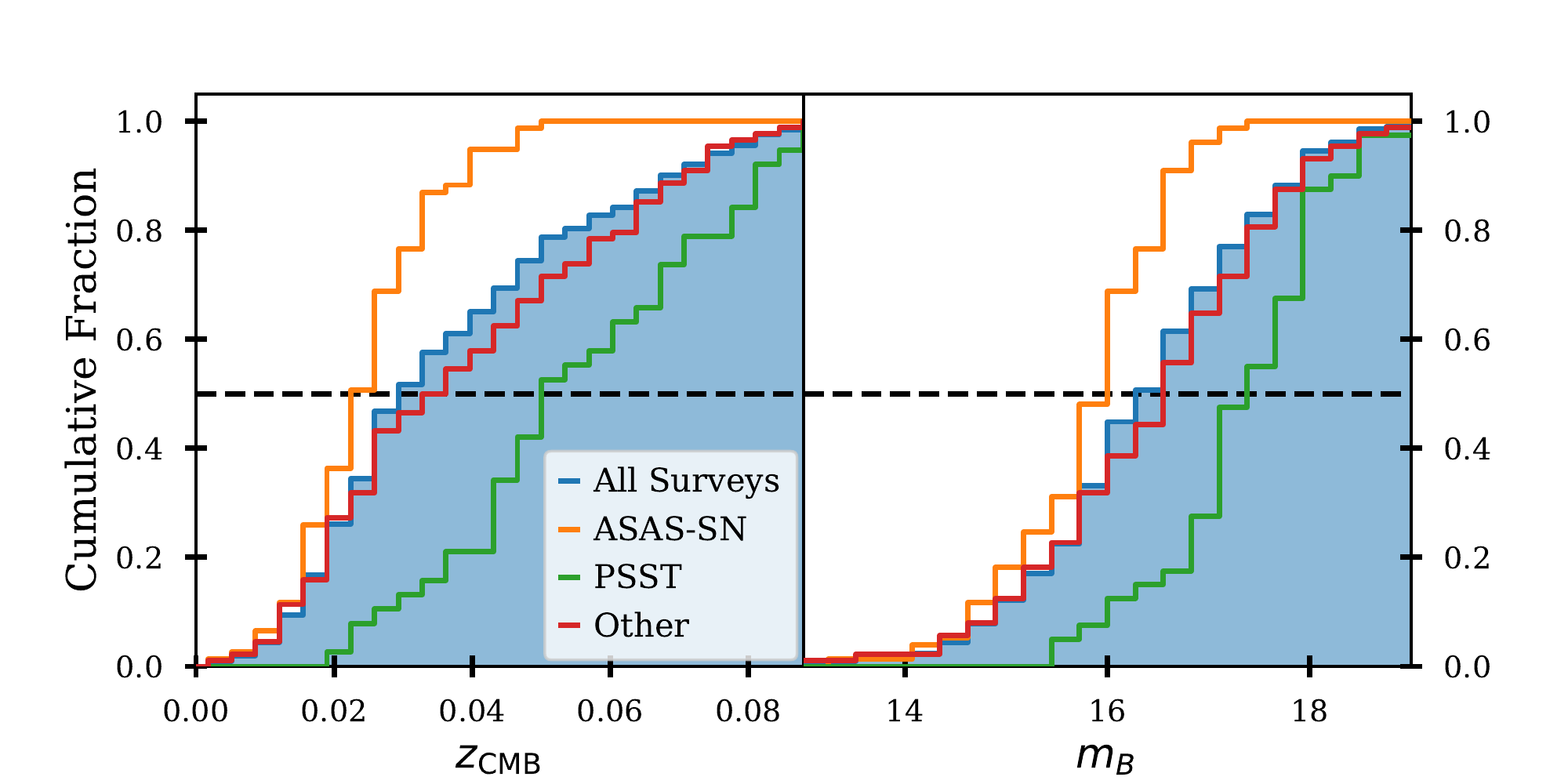}
\caption{Cumulative distributions of the redshift and
  peak magnitude of SNe\,Ia in the Foundation DR1 using
  data from \citet{Foley18}.  The two largest SN discovery
  surveys in this dataset are ASAS-SN (orange; 38\%)
  and PSST (green; 20\%).}
\label{fig:cdf}
\end{figure}

This paper presents the union of the Foundation
and PS1 MDS data sets,
both assembled using the PS1 telescope.
The extremely well-calibrated PS1 photometric system
\citep{Schlafly12} makes it ideal for cosmological
analyses of SN Ia, which typically have photometric calibration
as the dominant systematic uncertainty (e.g., \citealp{Sullivan11,Betoule14,Rest14,Scolnic14b}).

The PS1 MDS observed 70 square degrees
of sky over approximately 4 years, with a typical observing sequence of $g_{P1}$
and $r_{P1}$ on the same night, followed by $i_{P1}$ and $z_{P1}$ on the
second and third nights, respectively.  The $y_{P1}$ filter
was primarily used in bright time, and does not currently have
the depth or precise calibration necessary for a cosmology
analysis using high-$z$ SNe.  Further details about the MDS
strategy are given in \citet{Chambers16}.

The PS1 MDS discovered approximately 5,200 SNe, $\sim$350
of which are spectroscopically classified SNe\,Ia.  The spectroscopically
classified SNe\,Ia were used to measure cosmological parameters in 
\citetalias{Scolnic18} and \citet{Rest14,Scolnic14b}.  Additionally, we measured $\sim$3200
host-galaxy redshifts, and 1,169 of these SNe with either
spectroscopic or light-curve-based classifications were
used to measure cosmological parameters (\citealp{Jones17}; \citetalias{Jones18}).
In the present work, we use both spectroscopically and photometrically classified MDS data
to measure cosmological parameters.  The light curves and host-galaxy spectra for likely SNe\,Ia
are available online at
\dataset[10.17909/T95Q4X]{https://doi.org/10.17909/T95Q4X} and
through the Open Supernova Catalog \citep{Guillochon17}.  The
remainder, including likely CC\,SNe and noisy SN\,Ia,
will be published in future work.

The Foundation Supernova Survey uses the PS1 telescope to follow
SNe\,Ia found by surveys that quickly publish new SN discoveries.
To be followed by Foundation, SNe must be within the 3$\pi$
footprint ($\delta > -30^{\circ}$), have $z \lesssim 0.08$, and have Milky Way
$E(B-V) \lesssim 0.2$~mag.  To minimize peculiar velocity uncertainties,
the minimum redshift of Foundation is 0.015, unless the SNe are near enough
to potentially have a Cepheid or Tip of the Red Giant Branch (TRGB)
distance.

The calibration and sample selection of the Foundation SNe are
more similar to the high-$z$ SNe than to previous samples of
low-$z$ SNe.
The higher-$z$ samples have historically been observed on better-calibrated
photometric systems than low-$z$ samples.  Foundation, however,
uses the PS1 photometric system, which has systematic
uncertainties on the few mmag level
\citep{Schlafly12,Scolnic15}.
Similar to high-$z$ SNe, most Foundation SNe are also discovered
by surveying a given area on the sky rather than targeting a pre-selected set of
galaxies, with the majority of the sample coming
from ASAS-SN and PSST (Figure \ref{fig:cdf}).
In previous samples of low-$z$ SNe, the systematic error
  due to selection effects was greatly increased by uncertainty
  over whether the surveys were predominantly magnitude-limited or
  volume-limited (\citetalias{Jones18,Scolnic18}).  A simple yet important
  advantage of the Foundation Supernova Survey is that we understand
  that our sample is magnitude limited, which can reduce our final systematic
uncertainty due to selection effects by $\sim$50\% (\S\ref{sec:syserr}).

The first Foundation data release includes 225 SNe\,Ia.
SNe are observed in $griz_{P1}$ at each epoch, with a median cadence of 8 days overall
and 5.5 days within 10 days of peak.
180 pass the criteria presented in \citet{Foley18} for
inclusion in a cosmological analysis (see \S\ref{sec:selection})
and 175 of those are included here\footnote{Two SNe do not pass the sample cuts
  when the wavelength range of the SALT2 model is reduced
  to $<$7000 \AA, and three are below our minimum
  redshift of $z = 0.015$.}.  Redshifts and classifications
for Foundation SNe are given by \citet{Foley18} and references therein.
Foundation will eventually obtain light curves for $\sim$800 SNe\,Ia
to match the {\it WFIRST} low-$z$ sample requirement
\citep{Spergel15}.

\subsection{Photometric Pipeline Processing}

\begin{figure*}
\includegraphics[width=7in]{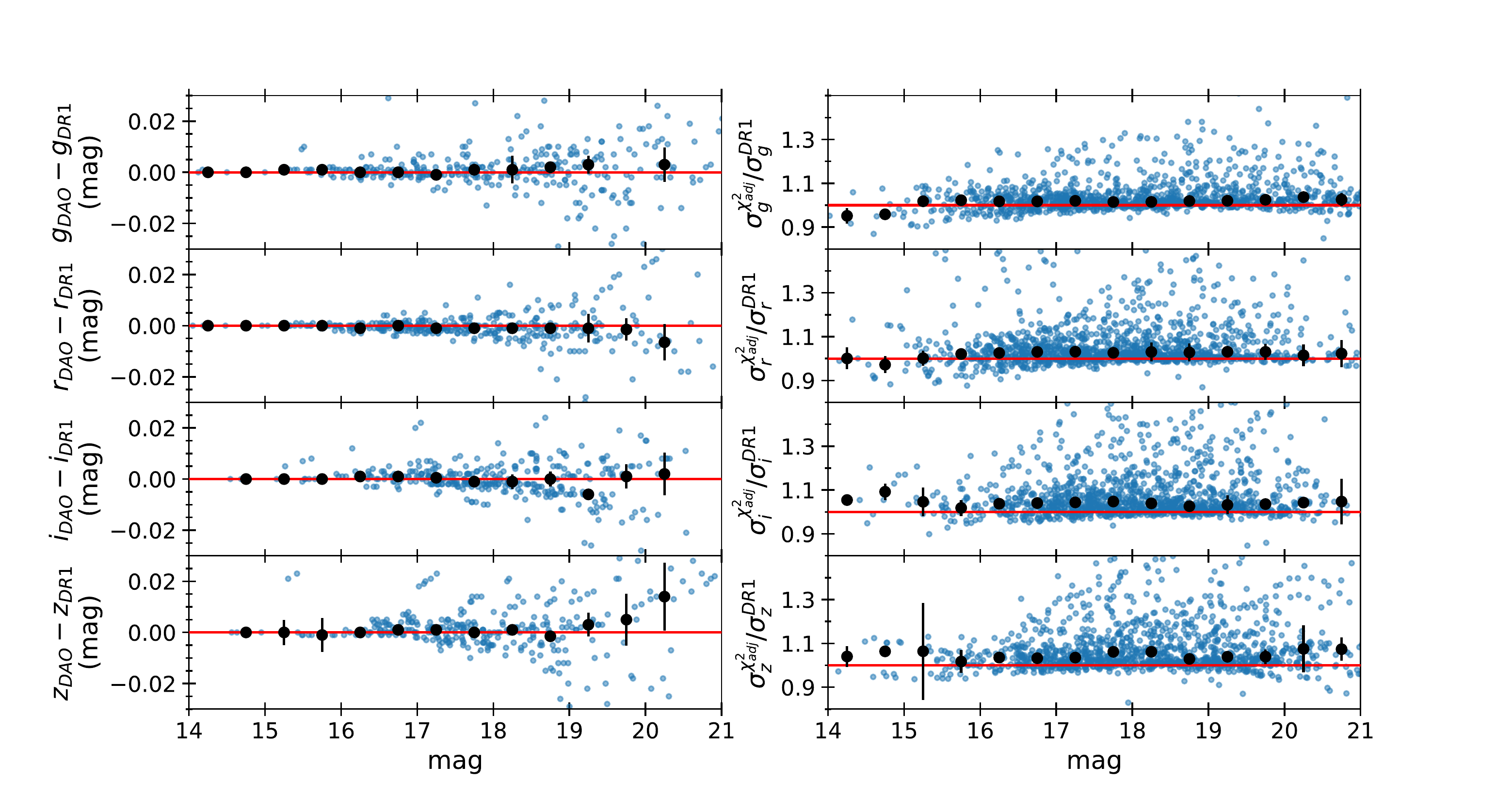}
\caption{Comparison of photometric measurement
  techniques for the low-$z$ Foundation sample to those
  used for the high-$z$ PS1 MDS. Left: For a
  subset of Foundation SNe, the difference between forced DAOPHOT photometry
  and unforced DOPHOT photometry as a function of magnitude.
  The median difference is $<$1 mmag.  Right: As a function of magnitude, the
  fractional increase in uncertainty for Foundation SNe after empirically adjusting
  the photometric uncertainties
  using the noise at the SN location in pre-explosion epochs.  Performing
  this procedure on the Foundation DR1 photometry shows that DR1 flux
  uncertainties are underestimated by just 2-4\% on average.}
\label{fig:phot}
\end{figure*}

The photometric pipeline and template construction used to
measure the light curves of Foundation SNe is reviewed in \citet{Foley18} and is
in large part based on the \texttt{photpipe} pipeline \citep{Rest05,Rest14}.
The pipeline for the PS1 MDS is nearly identical, and is
reviewed in \citetalias{Scolnic18}.  In brief, Foundation image
templates were taken from stacked exposures created by the
PS1 survey while PS1 MDS templates were created from the medium
deep survey itself after excluding images taken during the season
a given SN was observed.  Both images and templates are re-sampled
and astrometrically aligned to match a skycell in the
PS1 sky tessellation. An image
zeropoint is determined by comparing PSF photometry
of the stars to updated stellar catalogs of
PS1 observations \citep{Chambers16}.  The PS1 templates are
convolved to match the nightly images and the convolved
templates are subtracted from the nightly images with
HOTPANTS \citep{Becker15}.  Finally, a flux-weighted
centroid is found for each SN position and PSF
photometry is performed using ``forced photometry'': the 
centroid of the PSF is forced to be at the SN position.  The
nightly zeropoint is applied to the photometry to
determine the brightness of the SN for that epoch.  

There are multiple systematic uncertainties related
to this process, but all of them are on the milli-mag (mmag) level.
Foundation templates are $\sim$2 mag deeper than the
individual exposures and are created from PS1 3$\pi$ survey data. 
Any expected systematic uncertainties from remaining SN
light or the manner in which templates are subtracted (e.g., image
subtraction versus scene modeling) are $\sim1$ mmag (\citetalias{Scolnic18}).
There is an additional $1$ mmag systematic uncertainty due to the fact
that, for Foundation, forced photometry is performed on the SNe but
not on the stars (Figure \ref{fig:phot}).
This systematic uncertainty is relatively low compared to \citet{Rest14} because $>$90\%
of the observations of any given Foundation SN have multiple
exposures with signal-to-noise ratios $\mathrm{S/N}>20$.

The processing of MDS versus Foundation data has
only a few subtle differences (below).  Figure \ref{fig:phot} shows
that the expected bias from these differences is negligible.

\begin{enumerate}
\item The photometry of Foundation DR1 light curves was measured
  using DOPHOT \citep{Schechter93}, while DAOPHOT \citep{Stetson87}
  was used for the PS1 MDS.
  The primary difference between the two methods is that
  DOPHOT uses a Gaussian model to fit the point spread
  function (PSF) of each image, while DAOPHOT
  uses a Gaussian with an empirical lookup table
  for fit residuals.  We
  have verified that the two methods have a median difference
  of less than 1 mmag (Figure \ref{fig:phot}, left), and
  anticipate updating the photometry
  with DAOPHOT in the Foundation second data release.
\item In processing the MDS images, \texttt{photpipe} used
  an astrometric alignment algorithm to align the images.  The astrometry of
  the PS1 warp images $-$ single-epoch images that have been
  re-sampled and aligned into a skycell of the PS1 sky
  tessellation $-$ is currently much more accurate than it was
  during the MDS, and the \texttt{photpipe} astrometric alignment
  stage is no longer necessary.
\item The MDS used forced photometry
  to compute individual image zeropoints, while the
  Foundation pipeline uses photometry with
  a floating centroid (unforced photometry).  We have
  verified that using unforced photometry (in conjunction
  with DOPHOT) does not bias the data (Figure \ref{fig:phot}, left)
  and have estimated just a 1 mmag systematic uncertainty
  due to this effect (Scolnic et al.\ in prep).
\item Light curves from the MDS have an additional correction
  for the host-galaxy background noise. Photometric uncertainties
  are increased such that epochs with
  no SN light have a reduced $\chi^2$
  of 1 and a baseline flux value is added such that
  the weighted average of the flux measured from these epochs is 0.
  SNe in the MDS are often much fainter
  than their host galaxies, and a proper accounting for
  host galaxy background noise can increase photometric uncertainties by up to a
  factor of $\sim$2-3 \citep{Kessler15,Jones17}.  However, the effect
  on Foundation SNe is much smaller.
  
  Host-galaxy noise must be added in quadrature
  to the light-curve flux uncertainties.  Because Foundation SNe have
  much larger fluxes than MDS SNe, they also have larger Poisson
  uncertainties $-$ a factor of 14 larger within 15 days of peak $-$
  and host-galaxy noise is therefore a much smaller effect for Foundation.
  However, we tested the size of the host-galaxy noise effect for
  a subset of the Foundation sample.  To do this, we used the fact that
  each Foundation template is comprised of $\gtrsim$10 individual
  PS1 single-epoch images.  Therefore, each SN has $\gtrsim$10 measurements of the
  of the background prior to the SN explosion from which the measured uncertainties
  are compared to the variance of the background.
  From these measurements, we find that the Foundation DR1 flux uncertainties
  are underestimated by only 2-4\% on average (Figure \ref{fig:phot}, right).
  Because we use
  PS1 3$\pi$ images as templates for difference imaging, there is
  correlation between the image flux and the template flux.  But, given that
  there are $\gtrsim$10 warps per template, this correlation is small.
  
\end{enumerate}

In addition to these subtle differences, the photometry of
\textit{both} Foundation and the MDS has been corrected
for a subtle bias introduced because the shape of
the PS1 PSF (and all PSFs) is dependent on the
color of the source.  Using the formalism of \citet{Guy10},
this bias can be corrected
empirically by a slight linear, wavelength-dependent
adjustment to the PS1 $g$-band
throughput $T$ when PSF photometry is used:

\begin{equation}
\tilde{T} = T \times [1+0.065\times(\lambda-4979)/1000]
\end{equation}

\noindent The full procedure for improving
the color-dependent photometric measurements will
be presented in Scolnic et al.\ (in prep), and further
details about this bias are given in
\citet[see their Figure 4]{Guy10}.

\subsection{Host-Galaxy Masses and Star-Formation Rates}
\label{sec:host}

\begin{figure*}
  \includegraphics[width=7in]{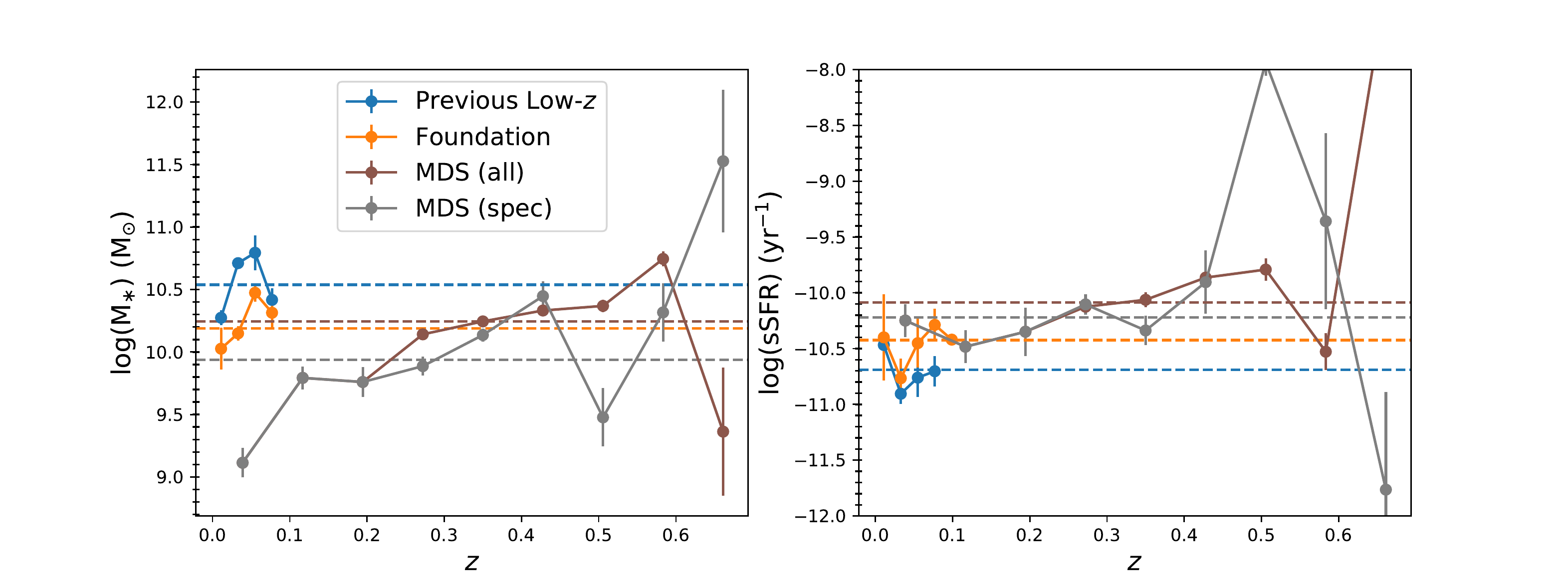}
  \caption{Host-galaxy mass and sSFR
    as a function of redshift for the previous low-z sample (blue), the Foundation sample (orange),
    the full MDS sample (brown) and the spectroscopically classified subset of MDS SNe\,Ia (grey).
    Dashed lines indicate the median host mass and sSFR of each sample.  We show median bins with uncertainties
    estimated from the median absolute deviation.  Foundation is a closer match to
    the host-galaxy properties of the high-$z$ PS1 data than to those of the
    low-$z$ sample.  In the MDS sample, low-$z$ SNe in faint hosts may have been favored for follow-up as
    candidate higher-$z$ objects.}
  \label{fig:massvz}
\end{figure*}

Accurate SN\,Ia distance measurement requires a correction
based on the mass of each SN\,Ia host galaxy
(e.g., \citealp{Kelly10,Lampeitl10,Sullivan10}).  There has been some
additional evidence that the star-formation rate (SFR) or
specific star-formation rate (sSFR) near the site of
the SN explosion might
correlate with measured SN\,Ia distances
\citep{Rigault13,Rigault15,Rigault18,Kim18,Roman18},
although these dependencies may be partially included in
a correction based on the host-galaxy mass.
Some of this evidence was also disputed in \citet{Jones15}.
However, given that we have the largest current sample of cosmologically
useful SNe\,Ia, it makes sense to simultaneously
measure masses and sSFRs both to test for correlations
and simultaneously constrain cosmological bias.

To measure host-galaxy masses and sSFRs, we measure
PS1 $grizy_{P1}$ photometry and SDSS $ugriz$ for each
host galaxy.  For Foundation, we use PS1 3$\pi$ images
from the publicly available first data release \citep{Flewelling16}.  For the PS1
MDS data however, 3$\pi$ images are contaminated by SN
light and therefore we use single-season templates from
the PS1 MDS fields as discussed in \citetalias{Jones18} and \citetalias{Scolnic18}.
The most likely host galaxy for all SNe is determined
by using the $R$ parameter method \citep[their Equation 1]{Sullivan06} (see also \citetalias{Jones18}),
which uses the proximity,
size, and orientation of the galaxy.  Nearly all low-$z$
hosts were validated by ensuring that the host redshift matches
the SN redshift.  For the high-$z$ sample, \citet{Jones17}
estimated that $1.2 \pm 0.5$\% of these host galaxy determinations
were incorrect, an effect that will not significantly bias
measuring the correlation of Hubble residuals
with host mass and sSFR (see also \citealp{Gupta16}).

We measure photometry within
a fixed elliptical aperture given by the SExtractor \citep{Bertin96}
isophotal radius in the PS1 $r$ band.  We increase the
radius to account for the PSF sizes of each filter
and instrument and we mask out SExtractor-identified
contaminating objects in each image.  We verify that
our measured aperture magnitudes are consistent with
the catalog magnitudes provided for each instrument,
finding an average offset of $<$0.05 mag.  Some offsets can occur
because the catalog magnitudes do not always capture
all of the flux from each galaxy and use different
effective galaxy radii for different photometric filters.

We use LePHARE \citep{Arnouts11}
with \citet{Bruzual03} spectral templates
to estimate galaxy stellar masses and sSFRs.
Galaxy SED templates
correspond to spectral types SB, Im, Sd, Sc, Sbc, Sa, S0 and E.
$E(B-V)$ is allowed to vary from 0 to 0.4 mag during the fit.
We estimate uncertainties by generating Monte Carlo realizations
of the host-galaxy photometry.  For each filter, we generate
mock photometric points from
a normal distribution with a standard deviation equal to the
photometric uncertainties.  We then use LePHARE to fit SEDs to each
realization of the photometric data and the uncertainty
on the host mass and SFR is given by the spread in
output values.

The average host mass and sSFR are shown as a function of redshift in
Figure \ref{fig:massvz} for the Foundation sample, previous low-$z$ samples, and the MDS
samples.  The Foundation sample is significantly less biased
toward high-mass and low-sSFR hosts than previous low-$z$ samples.
If SN\,Ia depend on their host galaxies in an unexpected way, an analysis with Foundation
SNe will not be as systematically affected as one using previous low-$z$ data.
See also Figure 8 of \citet{Brown18}
for a comparison between SN\,Ia host-galaxy masses in the targeted
LOSS sample \citep{Li11} and SN\,Ia host masses in the volume-limited ASAS-SN
sample, as ASAS-SN discovers many SNe that are followed by Foundation.
We also note that the full MDS sample selection is less biased than the spectroscopically classified
  sample, in which faint, low-$z$ galaxies may have been thought to be at higher redshift
  and therefore targeted for followup.  The full MDS sample necessarily excludes most
  apparently hostless SNe, as a redshift cannot be obtained, but the redshift
  followup discussed in \citet{Jones17} is otherwise complete to $z \approx 0.3$.
  At $z > 0.3$, the sample is biased toward brighter, more massive host galaxies for
  which redshifts can be more easily obtained.
The correlation between Hubble residuals and host-galaxy mass/sSFR
are given for the Foundation sample in \S\ref{sec:results} and \S\ref{sec:discussion}.

\subsection{SALT2 Distances and Selection Criteria}
\label{sec:selection}

We use the SALT2.4 light-curve fitting method \citep{Guy10,Betoule14}
as implemented in the SuperNova ANAlysis software (SNANA;
\citealp{Kessler09}) to measure the shape, color, and flux parameters
of SNe\,Ia in this sample.  The relationship between
SALT2 parameters and the SN\,Ia distance is given by the Tripp formula
\citep{Tripp98}:

\begin{equation}
  \mu = m_B - \mathcal{M}\ + \alpha \times x_1 - \beta \times c + \Delta_B(z) + \Delta_M.
  \label{eqn:salt2}
\end{equation}

\noindent where the light-curve stretch parameter, $x_1$,
the color parameter, $c$, and the amplitude parameter, $m_B$,
are measured from each SN light curve.  Nuisance parameters
$\alpha$ and $\beta$ are free parameters that we will estimate
simultaneously with SN distance in \S\ref{sec:cosmoest}.
$\mathcal{M}$, a combination of the absolute SN\,Ia magnitude in the $B$ band
at peak and the Hubble constant, is marginalized over during cosmological
parameter estimation.
$\Delta_B$ is the correction for the redshift-dependent distance bias, which is computed
from simulations of our sample (\S\ref{sec:foundsims}) and $\Delta_M$,
the ``mass step'' is a step function that depends on the host-galaxy
mass for each SN (see \S\ref{sec:hosts} for alternate parameterizations).
SN\,Ia uncertainties include
redshift uncertainty and lensing uncertainty
($\sigma_{lens} = 0.055z$; \citealp{Jonsson10}).

We use these SALT2 parameters to apply the
standard sample selection criteria
used by \citetalias{Jones18}, which in turn are based
on \citet{Betoule14}.  These selection criteria
include cuts on the SALT2 shape ($-3 < x_1 < 3$) and
color ($-0.3 < c < 0.3$)
that ensure the SNe\,Ia are within the parameter
space covered by the SALT2 model training.
Cuts on the shape and time of maximum light uncertainty
ensure that these parameters are relatively well-measured
($\sigma_{t0} < 2$ rest-frame days and $\sigma_{x1} < 1$).

We remove light curves without any epochs between
5 and 45 days after maximum light to avoid multi-modal
probability distribution functions in the
light-curve parameters \citep{Dai16}.  We also
apply a relatively loose cut based on the $\chi^2$
and number of degrees of freedom-based
fit probability that the data are consistent
with the SALT2 model (\fp\ $>$ 0.001).
Finally, after fitting with the SALT2 model, we remove
up to two photometric outlier points lying $>$3$\sigma$
from the model fit and then re-run the fitting.
Because SALT2 fits a median of 35
light-curve points for MDS SNe, removal of up to two outliers
does not greatly affect the fitting.
In cases where there are more than two outliers, we remove
the two most extreme outlier points.
These outliers can be
caused by image defects or poor subtractions;
however, removing too many epochs could
make CC\,SNe appear more like SNe\,Ia, which
would negatively affect our ability to classify
them in the future (\S\ref{sec:priors}).
Even when SNe are spectroscopically classified
as Type Ia, we only remove a maximum of two outliers
for consistency across the sample.

For the Foundation sample, the \citet{Foley18}
cuts also remove spectroscopically peculiar SNe\,Ia and require the
first light-curve point to have a phase of less than $+$7 days.  We require
a minimum of 8 Foundation light-curve observations in the $gr$ bands\footnote{\citet{Foley18} require 11
  total light-curve points in $gri$.  However, for the majority of the sample we use only $gr$,
  so we slightly loosen this requirement}.  The uncertainty on the
time of maximum light must be less than 1 day, rather than the
looser requirement of 2 days for the MDS sample.

The effect of these cuts on the final sample
size are given in \citetalias{Jones18}, Table 1 and \citet{Foley18}, Table 7.
For the high-$z$ sample, some of the most significant cuts are the
$x_1$ uncertainty cut, which reduces the sample by 27\% and the $x_1$ and $c$
cuts, which each reduce the sample size by $\sim$20\%.  For the Foundation sample,
all SNe are spectroscopically classified SNe\,Ia and the S/N of all data are higher.
Because of this, only a single SN fails the $x_1$ uncertainty cut, while 6\% and 1\%
fail the $x_1$ and $c$ cuts, respectively.

\section{Cosmological Parameter Estimation}
\label{sec:cosmoest}

The steps for estimating cosmological parameters
from Foundation and MDS SNe are presented in this section:

\begin{enumerate}
\item \S\ref{sec:foundsims}:  We
  correct SN\,Ia light-curve parameters, measured
  by fitting with SALT2 as discussed above, for
  sample selection or distance-dependent biases
  (often referred to as Malmquist biases).  Simulations of the SN\,Ia sample
  give a prediction for the bias in SN distance measurement
  as a function of redshift.
\item \S\ref{sec:priors}: Measuring distances from a sample
  with CC\,SN contamination
  requires an estimate of the
  probability that each SN is CC\,SN or
  Type Ia.  These must be estimated from SN
  light-curve classification (e.g., PSNID; \citealp{Sako11}),
  and are used as priors in the SN likelihood formalism.
\item \S\ref{sec:likelihood}: We apply a likelihood function for
  estimating distances from a sample with both SNe\,Ia
  and CC\,SNe.  We estimate
  distances and uncertainties at a set of 25 log($z$)-spaced redshift
  ``control points'' from $0.01 < z < 0.7$.
\item \S\ref{sec:sys}: For each systematic uncertainty, we
  repeat step three after adjusting the SN light curves for that systematic uncertainty.
  We generate a combined covariance matrix from all systematic
  and statistical uncertainties.
\item \S\ref{sec:results}: After including constraints from
  CMB measurements, BAO measurements,
  and the local measurement of H$_0$, we
  estimate final cosmological parameters with CosmoMC \citep{Lewis02}.
  
\end{enumerate}

We use two different algorithms to correct
for selection biases (\S\ref{sec:foundsims})
and marginalize over CC\,SNe (\S\ref{sec:likelihood}).
These algorithms are based on the Bayesian Estimation
Applied to Multiple Species (BEAMS; \citealp{Kunz07})
method of marginalizing over CC\,SNe, which is
discussed in \S\ref{sec:likelihood} below.
The first is the approach given in \citetalias{Jones18}, which we will refer
to as the Photometric Supernovae with BEAMS method (PSBEAMS).  The second
is the BEAMS with Bias Corrections method (BBC; \citealp{Kessler17}).
We will discuss each of these methods below and present separate measurements
of $w$ using each algorithm.  These algorithms differ in the 
implementation of bias corrections (1D versus 5D), in the modeling of 
CC\,SNe, and the ways in which events in different redshift bins are combined.
The final constraints on $w$ will use
the BBC method, as discussed in \S\ref{sec:results}.

\subsection{Simulating the SN\,Ia Sample and Correcting for Selection Biases}
\label{sec:foundsims}

\begin{figure*}
\includegraphics[width=7in]{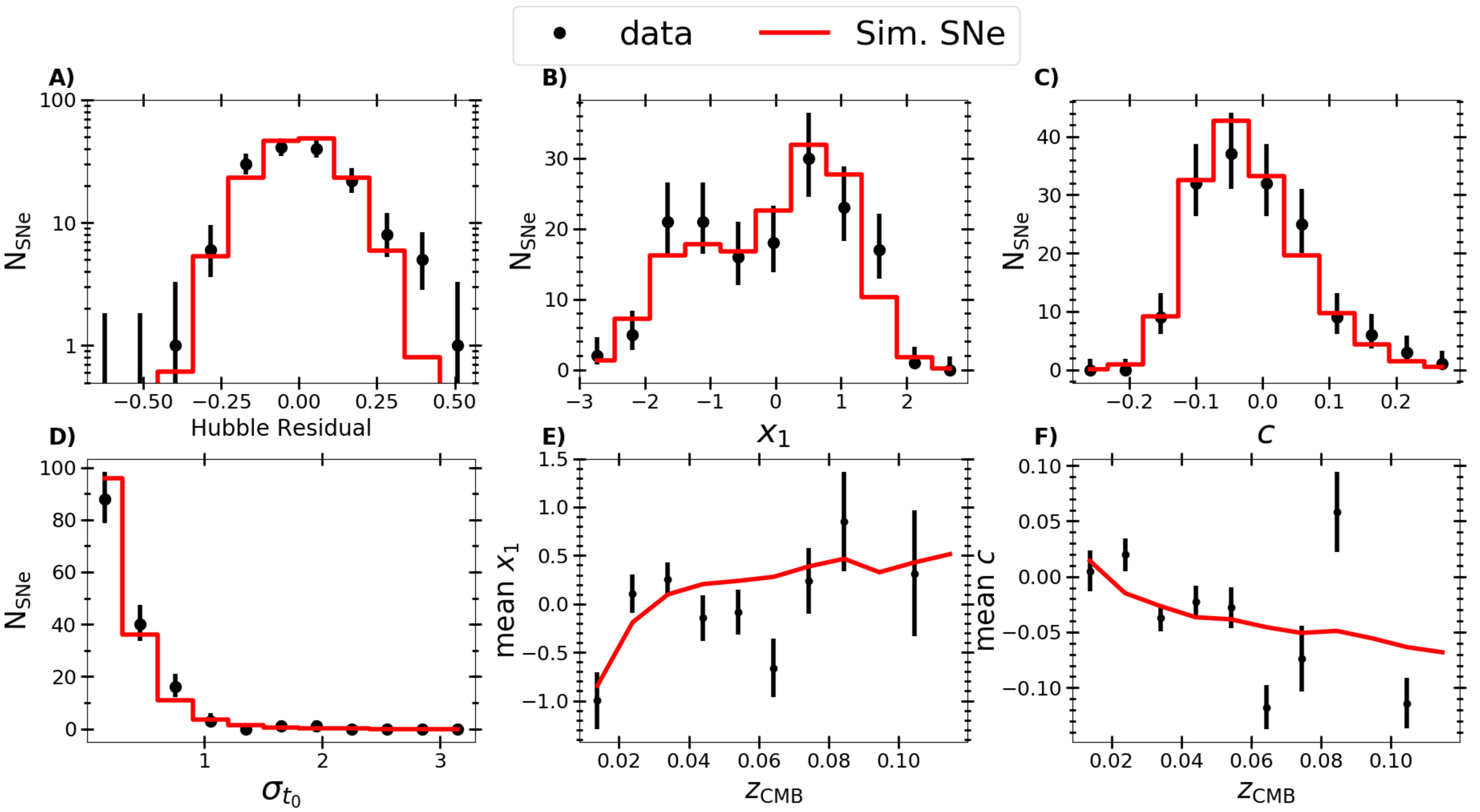}
\caption{Foundation data (black points) compared to G10 SNANA simulations (red).  Panels A-D
  show Hubble residuals, $x_1$, $c$ and uncertainty in the time of maximum light, respectively.
  Panels E and F show the average $x_1$ as a function
  of redshift, and average $c$ as a function of redshift, respectively.}
\label{fig:foundsim}
\end{figure*}

\begin{figure}
\includegraphics[width=3.5in]{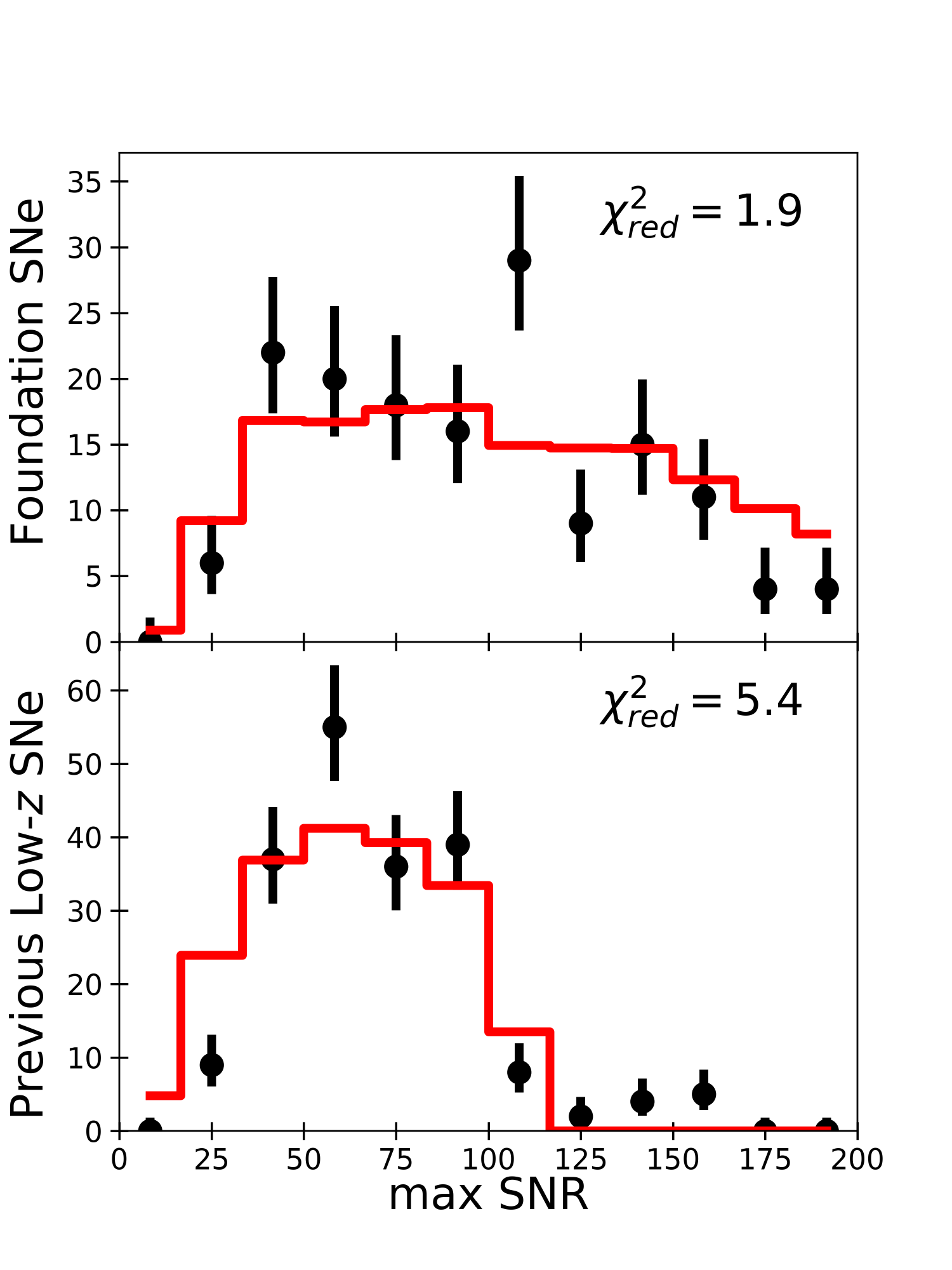}
\caption{Number of SNe in the Foundation sample (top panel) and previous low-$z$ sample
  (bottom panel) as a function of the peak-brightness S/N
  (black points), with Poisson uncertainties displayed for each S/N bin.
  In red, we display the best-fit G10 simulation for each sample.  Comparing
  the data to the simulations, the reduced $\chi^{2}$ for the Foundation
  and previous low-$z$ samples are 1.9 and 5.4, respectively.}
\label{fig:foundsimcomp}
\end{figure}

Measurements of SNe\,Ia in magnitude-limited samples will have
distance biases that are caused by SN selection effects.
Primarily, these are due to the fact that it
becomes more difficult to detect fainter SNe
at greater distances.  As the S/N
of the photometric measurements become lower, secondary effects
such as biases caused by the $-0.3 < c < 0.3$ and $3 < x_1 < 3$
box cuts come into play.

We predict the distance biases affecting our sample
using SNANA simulations.  The simulations generate realizations of
the SALT2 model after applying real detection efficiencies,
survey zeropoints, PSF sizes, sky noise, host-galaxy noise,
and other observables from the real survey.  Simulations
must be tuned so that the uncertainties as a function
of magnitude match the data.  One of the primary underlying
reasons is that in SNANA simulations, the entire
PSF contributes to the noise computation, whereas in the
data we must choose a fixed radius within which to fit the PSF
and estimate the uncertainties.
Simulations also require that the underlying distributions
of SN light-curve shapes and colors be robustly determined.  These
distributions were inferred using the method of
\citet{Scolnic16}, and will be discussed in more detail
in Scolnic et al.\ (in prep).

The SNANA simulations are complex,
and detailed discussions of the MDS
simulations are given in \citet{Jones17} and
\citetalias{Jones18}.  Here, we focus on
the Foundation simulations, which are
illustrated in Figure \ref{fig:foundsim}.
See also \citet{Kessler18} for a detailed description
of the SNANA simulation process (an overview is
given in their Figure 1).

Approximately $60$\% of the Foundation DR1 data are discovered by two magnitude-limited
surveys: ASAS-SN \citep{Holoien17} and the Pan-STARRS Survey for
Transients (PSST; \citealp{Huber15}).  The spectroscopic
follow-up observations of ASAS-SN-discovered SNe are
100\% complete \citep{Holoien17}, while the PSST follow-up
observations are incomplete and must be modeled empirically.
In addition to biases from the photometric discovery surveys,
  there are additional complex biases from spectroscopic followup.
  ASAS-SN, for example, is 100\% spectroscopically complete, but
  Foundation can only follow SNe Ia if the spectra are obtained
  before maximum light. The other surveys have similar complexities.

We must therefore determine the selection efficiency empirically.  We first generate
a Foundation simulation without any spectroscopic
selection cut and compare the observed peak magnitude distribution
to that of the data.  Dividing the two distributions gives
the efficiency of follow-up observations as a function of magnitude.
This procedure follows the one used for the Pantheon sample \citepalias{Scolnic18},
and additional details about the Foundation spectroscopic
efficiency will be given by Scolnic et al.\ (in prep).
When the Foundation sample is larger it will be possible
  to empirically model the samples from each discovery
  survey individually, but currently this approach is 
  limited by statistical uncertainty.

  Survey of origin likely contributes to some of the dispersion
    in SN properties seen in Figure \ref{fig:foundsim}, but we also see that the distance bias
  is most strongly affected by the need to spectroscopically
  classify SNe\,Ia rather than by the detection limit of
  the individual surveys.  At higher-redshift, PSST has a deep
  detection threshold and is the dominant discovery survey, but
  the SNe included in Foundation are still preferentially brighter, bluer, and with
  broader light curve shapes than the lower-$z$ samples such as
  ASAS-SN.  ASAS-SN SNe are in fact a bit redder and lower-stretch
  than the average Foundation SN, even though ASAS-SN has a relatively
  bright detection limit.  A Kolmogorov-Smirnov test finds p-values
  of 0.30 and 0.32 for the ASAS-SN $x_1$ and $c$ distributions compared
  to the rest of the Foundation sample, showing that these distributions
  are consistent with those of Foundation as a whole.

Additionally, ASAS-SN is more efficient at finding SNe
  near the cores of their host galaxies than other
  surveys \citep{Holoien17}.  There has been
some investigation of whether SNe near their host galaxy
centers have biased distance measurements \citep{Hill18},
but no evidence has yet been found.  We explore the possibility
of this bias using the public data
of \citet{Jones18b}, which includes much of the Foundation
sample; we use the SN/host galaxy $R$ parameter to divide the sample into low-$R$
and high-$R$ subsets (see the discussion of $R$ in \S\ref{sec:host}).  We find the measured Hubble residuals
are largely insensitive to $R$, with a maximum 1.6$\sigma$ difference
of $0.058\pm0.035$ mag between SNe\,Ia with $R > 0.5$ and SNe\,Ia
with $R < 0.5$ (likely even less significant than 1.6$\sigma$ as we have
not accounted for the look-elsewhere effect).  Because 14\% of the SN\,Ia in the high-$z$ sample
have $R < 0.5$ and just 7\% of the SNe in the low-$z$
sample have $R < 0.5$, we also do not expect a strong redshift dependent bias
even if this effect is measured to be significant in the
future.  Na\"{\i}vely, the $z$-dependent bias would be
the difference in percent of $R < 0.5$ SNe between low-$z$ and high-$z$ samples multiplied by the
size of the difference, which gives a bias of $\sim 3$ mmag.  We do not model this aspect of possible
distance measurement bias, whether due to statistical fluctuation, SN physics, host
galaxy dust, or photometric measurement bias, but note
that this may be a necessary area to explore as measurements
of $w$ with SNe\,Ia become increasingly precise.

We simulate two models for the scatter of SNe Ia, with
distributions of $x_1$ and $c$ for each model given
by \citealp{Scolnic16}.  The standard SALT2 error model \citep[hereafter G10]{Guy10} attributes
$\sim$75\% of SN dispersion to variation in SN luminosity
that is uncorrelated with color.  Approximately 25\% of
the dispersion is given by wavelength-dependent flux variation
that is uncorrelated with luminosity.  The \citet{Chotard11}
model (hereafter C11), on the other hand, attributes most dispersion to variation
in color that is uncorrelated with luminosity.
\citet{Kessler13} translated the G10 and C11 models of broadband
covariance into wavelength-dependent models that can be simulated
as spectral variations, and these model are the basis for our C11 simulations.
We simulate
both models, as the data are unable to distinguish between them.
The most significant consequence of changing between these
two models is that the inferred $\beta$ changes
by $\sim$0.5 (noted by \citealp{Scolnic14}).

Figure \ref{fig:foundsim} compares the Foundation simulations
to the data.  In panel A, there is a slight discrepancy on the
faint side of the Hubble residuals; the data show 5 SNe in this
bin, while the simulations predict just one.
However, the data in the fainter bins have somewhat larger uncertainties
in the x-direction, which makes the number of SNe in the fainter
bins more subject to statistical fluctuation.

In Panel C, we note that simulating just a single value
of $\beta$ appears to reproduce the observed data well.
However, there is substantial evidence from UV to NIR observations
that SN\,Ia extinction laws vary significantly from SN to SN
(e.g., \citealp{Amanullah15}), which could be partly explained by SN
radiation pressure impacting the grain sizes of the dust distribution near the SN \citep{Bulla18}.
There is also evidence that the $\beta$ parameter similarly varies
in different types of host galaxies (e.g., \citealp{Jones18b}).
It is therefore likely that some of the scatter of SN\,Ia about the Hubble diagram
is due to variation of dust properties.  Although we do not fit for $\beta$
for individual SNe (e.g. \citealp{Burns14,Burns18}),
for measurements of $w$ we are concerned
primarily with whether the average value of $\beta$ $-$ due to the redshift dependence
of dust properties or SN\,Ia properties $-$ is changing with redshift.
In \S\ref{sec:syserr}, we
therefore test whether allowing $\beta$ to evolve with redshift could be systematically
biasing our cosmological results.  Because SN\,Ia host galaxies in Foundation are
more similar to high-$z$ host galaxies as discussed in \S\ref{sec:host}, we anticipate that
this analysis is less sensitive to this type of potential bias than previous analyses.

Finally, in panels E and F there are occasionally large bin-to-bin
jumps in average $x_1$ and $c$ as a function of redshift.  While
some of these jumps may be due to statistical fluctuation, it is
also likely that our simulations are not a perfect description of
the underlying data due to approximations in modeling the multiple
sub-surveys that comprise Foundation.  To explore whether changing
the $z$-dependence of $x_1$ and $c$ affects the predicted distance biases, we
simulated a Foundation-like survey with an extremely $z$-dependent mean $x_1$ and $c$.
We smoothly evolve the mean $x_1$ from $-1$ at $z = 0$ to $+2$ at $z = 0.08$ and $c$ from
a mean of $0$ at $z = 0$ to a mean of $-0.15$ at $z = 0.1$.
Even these extreme variations in the average color result in 
distance biases that change by a maximum of 3 mmag with no significant systematic trend.

Figure \ref{fig:foundsimcomp} demonstrates that the S/N of SNe in Foundation
simulations are a better match to the
real data than the previous low-$z$ sample simulations
were to the previous CfA/CSP data
\citep{Riess99,Jha06,Hicken09a,Hicken09b,Hicken12,Contreras10,Folatelli10,Stritzinger11}.
In particular, the highest-S/N SNe from CfA/CSP
are not represented in the simulations and the lowest-S/N
data does not precisely follow the prediction given by
the simulations.  The reduced $\chi^2$ of the CfA/CSP
data compared to the simulations is 5.4
(although the exact value depends on the bin sizes).
The large reduced $\chi^2$ is largely because of
the heterogeneous nature of the low-$z$ SN compilation,
which came from a large number of separate surveys and the
sample selection criteria of those surveys are varied.
The Foundation data agree more
closely with the Foundation simulations, with a reduced $\chi^2$ of
1.9 when compared to the data.

These simulations are used to predict the distance bias as a function of redshift for each SN sample.
The PSBEAMS method of correcting for distance biases,
the standard approach
prior to the Pantheon analysis (e.g., \citealp{Conley11,Betoule14}),
used SNANA simulations to generate a one-dimensional correction to the SN distance
as a function of redshift.
An alternative approach is the BBC method
\citep{Kessler17}, which was used to derive cosmological
parameter measurements from the Pantheon sample \citepalias{Scolnic18}.
BBC applies bias corrections to three parameters
$-$ $m_B$, $x_1$, $c$ $-$ and those bias corrections
are also treated as a function of nuisance parameters $\alpha$ and $\beta$.
The $\alpha$ and $\beta$ dependence of the bias corrections
are included in the BBC likelihood.

The BBC approach reduces the scatter about
the Hubble diagram while explicitly correcting
the known dependence of Hubble residuals on $x_1$ and $c$.
The BBC method is also used to marginalize
over CC\,SNe in the present analysis, but BBC improves
the precision of cosmological parameter measurements even
when used in an analysis restricted only to spectroscopically
classified SNe Ia such as Pantheon \citepalias{Scolnic18}.
We use both the PSBEAMS method and the BBC method in this analysis.

\subsection{Simulating the CC\,SN Sample and Generating Prior Probabilities}
\label{sec:priors}

\begin{figure*}
\includegraphics[width=7.3in]{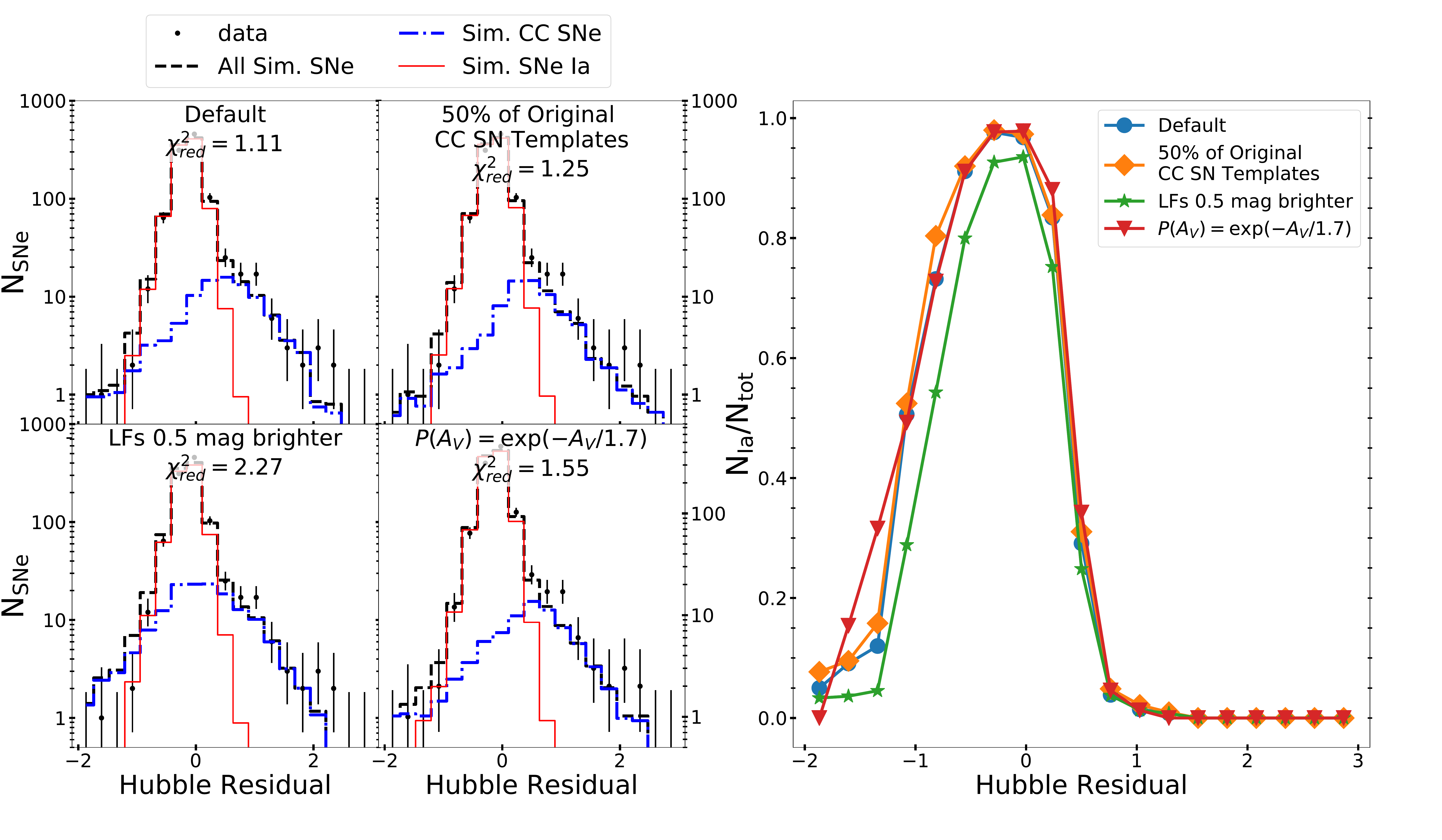}
\caption{Left: Adjusted CC\,SN simulations (blue) compared to the data (black points).
  Simulated SNe\,Ia are in red and the cumulative simulated CC$+$Ia distribution is in black.
  From top left to bottom right,
  we show the default simulations, simulations that reduce by 50\% the
  number of CC\,SN templates used to simulate the sample,
  simulations that brighten the intrinsic luminosity of the
  simulated CC\,SNe by 0.5~mag, and simulations that add additional extinction
  to the CC\,SN templates.  Right: The fraction of SNe\,Ia as
  a function of Hubble residual for each simulation.}
\label{fig:ccsntweaks}
\end{figure*}

\begin{figure}
\includegraphics[width=3.5in]{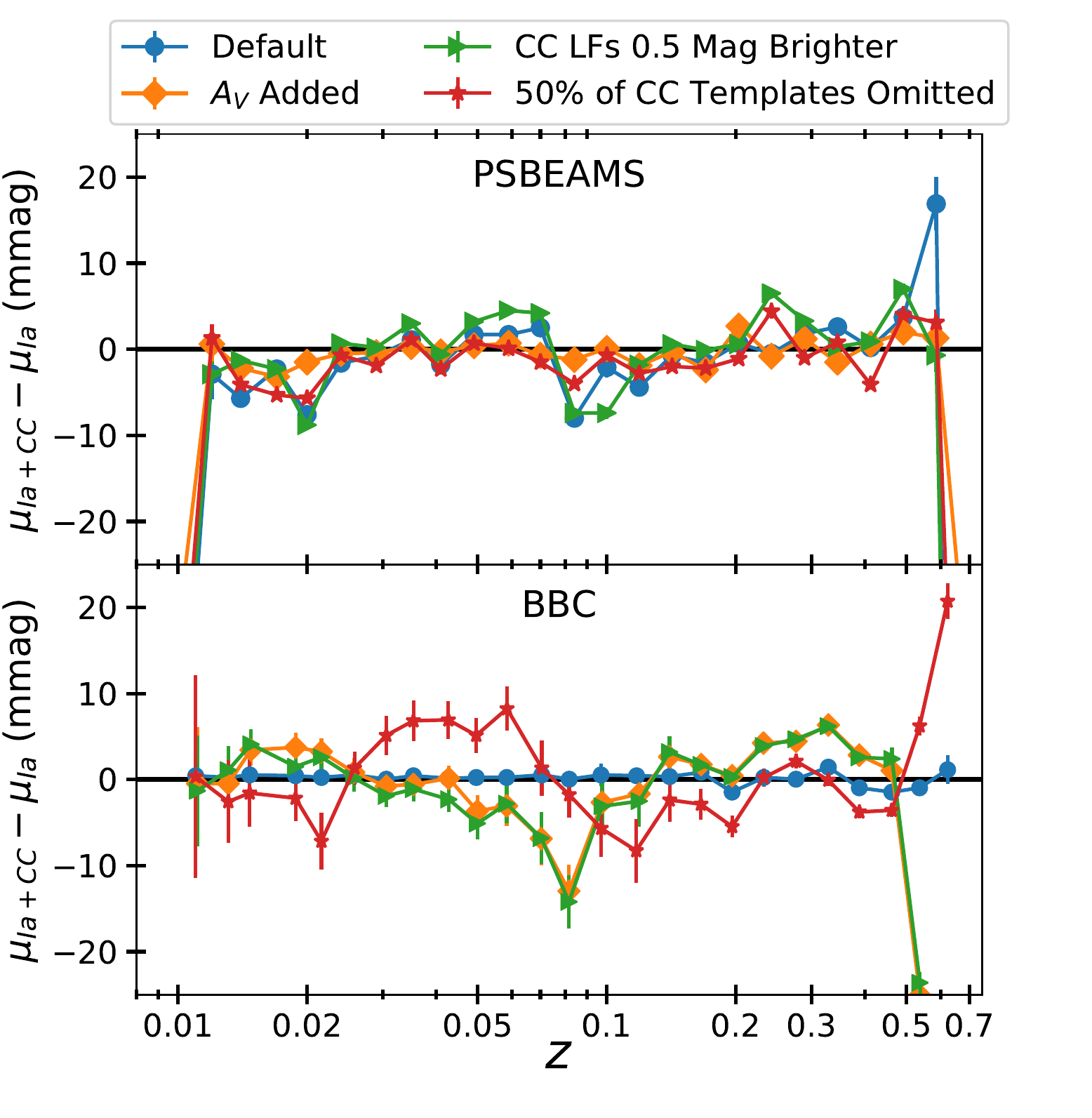}
\caption{For the PSBEAMS method (top) and BBC method (bottom), bias in binned SN\,Ia distance measurements due to CC\,SN
  contamination as a function of redshift.
  Distances are measued from the real
  spectroscopically classified sample combined with
  a simulated photometric (CC\,SN-contaminated) sample.  The simulated photometric sample uses
  one of 4 different CC\,SN simulations, which are discussed in
  \S\ref{sec:sys}.  Biases are typically $\lesssim$5~mmag for all
  simulated CC\,SN populations.}
\label{fig:ccmodbias}
\end{figure}

\begin{figure*}
\includegraphics[width=7in]{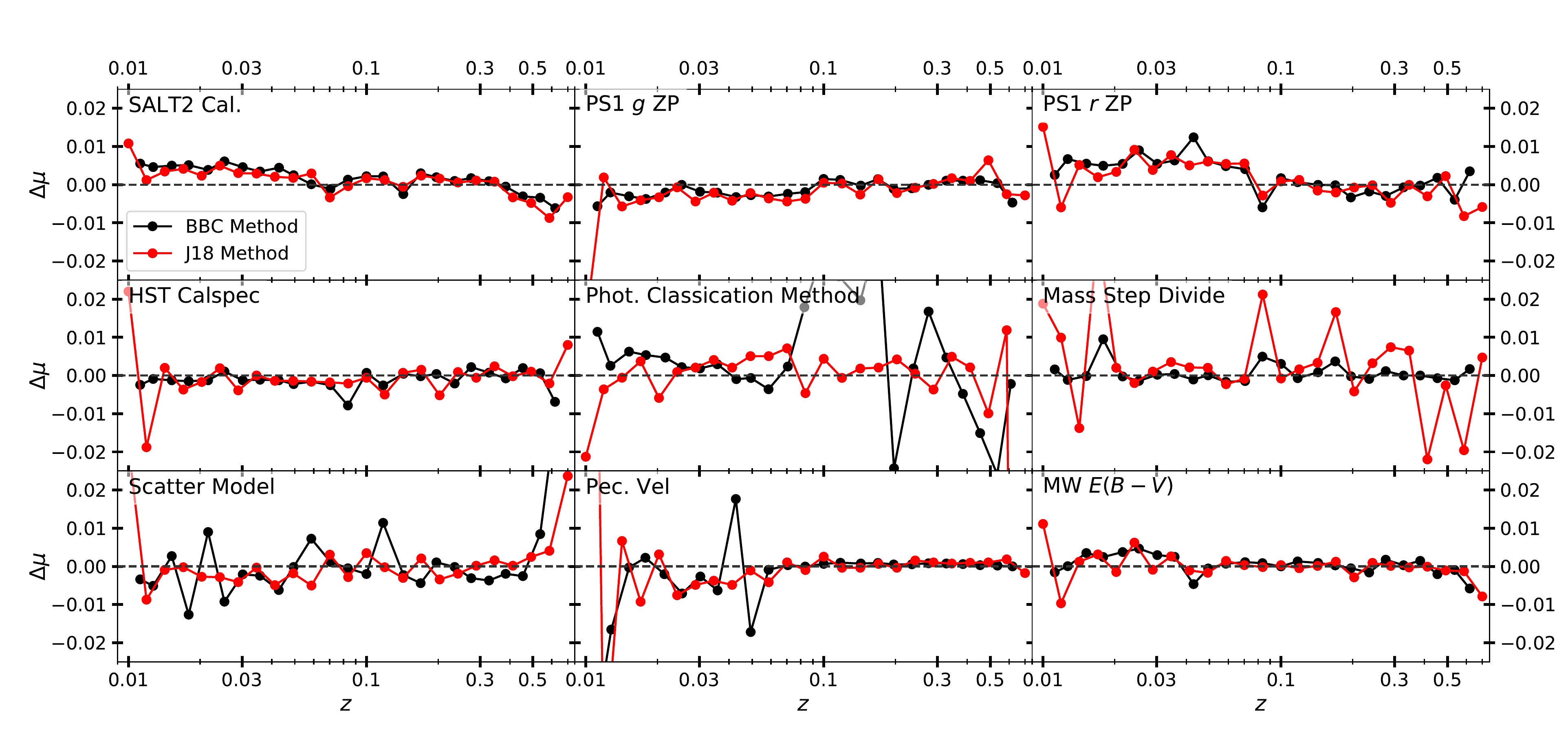}
\caption{Comparison of BBC systematic uncertainties (black) to systematic uncertainties
  from the PSBEAMS method used in \citetalias{Jones18} (red).  For several of the largest sources
  of systematic uncertainty, we show how the distances are affected as a function of
  redshift by that systematic uncertainty (we show a 1$\sigma$ shift).
``SALT2 Cal.'', ``PS1 $g$ ZP'', ``PS1 $r$ ZP'' and ``HST Calspec'' refer
  to the SALT2 calibration uncertainty, PS1 $g$- and $r$-band zeropoint uncertainties, and calibration uncertainty in the
  Calspec system as defined through {\it HST} observations,
  respectively.  ``Phot.\ classification method'' is the shift in distances
  caused by using a different method of estimating the prior
  probabilities that each SN is of Type Ia and ``Mass Step Divide'' refers
  to the systematic uncertainty if the divide between what is defined as low- and high-mass host galaxies
  is shifted by 0.15~dex (within current observational constraints).
  ``Scatter Model'' is the systematic uncertainty caused by the difference between the G10/C11 models.
  ``Pec.\ Vel.'' and MW $E(B-V)$ are uncertainties in the peculiar velocity correction and
  the Milky Way extinction.}
\label{fig:bbcsys}
\end{figure*}

\begin{figure}
\includegraphics[width=3.3in]{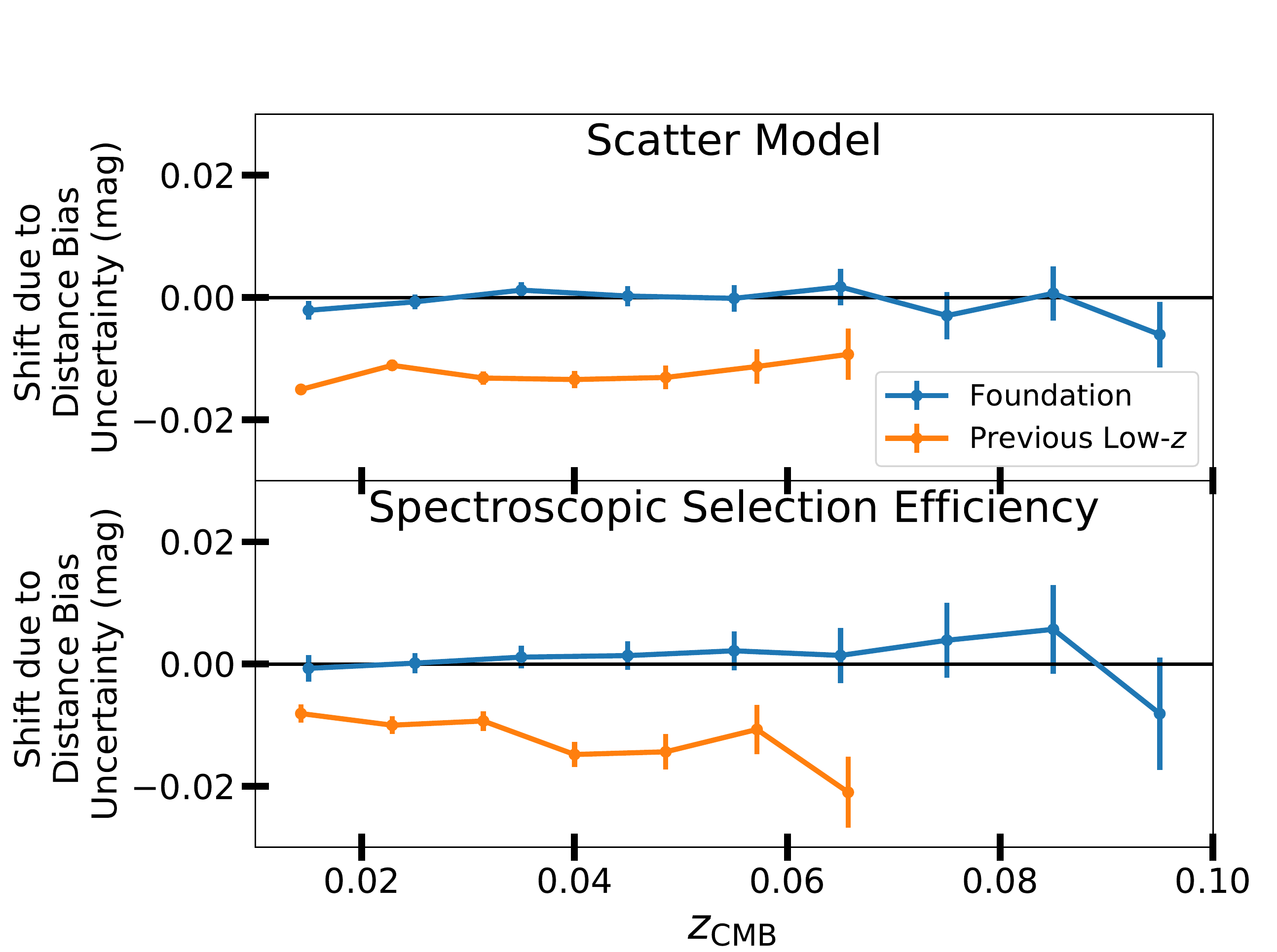}
\caption{Comparison of systematic uncertainties due to bias correction as a function of redshift
  between the Foundation sample (blue) and the previous low-$z$ SN sample (orange).  We
  show the shift in distance
  due to the difference between the distance bias predictions of the G10 and C11 models (top)
  and due to adjusting the uncertain spectroscopic selection efficiency (bottom).}
\label{fig:scatbias}
\end{figure}

Bayesian likelihood models such as those used in this work require
accurate simulations of the CC\,SN contamination in order to
  validate the method and yield estimates of the prior
  probability that each SN is of Type Ia.
In previous work, we estimated prior probabilities from
one of four different methods.  The most reliable of these
methods was the PSNID classifier \citep{Sako11}, which
compares noise-free simulations of CC\,SNe and SNe\,Ia
to the observed photometric data and gives a Bayesian probability
that each SN is of Type Ia.  The second-most reliable
method was Nearest Neighbor (NN; \citealp{Kessler17,Sako18}),
which compares the redshift and the SALT2 $x_1$ and $c$ parameters
measured for each SN
to simulated distributions of these
parameters.  NN gives a distance metric from data to simulated SNe\,Ia
and CC\,SNe to determine the classification.  The distance metric
relies on three free parameters, one each for $z$, $x_1$, and $c$,
which are determined during a training stage.

In \citetalias{Jones18} we also employed two classifiers that were independent of
a number of biases in CC\,SN simulations \citep{Jones17}.
The first, which we call \fp, used the $\chi^2$ and
degrees of freedom of the SALT2 model fit to estimate
the probability that the SALT2 model matches the data.
The second, GalSNID \citep{Foley13,Jones17}, uses the fact
that CC\,SNe are not found in old stellar populations to 
estimate the SN\,Ia probability from host-galaxy spectroscopy
and imaging alone.

Using simulations, we found that
GalSNID and \fp\ could bias $w$ by $\sim$0.02, but we
used them as systematic uncertainty variants due to their independence
of CC\,SN simulations.
However, these known biases are larger than necessary for the present analysis,
and therefore we instead use simulations of different CC\,SN
distributions to predict the small distance biases caused by
imperfections in our CC\,SN classification methods
(\S\ref{sec:priors}).

These methods were validated in \citet{Jones17}, but
  require separate validation for each sample and analysis as
  subtle differences in sample selection can change the effectiveness of our method of
  marginalizing over CC\,SNe.  These validation tests
  will be used as part of our systematic uncertainty budget (\S\ref{sec:sys}).
  
  We design three SNANA simulations to encapsulate the uncertainty
  in CC\,SN rates, luminosity functions, and the representativity of the available templates.
  We note that these simulations have already been tuned to match the MDS data
  by \citet{Jones17}, who found that the bright tail of the CC\,SN distribution
  is poorly constrained by \citet{Li11}, requiring the peak of the
  CC\,SN LFs to be adjusted by $\sim$1~mag\footnote{\citet{Jones17} do
    not suggest that the peak of the CC\,SN LFs as measured by \citet{Li11} is incorrect by $\sim$1 magnitude,
    but rather that an ad-hoc procedure to brightens LFs is capable of reproducing the
    observed bright tail of the CC\,SN distribution.  The MDS survey preferentially detected
    bright CC\,SNe.}.
  In the first simulation, we remove 50\% of the CC\,SN templates from the simulation to evaluate
  the effect of having an incomplete CC\,SN template set.  We randomly removed exactly 50\% of the templates
  for each CC\,SN subtype, so that all subtypes were still represented in the
  simulations.  In the second simulation,
  we increase the peak CC\,SN luminosity functions by an additional
  0.5~mag, which in turn makes them a higher fraction of the total SN population.
  In the third simulation, we apply a strong $A_V$ distribution to the data to account for the
  fact that templates may be preferentially unreddened compared to the
  real data.  These adjusted CC\,SN models are shown in
  Figure \ref{fig:ccsntweaks} along with the reduced $\chi^2$ of each model compared
  to the data.  Even though these adjustments are relatively drastic,
  all CC\,SN simulations have a reduced $\chi^2 < 2.5$, and therefore represent conservatively large
  adjustments to the CC\,SN simulation parameters that conservatively account for the uncertainties in
  modeling CC\,SN contamination.

  We then validate the method by replacing our photometrically classified SNe with each simulation, keeping
  the real spectroscopically classified SNe.
  We use PSBEAMS and BBC to measure distances from the sample (see below) both
  with and without including simulated
  CC\,SNe.  The difference in these two sets of distances is the distance biases
  that are introduced by our method of marginalizing over the CC\,SNe.
  We correct for the distance biases predicted for
  the baseline simulation and treat the biases from each of the three
  simulation variants, relative to the baseline biases, as systematic uncertainties.
  These systematic uncertainties are typically less than 5~mmag and are shown in
  Figure \ref{fig:ccmodbias}.  We note that the BBC method gives large biases at high-$z$
  for some simulations, but those biases primarily affect some of the largest-uncertainty bins.  The
  PSBEAMS points at $z = 0.7$ also have large uncertainties (as no SNe at $z > 0.7$ are used)
  and bias in these bins has minimal impact on cosmological parameter estimation.
  These offsets are included in the systematic uncertainty budget.

\subsection{The Likelihood Model}
\label{sec:likelihood}

Both frameworks that we use to analyze our
sample (PSBEAMS and BBC) are based on the BEAMS method, first presented
by \citet{Kunz07} and refined by \citet{Hlozek12}.
BEAMS is a Bayesian method for simultaneously modeling
multiple ``species'' that are partially overlapping in some
parameter space.  In this case, BEAMS models
SNe\,Ia and CC\,SNe, which overlap on the Hubble diagram.
A brief overview of the method is given in the appendix,
with the full mathematical formalism for the PSBEAMS and
BBC methods given in \citetalias{Jones18} and \citet{Kessler17}, respectively.

The primary difference between the methods, apart from the bias correction
differences discussed above, is that BBC uses SNANA simulations of the CC\,SN
distribution to put a prior on the 
$z$-dependent Hubble residuals expected from CC\,SN contamination.
This avoids the assumption of a redshift-dependent functional form for the CC\,SN distribution.
However, a parametric CC\,SN
model from \citet{Hlozek12} can also be specified within BBC, which allows more flexibility.
In contrast to the PSBEAMS method, which uses a parameterized
CC\,SN model that is linearly interpolated in log($z$) space,
\citet{Hlozek12} treats the CC\,SN distances and dispersion
as polynomial functions of $z$.

For the PSBEAMS method we remove likely CC\,SNe ($\mathrm{P(Ia) < 0.5}$),
following \citetalias{Jones18}, finding that the results are more robust and less affected by systematic uncertainty
variants after
this cut is applied.  This cut reduces the sample to 1085 SNe but does not significantly affect
the precision of the cosmological constraints.  For the BBC method we include likely CC\,SNe,
as our preference is to use all available data, even those data with low weight, when possible.

In addition,
unlike the PSBEAMS method, the BBC method does not have a parameter to
linearly shift $\mathrm{P(Ia)}$ values to adjust the prior probabilities
that a SN is Type Ia (Eq. \ref{eqn:norm}).  This parameter helps to correct
inaccurate light-curve classifications.  Because of this difference, for PSBEAMS we use PSNID as the
baseline classifier for determining prior probabilities that
a given SN is of Type Ia and NN as the baseline classifier for BBC,
as we found that excluding a linear shift parameter can give biased results when using
PSNID probabilities.
See the appendix for further explanation.

\subsection{Systematic Uncertainties}
\label{sec:sys}

The systematic uncertainties in this analysis are largely
unchanged from the MDS analysis of \citetalias{Jones18}.
Therefore, we summarize them here and direct the reader to \S4
of \citetalias{Jones18} for a more detailed description of
each systematic uncertainty.

The systematic uncertainties affecting SNe in this sample can be
attributed to 7 broad categories: Milky Way extinction,
distance bias/selection effect correction, photometric calibration,
SALT2 model calibration, sample contamination by CC\,SNe or incorrect redshifts,
low-$z$ peculiar velocity corrections,
and the dependence of SN\,Ia luminosities on their host galaxies.
Additionally, we check for the redshift dependence of
$\Delta_M$ and $\beta$, but find no significant evolution
in their values.
An example from each type of systematic uncertainty is
shown in Figure \ref{fig:bbcsys}, using both the PSBEAMS
method and the BBC method.

Replacing the previous low-$z$ sample with the Foundation sample has
reduced the low-$z$ distance bias systematic uncertainty from $\sim$1--1.5\% to
just a few mmag.
The reduced distance bias systematic uncertainty can be attributed to two effects,
both shown in Figure \ref{fig:scatbias}.  First,
\citetalias{Scolnic18} and \citetalias{Jones18} corrected
for distance biases considering that $z < 0.1$ SNe may either be from a volume-limited or
magnitude-limited sample, as the data come from surveys that
often targeted nearby, massive galaxies.  The simulations from \citetalias{Jones18} and the Pantheon analysis (\citetalias{Scolnic18}) indicated that
even some $z \approx 0.03$ SNe may have been missed if the survey was magnitude-limited.  For
the Foundation Supernova Survey,
we understand that our sample is dominated by magnitude-limited data
and therefore we do not include a variant that bias-corrects the
sample as though it were volume-limited.  Second, likely because
the Foundation Supernova Survey is not a targeted search, we have a sample
with a bluer distribution of SN colors that more closely matches
the high-$z$ sample.  Foundation SNe have a median $c$ parameter of -0.020
while SNe in the previous low-$z$ sample have a median $c$ of 0.004.
The high-$z$ PS1 data have a median $c$ of -0.017.  Comparisons of $x_1$
and $c$ for the Foundation sample compared to previous low-$z$ data
are shown in \citet{Foley18}, their Figure 7.
Because the average SN color varies less with redshift,
the difference between the distance bias
predictions from the G10 and C11 scatter models is much smaller.

The size of the photometric calibration systematic uncertainty has increased,
which is due to the fact that we have only a single
photometric system in this analysis.  Although PS1
calibration uncertainties are low $-$ just 3~mmag
per filter (Scolnic et al.\ in prep; \citealp{Scolnic15}) $-$ multiple uncorrelated
photometric systems would reduce the calibration systematic
uncertainty further, as would the ability to include the Foundation $iz$ observations.
In addition, because the bluest band in the sample is $g$,
high-$z$ SN observations measure much bluer rest-frame wavelengths
than the low-$z$ observations.  This increases the impact
of the PS1 calibration systematic uncertainties, the SALT2 calibration
systematic uncertainties, and the 0.5\% slope uncertainty in the {\it Hubble Space Telescope (HST)}
CALSPEC system \citep{Bolin14}.
For context, a 3~mmag change in distance modulus from
the median redshift of the Foundation sample ($\sim$0.035) to the median redshift of the MDS sample ($\sim$0.35) would
shift $w$ by approximately 1\%; however a shift in a single filter can also bias SN color measurements,
and therefore the shift in distance can be larger than 3~mmag in practice (Figure \ref{fig:bbcsys}, top right panel).
The systematic uncertainty associated with
the Supercal correction is not included in this analysis (and was
included in \citetalias{Jones18}), because all of our data are already on the
PS1 photometric system.

We have slightly altered the \citetalias{Jones18} method of accounting for
the systematic uncertainty caused by marginalizing over
CC\,SNe.  We first use an alternate method of estimating
prior probabilities (either NN or PSNID, depending on the
baseline choice for each analysis framework) and treat
the change in distances as a systematic uncertainty.
Second, for the PSBEAMS framework, we use two alternate
models of the CC\,SN distribution: a skewed Gaussian or two-Gaussian
parameterization of the CC\,SN distribution, and treat the average
of these two parameterizations relative to the baseline parameterization
as a systematic uncertainty.
When using the BBC framework, we use a simulated CC\,SN distribution
as the baseline CC\,SN distribution and
use the \citet{Hlozek12} parameterized CC\,SN model as a systematic uncertainty variant.
Additionally, using the four CC\,SN simulations discussed above,
we correct for the distance biases predicted for the baseline simulation
and treat the biases from each of the three simulation variants,
relative to the baseline biases, as systematic uncertainties.
Finally, we test the effect of fixing nuisance parameters to
  the values determined from just the spectroscopically classified data
and include the differences in distance in our systematic uncertainty budget.

Incorrect redshifts due to noisy spectra and mismatched
host galaxies are also a source of contamination
that is discussed in detail in \citet{Jones17}.
In the high-$z$ sample presented here, we expect of order $1.2\pm0.5$\%
of the redshifts to be incorrect.  We implicitly model
these as part of the CC\,SN distribution, as simulations
show that they are nearly always outliers on the Hubble
diagram and do not follow the same distribution as SNe\,Ia.
As discussed in \citet{Jones17}, simulations show
that these outliers are not a major source of systematic error.
See \citet{Roberts17} for an alternative method of accounting
for redshift contamination that includes incorrect redshifts
in the BEAMS likelihood itself.

Finally, extinction by dust in the IGM is a potential source of systematic
uncertainty \citep{Menard10,Goobar18}.  \citet{Goobar18} argue that a
slight discrepancy in the low-$z$ data used by the Pantheon sample
could be due to IGM dimming.  However, we note that the Foundation sample is slightly fainter than $\Lambda$CDM (\S\ref{sec:results})
whereas the Pantheon low-$z$ data were slightly brighter, and therefore it appears more likely that
the differences in distance between Pantheon and $\Lambda$CDM at low-$z$
are due to survey-specific systematic uncertainties.  We do not include this effect
in our systematic uncertainty budget as it is not detected in the Foundation sample, but
note that future Foundation analyses, when
the $iz$ bands are included in the analysis, will be better able to put
constraints on the ways in which IGM dust and variable $R_V$ affect SN\,Ia distances.

\section{Results}
\label{sec:results}

\begin{figure*}
\includegraphics[width=7in]{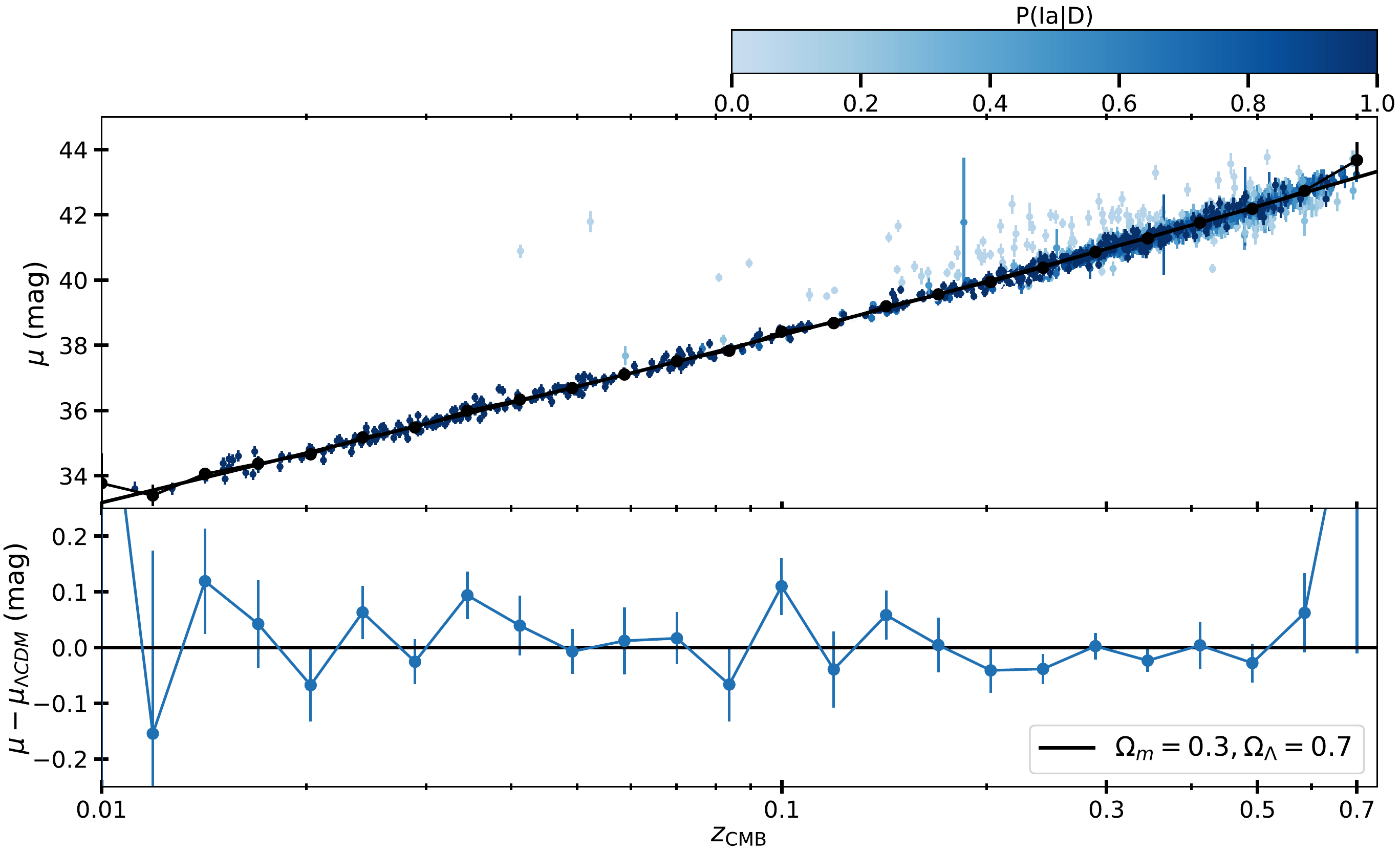}
\caption{Hubble diagram (top) and Hubble residuals (bottom) from the
  combined Foundation and MDS sample.  In the top panel, opacity is set using
  the approximate posterior probabilities, $\mathrm{P(Ia|D)}$, for the photometrically
  classified data.  In the bottom panel, the points and the lines connecting the points
  represent the piecewise-linear function of log($z$) that
  we use to fit the SN\,Ia distances (see appendix).
  Note that the highest- and lowest-redshift control
  points have extremely high uncertainties
  as no SNe are above or below them in redshift, respectively.  Residuals
  are shown compared to a nominal flat $\Lambda$CDM model with
  $\Omega_m = 0.3$ and $\Omega_{\Lambda} = 0.7$.}
\label{fig:foundhubble}
\end{figure*}

\begin{figure*}
  \centering
  \includegraphics[width=7in]{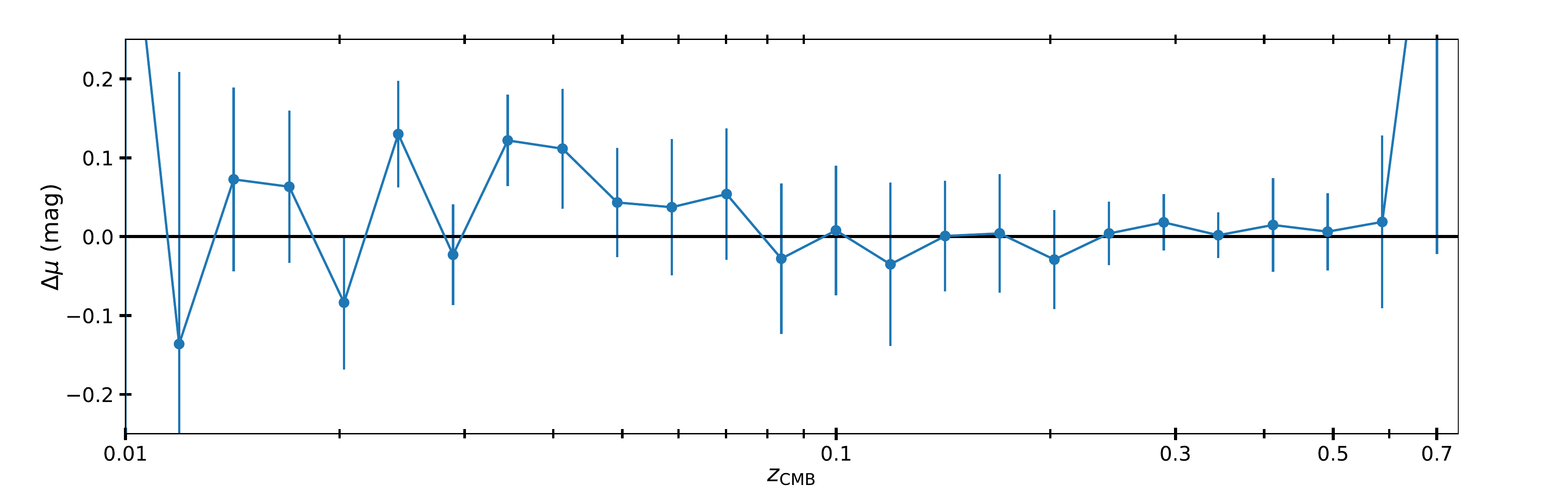}
  \caption{As a function of redshift, Hubble residuals from the
    combined Foundation$+$MDS sample subtracted by those from the previous low-$z$$+$MDS sample (\citetalias{Jones18}).
    Foundation distances are \lowzdistoff~mag
    fainter than those from the previous low-$z$ sample, which gives a positive
    change in measured $w$.  The highest- and lowest-redshift points have extremely high uncertainties
    as no SNe are above or below them in redshift.}
  \label{fig:foundhubbleres}
\end{figure*}

In this section, we begin by discussing the change in distances
when the previous sample of low-$z$ SNe\,Ia (CfA/CSP SNe) are replaced by Foundation SNe\,Ia.
Distances and cosmological parameters for the combination of CfA, CSP and MDS
SNe were reported in \citetalias{Jones18}; therefore, we begin with the
PSBEAMS method from \citetalias{Jones18} of correcting for distance biases and marginalizing
over CC\,SNe to allow a direct comparison to the \citetalias{Jones18} results.
We then discuss the results from the alternate BBC method, which
reduces the dispersion of the SN\,Ia sample and therefore improves the
precision of the results, albeit with additional differences
in methodology.

The full SN\,Ia sample used to measure cosmological parameters
from the Pan-STARRS1 telescope is shown in Figure \ref{fig:foundhubble}.
The binned distance residuals from SNe\,Ia
(after marginalizing over all nuisance parameters) are shown
relative to the binned distance residuals from the previous low-$z$ SN sample in
Figure \ref{fig:foundhubbleres}.  For a given redshift, Foundation SNe\,Ia have
larger distances on average, and a corresponding positive shift in Hubble
residuals, compared to SNe Ia from the previous low-$z$ sample.  The change
in distances from the $z < 0.1$ sample is
\lowzdistoff~mag (a weighted average including systematic
uncertainties and covariances), a significance of \lowzdistsig.  We note that
this difference is assuming the average of the C11 and G10 scatter models.
If only the G10 scatter model is used (neglecting the scatter model systematic uncertainty),
the difference would be
$0.030 \pm 0.023$~mag, which has a lower significance of 1.3$\sigma$.  Similarly, using only
the C11 model gives a $0.062 \pm 0.023$~mag difference, which may suggest that
the G10 model is favored by these data.  Systematic and statistical uncertainties
on this difference are approximately equal.  We note that the value of $\Delta_M$ measured
by PSBEAMS in this work ($0.088 \pm 0.013$) is 0.014~mag lower than than the
$\Delta_M$ value measured in \citetalias{Jones18}, too small of a shift to
cause the observed shift in distances.

To test if these low-$z$ distances could be systematically affected by fitting light curves
  where the bluest band is rest-frame $g$, we re-fit the CSP data with $gr$ photometry alone (neglecting $BV$ data),
  finding that distances were an average of 1.5\% fainter at 2.8$\sigma$ significance
  from statistical uncertainties alone.  However, this possible bias is just one
  third of the total shift in SN\,Ia distance when the previous low-$z$ data are replaced
  by Foundation.  Furthermore, as SNe\,Ia are better-calibrated and better standardizable candles at redder
  wavelengths, it may be that the $gr$-only results are less subjected to systematic
  SALT2 training or calibration uncertainties than the CSP $B$ measurements.

\begin{table}
  \centering
  \caption{\fontsize{9}{11}\selectfont Dependence of Nuisance Parameters on
    Analysis Method for MDS+Foundation}
  \begin{tabular}{lrrrrr}
\hline \hline\\[-1.5ex]
  &\multicolumn{2}{c}{PSBEAMS Method}&&\multicolumn{2}{c}{BBC Method}\\
  &&$\sigma_{\mathrm{stat}}$&&&$\sigma_{\mathrm{stat}}$\\*[2pt]
\hline\\[-1.5ex]
$\alpha$&0.162&0.007&&0.143&0.005\\
$\beta$&3.126&0.073&&3.218&0.071\\
$\sigma_{Ia}$&0.102&0.006&&0.086&\nodata\\
$\Delta_M$&0.088&0.013&&0.044&0.010\\
$\beta_{0}$&3.461&0.439&&3.362&0.277\\
$\beta_{1}$&$-$1.542&1.440&&$-$0.660&1.145\\
$\Delta_{M,0}$&0.065&0.019&&0.054&0.020\\
$\Delta_{M,1}$&0.082&0.054&&$-$0.038&0.060\\
\\[-1ex]
\hline\\[-1.5ex]
\multicolumn{6}{l}{
\begin{minipage}{6.8cm}
  Nuisance parameters measured using the PSBEAMS method
  compared to those measured using the BBC method.
  $\beta_0$, $\beta_1$, $\Delta_{M,0}$, $\Delta_{M,1}$ are intercept and slope
  parameters defining the linear $z$-dependence of $\beta$ and $\Delta_M$
  (Equation \ref{eqn:evol}).  $\beta$ evolution
  is detected with the PSBEAMS method, but is an artifact of the analysis
  method as shown in Figure \ref{fig:beta}.  Uncertainties on
  $\beta_0$ and $\beta_1$ are measured from
  the dispersion of simulations rather than the statistical uncertainties
  reported by the method.
\end{minipage}}
  \end{tabular}
\label{table:nuisancebbc}
\end{table}

\subsection{Nuisance Parameters}
\label{sec:nuisanceresults}

Nuisance parameters from the PSBEAMS and BBC analysis frameworks are reported
in Table \ref{table:nuisancebbc}.
All PSBEAMS measurements are consistent with the measurements
from previous low-$z$ data combined with MDS (reported in \citetalias{Jones18}),
although the systematic uncertainties due to the intrinsic dispersion
are significantly lower in the present analysis.  This may be because Foundation
SNe\,Ia have a lower dispersion that is more similar to the MDS data \citep{Foley18}.

The SALT2 $\alpha$ and $\beta$ parameters from the BBC method
are slightly different than those from the PSBEAMS method.
With BBC we find $\alpha = 0.143 \pm 0.005$ and $\beta = 3.218 \pm 0.071$
(stat. error only);
compared to the PSBEAMS method, $\alpha$ is lower by 0.019 and
$\beta$ is higher by 0.092.  The difference in $\alpha$ is marginally
statistically significant, and must be driven by the Foundation sample as both
\citetalias{Jones18} and \citetalias{Scolnic18} measured the value of $\alpha$ from
the MDS sample to be $>$0.16.  With the BBC method, we measure $\alpha$ from the
Foundation sample alone to be $0.137 \pm 0.013$.

The reason why the Foundation sample
$\alpha$ would be particularly affected by observational biases
is unclear, and such a bias is not recovered in simulations
of the sample.
From simulations, we confirmed that both PSBEAMS and BBC recover $\alpha$ and $\beta$
accurately in a spectroscopically classified sample, though when CC\,SN contamination is
included $\beta$ is biased by an average of $-0.03$ by the BBC method
and by $+0.09$ by the PSBEAMS method (see \citealp{Jones17} for more discussion of these biases).
Similarly to the Pantheon results, BBC reduces the sample dispersion
substantially $-$ by 16\% in this analysis $-$ compared to 1D bias correction methods.

The mass step, $\Delta_M$, from BBC is $0.044 \pm 0.010$~mag,
lower than the PSBEAMS mass step by 0.044~mag.
We find that the difference between the BBC and PSBEAMS
mass step is driven by the strong implicit dependence of the average $x_1$ bias correction
on host-galaxy mass.  The photometric sample has a number
of high-mass host galaxies, as these are preferentially more
likely to yield redshifts, and SNe\,Ia in these galaxies have narrower light
curves on average than those in low-mass galaxies \citep{Howell01}.
SNe with narrower shape parameters tend to have somewhat negative
Hubble residuals due to observational bias \citep{Scolnic16},
an effect corrected by the BBC
framework and not PSBEAMS.  A lower mass step from the BBC
method was also seen in the Pantheon analysis.
These findings also agree with \citet{Brout18},
who see a positive correlation between the
size of $\Delta_M$ measured from a given sample
and the intrinsic dispersion of that sample.

\subsubsection{Nuisance Parameter Evolution}

We test for redshift dependence of $\beta$ and $\Delta_M$ by allowing
those parameters to evolve linearly with redshift:

\begin{equation}
  \begin{split}
  \Delta_M &= \Delta_{M,0} + \Delta_{M,1} \times z,\\
\beta &= \beta_0 + \beta_1 \times z.
  \end{split}
  \label{eqn:evol}
\end{equation}

\noindent Because the MDS sample has a median redshift of $\sim$0.35, the
redshift range that can be used to constrain these parameters is limited.
A large $\beta_1$ or $\Delta_{M,1}$ coefficient does not imply that
a trend is observed to the maximum $z = 0.7$.

\begin{figure}
  \includegraphics[width=3.5in]{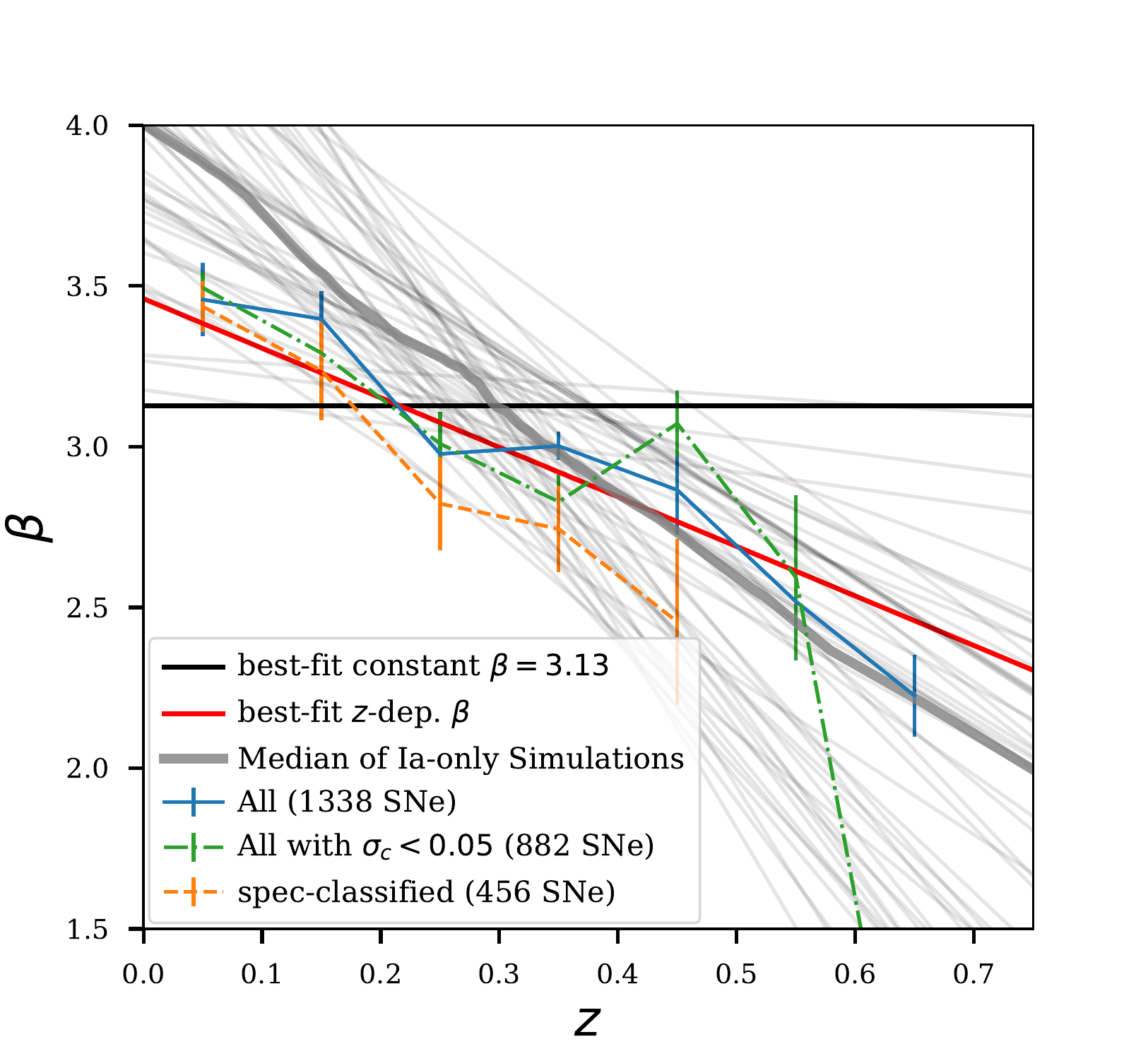}
  \caption{For the PSBEAMS method, evolution of the nuisance parameter $\beta$ in
    the full sample (blue), the full sample with a conservative $\sigma_C < 0.05$ (green),
    and the spectroscopically classified sample (orange).
    However, simulated SN samples (grey) with a constant simulated $\beta$
    also show evidence for $\beta$ evolution.  The BBC method does not find
    evidence for $\beta$ evolution, either in simulations or data.}
\label{fig:beta}
\end{figure}

From MDS, CSP and CfA SNe, \citetalias{Jones18} found a marginal ($1.6\sigma$) detection of $\beta$
evolution in the SN\,Ia data.  In this analysis, using a more homogeneous
data set, the PSBEAMS method gives a $\sim 3$-$\sigma$ detection of $\beta$ evolution
while the BBC method finds no evidence for $\beta$ evolution.
For this reason, the trend is likely due to observational biases
or an oversimplified analytic treatment.  However,
we investigated further for the PSBEAMS method by 
simulating SNe with a fixed input $\beta$ and then allowing $\beta$
to float in our likelihood model.  The results of this test are shown in
Figure \ref{fig:beta} along with the measured $\beta$ evolution for both
the full and spectroscopically-classified samples.

The simulations have a wide dispersion in measured $\beta$ evolution, which
may be due in part to the noisy high-$z$ data, or perhaps the limited
redshift range of the sample.  However, the general trend is negative,
and our results from the data are consistent with these simulations,
implying that the observed $\beta$ evolution is
unlikely to be a physical effect.  We therefore do not
include it in our systematic uncertainty budget.
Measurements of $\beta$ evolution with the BBC method
  do not have a statistically significant bias but do have
  wider dispersions than the measured 1$\sigma$ uncertainties by factors of 2$-$3 (Table \ref{table:nuisancebbc}).

We note that if $\beta$ is fixed as a function of redshift, the resulting
value is not biased by the PSBEAMS analysis method (but can be biased by inaccurate priors
when CC\,SN contamination is present).  Similarly, the BBC framework
explicitly corrects for both the bias in $\beta$ as well as
the $z$-dependent bias in distance when $\beta$ is allowed to vary with redshift.

We do not measure significant evolution of the
host-galaxy mass step $\Delta_M$ with either BBC or the PSBEAMS
method.  However, we report $\beta(z)$ and $\Delta_M(z)$ in Table \ref{table:nuisancebbc} and
in \S\ref{sec:syserr} estimate what their contribution to the systematic uncertainty budget 
would be had they been included in the error budget.

Hints of a non-zero $\beta(z)$ or $\Delta_M(z)$ were found in \citetalias{Jones18} and \citetalias{Scolnic18},
respectively, while no evidence for an $\alpha(z)$ term has been found.  There is also
somewhat more physical motivation for $\beta$ and $\Delta_M$ evolution:
dust properties are expected to evolve with redshift and progenitor ages, which also evolve with redshift, may drive
the $\Delta_M$ step \citep{Childress14}.
However, we also explore the possibility of $\alpha(z)$ here using the BBC method, finding that in our sample
$\alpha(z)$ is only significant at the 1.5$\sigma$ level using both the G10 and C11 scatter models.
We measure $\alpha(z) = 0.155\pm0.011 - 0.05740\pm0.038 \times z$ using the G10 model and a nearly identical
step with the C11 model.

We note that if the SN\,Ia luminosity were evolving independent of changes
in $\alpha$, $\beta$, or $\Delta_M$, we would be unable to distinguish this
evolution from changes in $w$.  However, if such evolution were physical it would
cause much larger discrepancies with $\Lambda$CDM, appearing
to favor exotic dark energy, in studies with a larger redshift baseline such
as \citet{DES18} or \citetalias{Scolnic18}.  No such effect has yet been observed.

\begin{figure}
\includegraphics[width=3.2in]{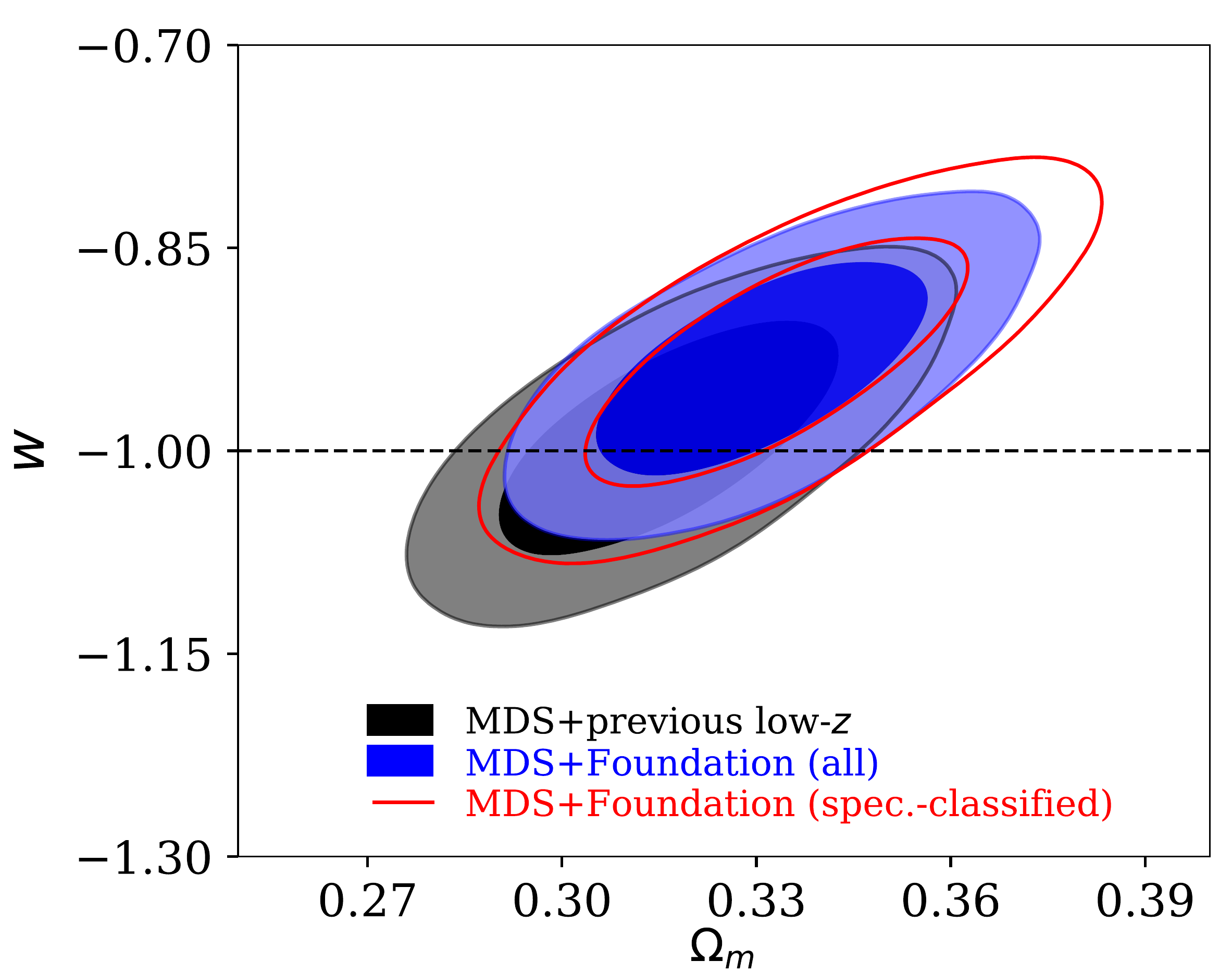}
\caption{Constraints on $w$ and $\Omega_m$ assuming a flat $w$CDM model. The Foundation$+$MDS sample (blue)
  and the combined sample of Foundation and spectroscopically classified MDS SNe (red) are
  compared to the \citetalias{Jones18} results that use the previous low-$z$ data instead of Foundation (black).}
\label{fig:cosmoparams}
\end{figure}

\setlength\doublerulesep{1mm}
\begin{table*}
\caption{\fontsize{9}{11}\selectfont $w$CDM and $w_a$CDM Parameters from MDS+Foundation SNe, BAO, CMB, and H$_0$}
  \centering
\begin{tabular}{lccccccccc}
  \hline \hline\\[-1.5ex]
&\multicolumn{9}{c}{$w$CDM Constraints}\\*[2 pt]
  &\multicolumn{3}{c}{$\Omega_m$}&\multicolumn{3}{c}{$w$}&&&\\
  \hline\\[-1.5ex]
SNe$+$CMB&\multicolumn{3}{c}{$0.331\pm0.017$}&\multicolumn{3}{c}{$-0.938\pm0.053$}&&&\\
SNe$+$CMB$+$BAO&\multicolumn{3}{c}{$0.316\pm0.009$}&\multicolumn{3}{c}{$-0.949\pm0.043$}&&&\\
SNe$+$CMB$+$H$_0$&\multicolumn{3}{c}{$0.295\pm0.012$}&\multicolumn{3}{c}{$-1.034\pm0.042$}&&&\\
SNe$+$CMB$+$BAO$+$H$_0$&\multicolumn{3}{c}{$0.301\pm0.008$}&\multicolumn{3}{c}{$-1.014\pm0.040$}&&&\\
\\[-1ex]
\tableline\\*[2 pt]
&\multicolumn{9}{c}{$w_a$CDM Constraints}\\*[2 pt]
&\multicolumn{3}{c}{$\Omega_m$}&\multicolumn{3}{c}{$w_0$}&\multicolumn{3}{c}{$w_a$}\\*[2 pt]
\hline \\*[-1.5ex]
SNe$+$CMB&\multicolumn{3}{c}{$0.314\pm0.025$}&\multicolumn{3}{c}{$-0.810\pm0.144$}&\multicolumn{3}{c}{$-0.791\pm0.785$}\\
SNe$+$CMB$+$BAO&\multicolumn{3}{c}{$0.321\pm0.010$}&\multicolumn{3}{c}{$-0.825\pm0.095$}&\multicolumn{3}{c}{$-0.570\pm0.401$}\\
SNe$+$CMB$+$H$_0$&\multicolumn{3}{c}{$0.280\pm0.011$}&\multicolumn{3}{c}{$-0.734\pm0.082$}&\multicolumn{3}{c}{$-1.541\pm0.374$}\\
SNe$+$CMB$+$BAO$+$H$_0$&\multicolumn{3}{c}{$0.305\pm0.008$}&\multicolumn{3}{c}{$-0.895\pm0.095$}&\multicolumn{3}{c}{$-0.597\pm0.439$}\\
\\[-1ex]
\hline\\[-1.5ex]
\multicolumn{10}{l}{
\begin{minipage}{12cm}
Constraints on $w$CDM and $w_a$CDM using the BBC analysis method.
\end{minipage}}
\end{tabular}
\label{table:wcdm}
\end{table*}

\begin{table}
\caption{\fontsize{9}{11}\selectfont Summary of Systematic Uncertainties for $w$}
    \begin{tabular}{lcp{0.4in}ccp{0.4in}}

\hline \hline\\[-1.5ex]
\multirow{2}{*}{Uncertainty}&\multicolumn{2}{c}{\multirow{2}{*}{MDS+Foundation}}&&\multicolumn{2}{c}{MDS+low-$z$}\\
&&&&\multicolumn{2}{c}{(\citetalias{Jones18})}\\
  \cline{2-3} \cline{5-6}
  &$\sigma_w^{\mathrm{sys}}$&Rel. to $\sigma_w^{stat}$&&$\sigma_w^{\mathrm{sys}}$&Rel. to $\sigma_w^{stat}$\\
\\[-1ex]
\hline\\[-1.5ex]
All Sys.&0.041&1.241&&0.043&1.144\\
Phot.\ Cal.&0.027&0.832&&0.019&0.495\\
$-$ SALT2 Model&0.023&0.699&&0.008&0.203\\
$-$ PS1 Cal.&0.016&0.496&&0.007&0.190\\
CC Contam.&0.013&0.381&&0.013&0.334\\
Bias Corr.&0.011&0.340&&0.020&0.520\\
MW E(B-V)&0.007&0.205&&0.014&0.379\\
Pec.\ Vel.&0.006&0.181&&0.007&0.197\\
Mass Step&0.000&0.000&&0.018&0.469\\
\\[-1ex]
\hline\\[-1.5ex]
\multicolumn{6}{l}{
\begin{minipage}{8cm}
Each systematic uncertainty as a fraction of the statistical
  uncertainties for the MDS+low-$z$ sample ($\sigma_{stat} = 0.038$)
  and the MDS+Foundation sample ($\sigma_{stat} = 0.034$).
\end{minipage}}
\end{tabular}
\label{table:syserr}
\end{table}

\begin{table}
\caption{\fontsize{9}{11}\selectfont Measurements of $w$ from Alternative Methods of Marginalizing Over CC\,SNe}
    \begin{tabular}{lrr}

\hline \hline\\[-1.5ex]
  &$w$&$\Delta w$\\
\\[-1ex]
\hline\\[-1.5ex]
Baseline&$-0.920\pm0.033$&\nodata\\
CC\,SN Simulations&$-0.924\pm0.033$&-0.004\\
CC\,SN Prior&$-0.938\pm0.033$&-0.018\\
Classification Prior$^{\mathrm{a}}$&$-0.886\pm0.036$&0.034\\
Nuisance Parameters Fixed&$-0.922\pm0.033$&-0.002\\
\\[-1ex]
\hline\\[-1.5ex]
    \end{tabular}
\begin{minipage}{8cm}
  For the BBC method, changes in $w$ (statistical uncertainties alone) after applying analysis variants
  related to the use of CC\,SN-contaminated data (see Table 7 of \citetalias{Jones18} for
  a similar table for the PSBEAMS method).
  We show the effect of using CC\,SN simulations
  with alternate LFs, dust distributions, or CC\,SN templates,
  the effect of using the parametric \citep{Hlozek12} prior
  on the CC\,SN distribution, the effect of a different light-curve
  classifier, and the effect of fixing the nuisance parameters to
  the values derived from the spectroscopically classified data.
  \begin{itemize}[leftmargin=*]
  \item[$^{\mathrm{a}}$] The large change in $w$ is due to the highest two redshift
    bins, which have CC\,SNe significantly blended with SNe\,Ia and are
    de-weighted when all systematic uncertainties are applied.
    The PSBEAMS method generally performs more consistently than BBC with the PSNID
    classifier as it has an additional parameter to scale the ``overconfident''
    P(Ia) probabilities \citep{Jones17}.
  \end{itemize}
\end{minipage}
\label{table:wphot}
\end{table}

\begin{table*}
\caption{\fontsize{9}{11}\selectfont Summary of $w$ Measurements and Systematic Uncertainties}
    \begin{tabular}{lrrrrrrrrrrrrrrr}
  \hline \hline\\[-1.5ex]
  &\multicolumn{5}{c}{Full Sample}&&\multicolumn{5}{c}{Spec Sample}&&\multicolumn{2}{c}{\citetalias{Jones18}}\\
\cline{2-6} \cline{8-12} \cline{14-15} \\[-1.5ex]
&\multicolumn{2}{c}{PSBEAMS}&&\multicolumn{2}{c}{BBC}&&\multicolumn{2}{c}{PSBEAMS}&&\multicolumn{2}{c}{BBC}&\\
\cline{2-3} \cline{5-6} \cline{8-9} \cline{11-12}\\[-1.5ex]
$w_{stat+sys}$&\multicolumn{2}{c}{$-0.918\pm0.057$}&&\multicolumn{2}{c}{$-0.938\pm0.053$}&&\multicolumn{2}{c}{$-0.955\pm0.063$}&&\multicolumn{2}{c}{$-0.933\pm0.061$}&&\multicolumn{2}{c}{$-0.990\pm0.057$}\\
$w_{stat}$&\multicolumn{2}{c}{$-0.961\pm0.036$}&&\multicolumn{2}{c}{$-0.920\pm0.033$}&&\multicolumn{2}{c}{$-0.954\pm0.050$}&&\multicolumn{2}{c}{$-0.936\pm0.047$}&&\multicolumn{2}{c}{$-1.022\pm0.038$}\\
\\
Sys. Error&$\Delta w$&$\Delta \sigma_w$&&$\Delta w$&$\Delta \sigma_w$&&$\Delta w$&$\Delta \sigma_w$&&$\Delta w$&$\Delta \sigma_w$&&$\Delta w$&$\Delta \sigma_w$\\
\cline{2-3} \cline{5-6} \cline{8-9} \cline{11-12} \cline{14-15}\\[-1.5ex]
Photometric Calibration&0.026&0.037&&$-$0.020&0.027&&0.013&0.037&&0.009&0.026&&0.012&0.019\\
$-$ SALT2 Model$^{\mathrm{a}}$&0.029&0.026&&$-$0.015&0.023&&0.006&0.026&&$-$0.000&0.021&&0.000&0.008\\
$-$ Supercal&0.001&0.001&&$-$0.003&0.000&&$-$0.001&0.000&&0.000&0.000&&$-$0.001&0.000\\
$-$ Filter Functions&$-$0.001&0.007&&$-$0.005&0.000&&0.004&0.008&&0.002&0.000&&$-$0.001&0.009\\
$-$ PS1 Zero Point&$-$0.005&0.024&&$-$0.011&0.016&&0.003&0.026&&0.008&0.017&&$-$0.002&0.007\\
$-$ {\it HST} Calibration&0.001&0.008&&$-$0.002&0.003&&0.003&0.000&&0.002&0.005&&0.002&0.009\\
Mass Step&0.012&0.009&&$-$0.001&0.000&&$-$0.000&0.000&&0.001&0.000&&0.005&0.018\\
CC Contamination&0.040&0.017&&$-$0.000&0.013&&0.000&0.000&&0.000&0.000&&$-$0.001&0.013\\
Bias Correction&0.012&0.009&&$-$0.009&0.011&&$-$0.007&0.004&&$-$0.008&0.019&&0.012&0.020\\
Peculiar Velocity&$-$0.002&0.010&&$-$0.002&0.006&&0.002&0.007&&0.004&0.007&&0.002&0.007\\
MW E(B-V)&$-$0.000&0.008&&$-$0.002&0.007&&0.000&0.000&&0.003&0.010&&0.009&0.014\\
$\beta(z)^{\mathrm{b}}$&$-$0.008&0.025&&0.004&0.018&&0.012&0.036&&$-$0.004&0.012&&0.012&0.016\\
$\Delta_M(z)^{\mathrm{b}}$&0.039&0.027&&$-$0.001&0.000&&$-$0.000&0.000&&0.003&0.000&&0.006&0.020\\
\hline
$\beta(z)^{\mathrm{b}}$&$-$0.008&0.024&&0.004&0.018&&0.012&0.036&&$-$0.004&0.012&&0.012&0.016\\
$\Delta_M(z)^{\mathrm{b}}$&0.040&0.026&&$-$0.001&0.000&&$-$0.000&0.000&&0.003&0.000&&0.006&0.020\\
\hline\\*[-1.5ex]
\multicolumn{16}{l}{
  \begin{minipage}{17.2cm}
    Summary of measurements and systematic uncertainties for each
  value of $w$ presented in this paper.
    \begin{itemize}[leftmargin=*]
\item[$^{\mathrm{a}}$] \citetalias{Jones18} reported a SALT2 systematic uncertainty that was smaller than the one used in this analysis and the Pantheon
  analysis.  Corrected, a SALT2 systematic shift with the same size as the one used in this analysis
  gives a systematic uncertainty of 0.016 for the \citetalias{Jones18} analysis instead of the reported 0.008.
  This increase does not significantly impact the final constraints on $w$ from \citetalias{Jones18}.
\item[$^{\mathrm{b}}$] As the $z$-dependence of $\beta$ and mass step have not been significantly detected in any analysis
  to date, we have not included
  them in the final systematic uncertainty error budget (also following \citetalias{Jones18} and the Dark Energy Survey
  analysis; \citealp{Brout18}).  However, we show their impact here.  Including them
  would not significantly increase the final BBC measurement uncertainty but would increase
  the uncertainties on $w$ from the PSBEAMS method by 25\%.
  \end{itemize}
  \end{minipage}}
    \end{tabular}

\label{table:allw}
\end{table*}

\subsection{Cosmological Parameters}
\label{sec:cosmoparam}

In this section, we use SNe\,Ia in combination with external data sets
to constrain three cosmological models:

\begin{enumerate}
\item A flat $\Lambda$CDM model.
\item The $w$CDM model.  A redshift-independent $w$ is
  allowed to vary from the cosmological constant
  value of $w = -1$.  We assume a flat universe.
\item The $w_a$CDM model. Using the \citet{Chevallier01,Linder03}
  formalism, $w$ is allowed to evolve with redshift: $w = w_0 + w_a z/(1+z)$.
  We again assume a flat universe.
\end{enumerate}

\noindent We use the more precise BBC distances to derive our baseline cosmological
parameter measurements.  BBC distance uncertainties are 18\% smaller, on average, than
PSBEAMS distance uncertainties (excluding the high-uncertainty $z = 0.01$ and
$z = 0.7$ bins).

First, when assuming $\Lambda$CDM we may derive useful, independent
constraints on the cosmic matter density from SNe alone.
We find the cosmic matter density $\Omega_m$ to
be \om\ when assuming $\Lambda$CDM (the curvature, $\Omega_k$, is fixed at 0).
This is consistent with the value of $\Omega_m = 0.315 \pm 0.007$ measured from the CMB
\citep{Planck18}.  \citetalias{Jones18} measured $0.319 \pm 0.040$ using the MDS high-$z$ data and
the previous low-$z$ sample.

Next, we combine the binned BBC distances with the 2015 CMB likelihoods from Planck
to constrain the $w$CDM model (\citealp{Planck16}; 2018 Planck likelihoods are currently unavailable).
We find $w = $ \w, consistent with a cosmological constant at the 1.2-$\sigma$ level for a flat universe.
Figure \ref{fig:cosmoparams} shows the change in the $w-\Omega_m$
plane when using Foundation SNe as the low-$z$ sample instead of the previous low-$z$ SNe.
From the spectroscopically-classified data alone, we measure
$w = $ \wspec, which is consistent at the 1.1-$\sigma$ level with
$\Lambda$CDM and demonstrates that the use of photometrically classified data
does not significantly change the measurement.

To increase the precision of these measurements, we
include BAO constraints
from \citet{Anderson14}, \citet{Ross15}, and \citet{Alam17}, which give measurements
of the BAO scale at $z = 0.15$, 0.32, 0.38, 0.51, 0.57, and 0.61.
The BAO scale is proportional to a combination of the angular
diameter distance to a given redshift and the Hubble parameter
at that redshift, and is measured from the signature of acoustic
waves on the cosmic matter distribution.
We also use H$_0$ constraints from \citet{Riess18}, which shift
the measured value of $w$ to be $\sim$6\% more negative
than SN$+$CMB$+$BAO due to their
3.7$\sigma$ inconsistency with the Planck results.

The full constraints on the $w$CDM model from this combination of
different probes are given in Table \ref{table:wcdm}.  For
the $w$CDM model, including BAO constraints moves $w$ slightly
closer to $\Lambda$CDM, but still $1.2\sigma$ from $w = -1$, and
including the local H$_0$ constraints moves $w$ to $-1.014 \pm 0.040$ (however,
the CMB and local H$_0$ measurements are internally inconsistent).

Allowing $w$ to evolve with redshift gives
$w_0 = $ \wo\ and $w_a = $ \wa\ from SNe$+$CMB, which is
consistent with $\Lambda$CDM.  Including H$_0$
constraints moves the value to nearly 3$\sigma$ from
$\Lambda$CDM, due to the internal inconsistency
of local H$_0$ measurements with Planck, but the
best measurement of SNe$+$CMB$+$BAO$+$H$_0$ is within
1$\sigma$ of $\Lambda$CDM.

\subsection{Systematic Uncertainties}
\label{sec:syserr}

\begin{figure*}
  \includegraphics[width=7in]{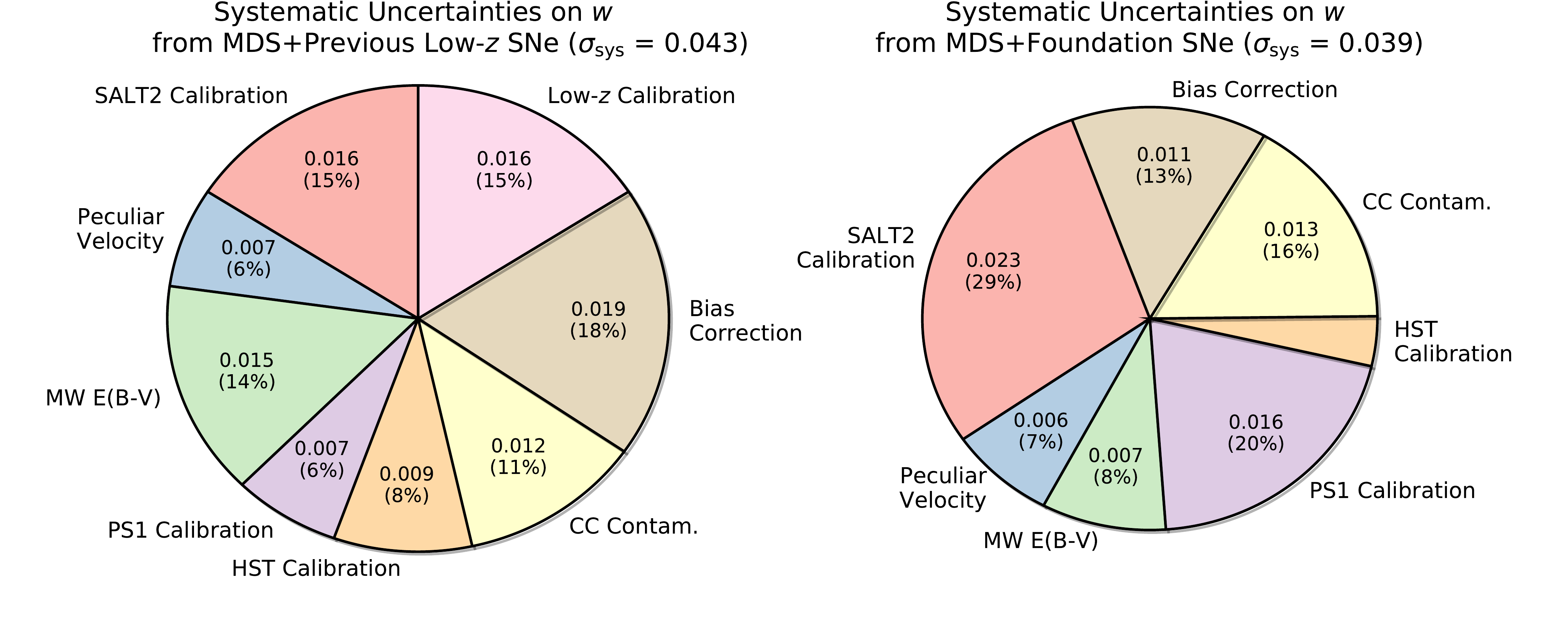}
  \caption{Systematic uncertainties on $w$ from \citetalias{Jones18} (left) compared to this analysis (right).
    The size of each chart is proportional to the size of the total systematic uncertainty
    budget for each analysis.  The size of each slice corresponds to the size of each
    systematic uncertainty as a fraction of the sum of all systematic uncertainties.
    The calibration systematic uncertainties are 42\% higher in this analysis as
    we have just one photometric system, but the bias-correction systematic
    uncertainty is \biascorsyspercent\ lower.  The overall systematic uncertainty is 9\%
    lower.}
  \label{fig:pie}
\end{figure*}

The systematic uncertainties from this analysis
compared to the \citetalias{Jones18} analysis are shown in Table
\ref{table:syserr} and Figure \ref{fig:pie} using the BBC framework.
A full table of systematic uncertainties for both the full and
spectroscopically classified samples, using both PSBEAMS and BBC,
is shown in Table \ref{table:allw}.
We focus on the systematic uncertainties from the BBC method here,
and discuss the difference between the methods in \S\ref{sec:bbccomp}.

With the reduction in the distance bias systematic uncertainty
and the use of just a single sample, photometric
calibration is the dominant systematic uncertainty (0.027)
in this sample.  This systematic uncertainty can be split into three
components: the SALT2 calibration,
{\it HST} CALSPEC calibration, and uncertainty in the PS1 calibration.
The SALT2 and PS1 calibration are the dominant components of the systematic uncertainty budget.

The bias correction systematic uncertainty has been reduced from 0.02
in \citetalias{Jones18} to just 0.011 in the current analysis, a result of the
better-understood selection effects in Foundation.  The
uncertainty due to marginalizing over CC\,SN contamination
remains approximately the same at $\sim$1.3\%.
This systematic uncertainty remains
subdominant to photometric calibration and will be improved
in future work.  Statistical uncertainty only measurements of $w$ after applying
  a number of different treatments of the CC\,SN contamination
  are shown in Table \ref{table:wphot}.  These include alternate
  light-curve classification methods, predicted biases from simulations, alternate parameterizations
  of the contaminating distribution, and nuisance parameters forced
  to be equal to those measured from the spectroscopically classified data.

Although we do not include possible redshift dependence of the
mass step or $\beta$ in our systematic uncertainty budget, we show their effect
in Table \ref{table:allw}.
Compared to the PSBEAMS method, BBC bias corrections
in shape and color bring the high-$z$ mass step, $\Delta_M = 0.043 \pm 0.013$, in better agreement with the
low-$z$ mass step, $\Delta_M = 0.060 \pm 0.024$, lowering the measured systematic uncertainty due to an evolving mass step.
We find that $\beta$ evolution is also a slightly lower systematic uncertainty in the BBC method
for similar reasons.

\subsection{Comparing BBC to the PSBEAMS Method}
\label{sec:bbccomp}

\begin{figure}
  \includegraphics[width=3.5in]{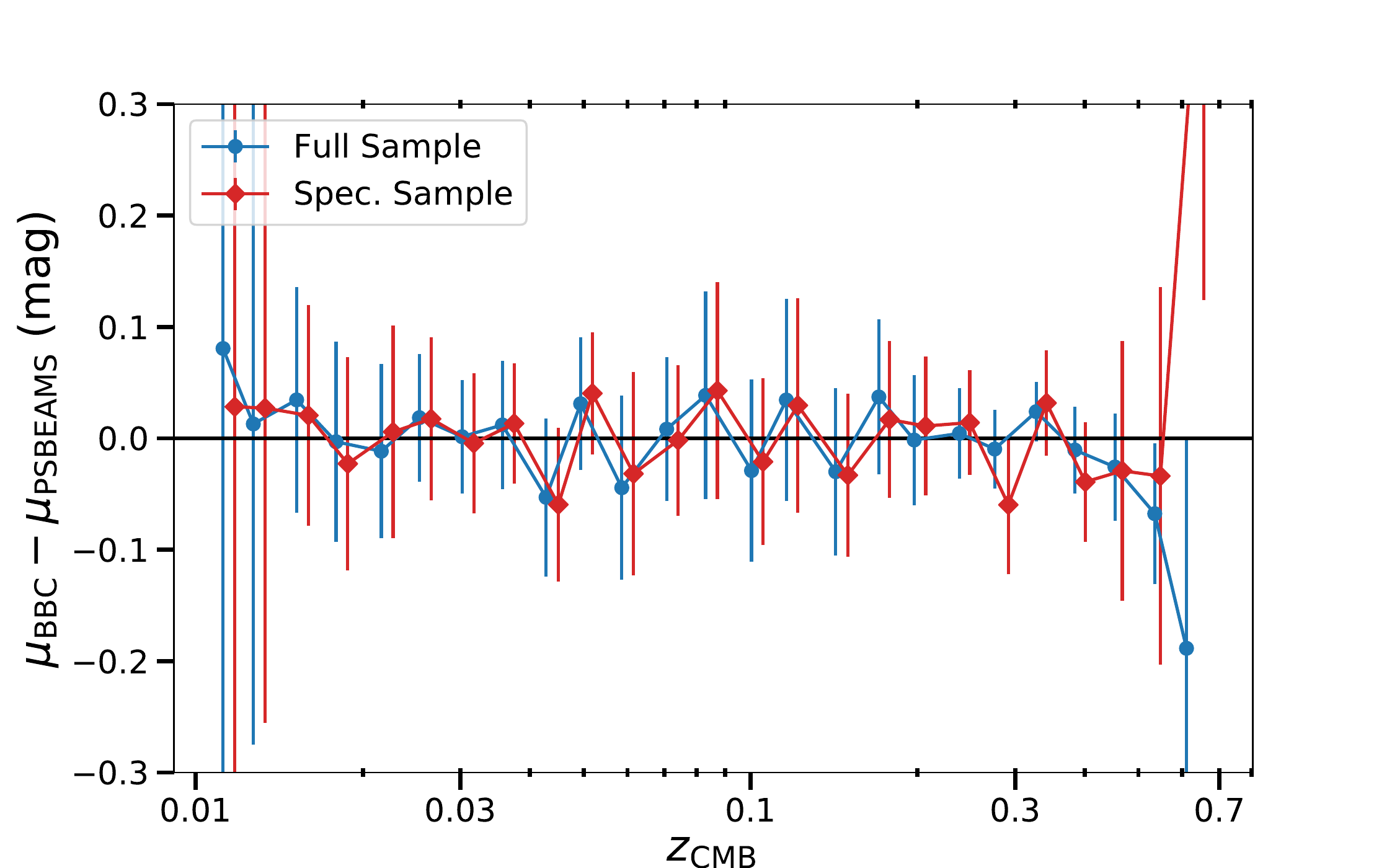}
  \caption{Difference between distances derived using the PSBEAMS and
    BBC methods, for both the full sample (blue) and spectroscopically
    classified sample (orange), as a function of redshift.
    Some modest discrepancies occur at high $z$
    where the only detectable CC\,SNe have brightnesses that are close
    to or greater than those of SNe\,Ia.  Points have been offset slightly
    for visual clarity.}
  \label{fig:bbccc}
\end{figure}
  
Use of the BBC method gives a more fine-grained approach to
bias corrections, as parameters $m_B$, $x_1$, $c$
are each corrected for
selection biases.  The results are consistent with
the values from the PSBEAMS method, but the final uncertainty on $w$ is reduced by
7\%.  We find that the final value of $w$ is lower by 0.020 when using
the BBC method; however, the value when excluding systematic uncertainties is
0.041 \textit{higher} when using the BBC method.
in the highest redshift bins.
The BBC results on spectroscopically classified data alone
are somewhat closer to results using the PSBEAMS method; $w$ is
higher by 0.022 when using BBC compared to the $w$ from the PSBEAMS method.

For the full sample, the difference in statistics-only $w$ values,
and the relatively large change of 0.043 between the PSBEAMS
measurement with statistical uncertainties only and
the PSBEAMS measurement when systematic uncertainties
are included, is primarily due to systematic uncertainties in
marginalizing over CC\,SNe.  Use of the NN classifier compared
to the PSNID classifier to assign the prior probabilities that
a given SN is of Type Ia, shifts $w$ by $+0.028$ (although, due to significant
covariances between bins, the uncertainty on $w$ does not increases by the
same amount).

With BBC, the method of marginalizing over CC\,SNe is somewhat different.
First, we use a different nominal classification method (\S\ref{sec:bbc}).  The PSBEAMS method
also includes a parameter that allows probabilities to be shifted linearly
(see Equation \ref{eqn:norm}),
while BBC does not.  Finally, the PSBEAMS method uses a point-to-point,
parameterized linear interpolation of the CC\,SN distribution, while
BBC uses the simulations themselves as a fixed prior on the distance.
The two methods yield distances consistent to $\ll$1$\sigma$ at $z < 0.4$, but disagree somewhat at
high-$z$ as shown in Figure \ref{fig:bbccc}.  CC\,SNe in the $z \gtrsim 0.5$
redshift range can be difficult to marginalize over as the CC\,SN distribution begins
to overlap substantially with the SN\,Ia distribution because of Malmquist bias.  As these
methods were developed and tested independently, they are
complimentary methods for measuring $w$ from photometrically
classified data.

\section{Discussion and Future Directions}
\label{sec:discussion}

\begin{figure*}
  \includegraphics[width=7in]{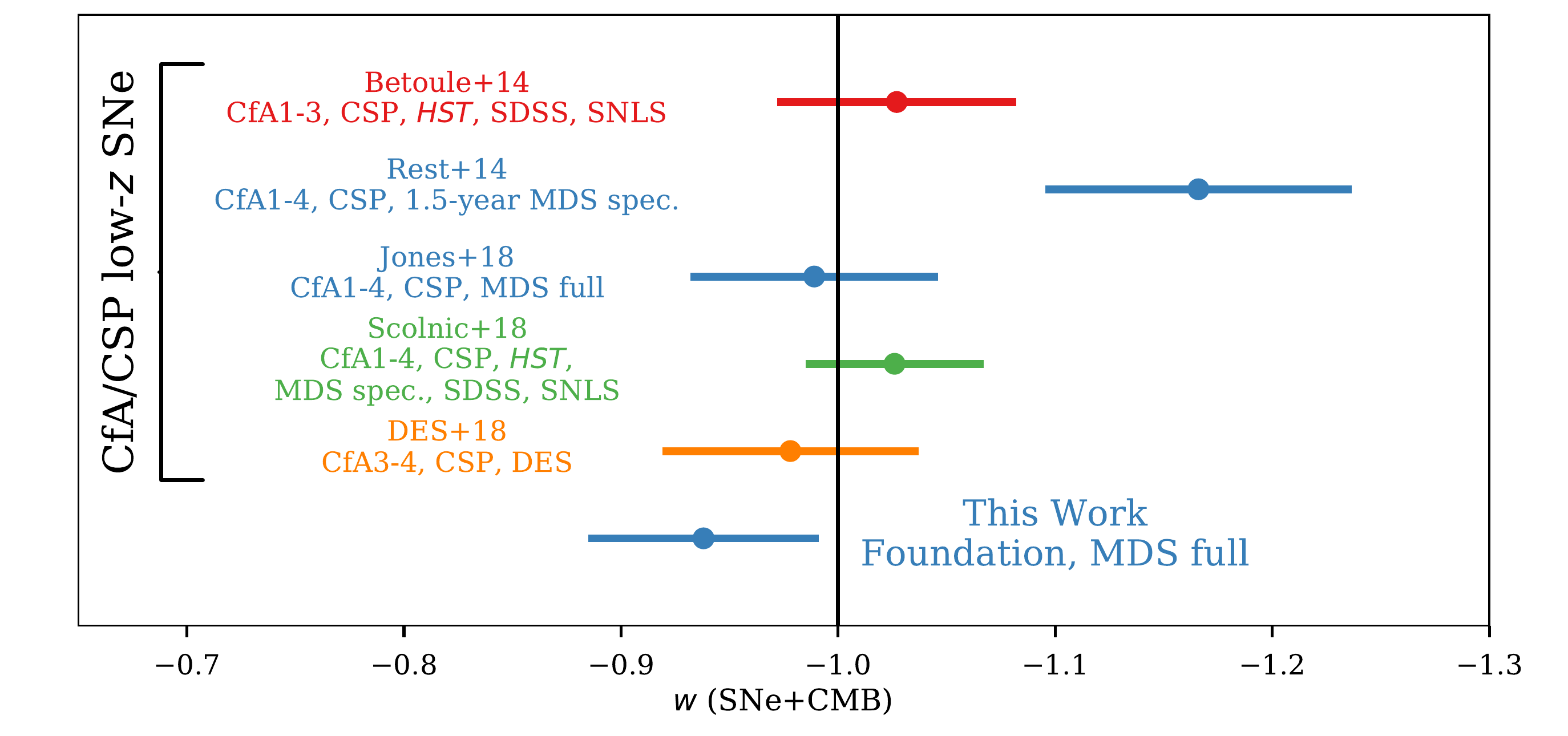}
  \caption{Measurements of $w$ using SNe with CMB constraints from
    Planck, published within the last five years.  From top to bottom, we
    show measurements from JLA \citep{Betoule14}, spectroscopically
    classified SNe\,Ia from the first 1.5 years of the MDS \citep{Rest14},
    the full MDS sample \citepalias{Jones18}, the Pantheon sample \citepalias{Scolnic18},
    spectroscopically classified SNe\,Ia from the first three years of DES \citep{DES18},
    and the present analysis.  The previous analyses share many of the same low-$z$
    and high-$z$ SN samples, which we indicate in the figure.  The ``spec.'' abbreviation indicates that
    only spectroscopically classified MDS SNe were used in these analyses.}
  \label{fig:cosmocomp}
\end{figure*}

\subsection{Differences between Foundation and Other Low-$z$ Samples}
\label{sec:found}

The measurement of $w$ presented here is independent of all previous
low-$z$ datasets and most previous high-$z$ datasets.
Figure \ref{fig:cosmocomp} compares the measurement of $w$ after including
the Foundation sample to other measurements of $w$ published within the
last five years.  All previous measurements use some combination
of CfA and CSP data at low $z$, and the \citet{Betoule14} and S18
analyses also use the same SDSS, SuperNova Legacy Survey (SNLS; \citealp{Astier06}), and {\it HST}
data.  \citetalias{Jones18} used the same high-$z$ sample
as this analysis does and after replacing the previous low-$z$
compilation with the Foundation sample, we find that $w$ is
shifted by $+$\wshiftpercent.

The total systematic uncertainty of our measurement of $w$
is 9\% smaller than in the \citetalias{Jones18} analysis;
although the total calibration systematic uncertainty is
increased by 42\%, the selection bias systematic uncertainties
are reduced by \biascorsyspercent\ (Figure \ref{fig:scatbias}).
This calibration systematic uncertainty is due in approximately
equal parts to the calibration of the SALT2 model and the
PS1 zeropoint calibration uncertainties.  The
uncertainties due to survey calibration in particular are larger because the previous low-$z$
sample was observed using $\geq$3 filters per telescope on up
to seven different photometric systems, while this analysis
uses only two filters on a single photometric system.

Reducing these systematic uncertainties will
require either adding SNLS, SDSS or DES high-$z$ data to the analysis,
adding additional low-$z$ samples, though these may have
lower weight than the Foundation sample when systematic
uncertainties are included, or using a re-trained
SALT2 model.  Because SN surveys are not consistently calibrated before training the SALT2 model,
a re-trained SALT2 model that includes Supercal-corrected photometry \citep{Scolnic15}
would substantially improve future analyses.
Similarly, a SALT2 model trained to use redder rest-frame wavelengths where SN\,Ia are
better standardizable candles (e.g., \citealp{Mandel11}) would
also improve distance measurements and improve the ability to plan future
NIR SN surveys such as {\it WFIRST} \citep{Pierel18}.
Though the MDS data are redshifted enough for all four filters to be used,
fifty percent of Foundation data (the $iz$ observations) are
not used in this work.  For this reason alone, a SN light-curve fitter with
redder wavelength coverage would provide enormous benefits by
(1) reducing the sample dispersion by using data at wavelengths where SNe\,Ia are better standard candles
and (2) reducing the single-filter calibration systematic uncertainties by including twice as many
filters at low-$z$.

The Foundation sample is much more
similar to the high-$z$ sample $-$ in sample selection,
photometric reduction, and photometric system $-$
than the previous low-$z$ samples.  With this
sample, we therefore expect any bias caused
by unforeseen systematic uncertainties (e.g., unexpected
dependence of SNe\,Ia on their host galaxies)
to be greatly reduced.  Unforeseen systematic
  uncertainties would increase the error on our measurement, but
would affect previous measurements more strongly.  Although
the shift in distances with these new data is only
marginally significant, it may well be that
this shift is hinting at an aspect of SN\,Ia 
physics that will be revealed in future work.
The slight change in $w$
found in this work is driven by the low-$z$ sample,
which also has SNe with masses and sSFRs that are significantly
shifted with respect to the previous low-$z$ sample
(Figure \ref{fig:massvz}).

The Foundation Supernova Survey aims to observe up to 800 cosmologically
useful SNe\,Ia.  With such large statistical
leverage, we may be able to better understand the ways in which
SN\,Ia distances may be affected by unexpected
systematic uncertainties.

\subsection{Photometric Classification}
\label{sec:photclassdisc}

We have shown that the $\sim$1$-$$\sigma$ difference between
the value of $w$ measured from this data and the previous (\citetalias{Jones18}) results
stems from the low-$z$ sample alone and not
from any biases caused by marginalizing over CC\,SNe.
First, our results are consistent
at $\ll$1$\sigma$ with the spectroscopically classified
data.  Second, we have employed 5 separate approaches to
modeling the CC\,SN distribution: the PSBEAMS method
uses a single Gaussian CC\,SN model, a two-Gaussian CC\,SN model and
an asymmetric Gaussian model, each of which have means and dispersions
that are linearly interpolated between 5 control points
across the redshift range of the sample.  With the BBC method, we use
a Monte Carlo simulation-based prior and a parameterized
single-Gaussian CC\,SN model that evolves as a second-order polynomial
across the redshift range.  Finally, we adopt three different methods of
classifying CC\,SNe.  All methods yield results $\sim$1$\sigma$
from $\Lambda$CDM.

CC\,SN contamination is not the dominant systematic uncertainty in this
work, but it remains a significant one at 1.3\%.  It will need to be
reduced in future work, and the true distribution of CC\,SNe
will need to be better understood.  However, we see no
evidence, either in this work or from previous recent studies of
using photometrically classified SNe for cosmology, that
our conservative estimation of this systematic uncertainty is unrealistic.

An additional consideration related to CC\,SN marginalization was found
by \citet{Knights13}, who note that the BEAMS formalism breaks down
in samples with large correlated systematic uncertainties.
However, their analysis explored correlations on the order of
$\sim$10\%, and only percent-level correlations exist in this sample.
We expect that a more sophisticated treatment of systematic uncertainties will
only become necessary in analyses with even larger samples (e.g., LSST).

\subsection{The Relationship Between SNe\,Ia and their Host Galaxies}
\label{sec:hosts}

\begin{table*}
  \centering
  \caption{\fontsize{9}{11}\selectfont Alternate Relationships between SN\,Ia Hubble Residuals and Host-Galaxy Properties and their Effect on $w$}
  \begin{tabular}{lrrrrr}
    \hline \hline\\[-1.5ex]
    &\multicolumn{2}{c}{PSBEAMS}&&\multicolumn{2}{c}{BBC}\\
&step size&$\Delta w$&&step size&$\Delta w$\\*[2pt]
    \hline\\[-1.5ex]
default mass step&$0.088\pm0.013$ (6.7$\sigma$)&\nodata&&0.044$\pm$0.010 (4.4$\sigma$)&\nodata\\*[2pt]
\hline
\multirow{2}{*}{$z$-dependent mass step}&$\Delta_{M,0} = 0.065\pm0.019$ (3.4$\sigma$)&\multirow{2}{*}{0.032}&&$\Delta_{M,0} = 0.054\pm0.020$ (2.7$\sigma$)&\multirow{2}{*}{0.016}\\
&$\Delta_{M,1} = 0.082\pm0.054$ (1.5$\sigma$)&&&$\Delta_{M,1} = -0.038\pm0.060$ (0.6$\sigma$)&\\*[2pt]
\hline
Global sSFR step$^{\mathrm{a}}$&$0.038\pm0.013$ (2.9$\sigma$)&$-0.014$&&$0.018\pm0.011$ (1.6$\sigma$)&0.049\\*[2pt]
\hline
SNe in locally massive regions$^{\mathrm{b}}$&\nodata&$-0.001$&&\nodata&0.029\\*[2pt]
\\[-2ex]
\hline\\[-1.5ex]
  \multicolumn{6}{l}{
  \begin{minipage}{17cm}
    \begin{itemize}[leftmargin=*]
    \item[$^{\mathrm{a}}$] The 0.049 shift in $w$ when using the BBC method is not
      physical, as we replace a highly significant mass-step correction
      with a marginally significant sSFR step.
    \item[$^{\mathrm{b}}$] PS1 does not have sufficient resolution to measure
      the local mass step at high-$z$, so we restrict to SNe in
      probable locally-massive regions.  \citet{Jones18b} measured
      the local mass step to be $0.067 \pm 0.017$~mag.
    \end{itemize}
  \end{minipage}}
  \end{tabular}
  \label{table:altsteps}
\end{table*}

\begin{figure*}
  \includegraphics[width=7in]{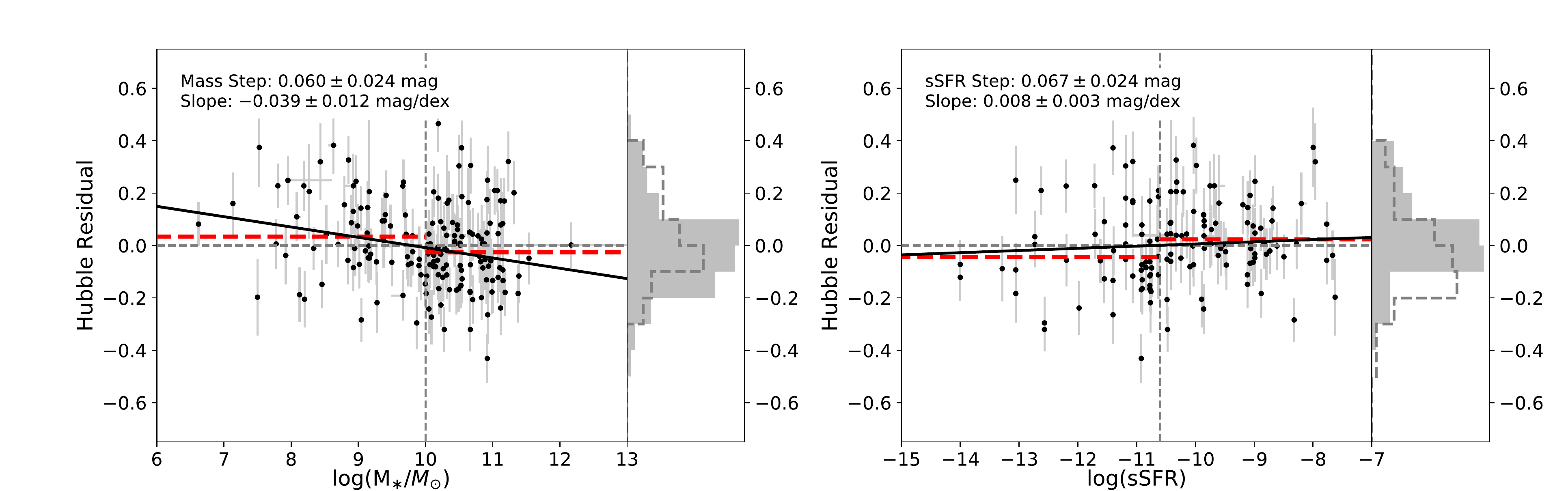}
  \caption{
    For the Foundation sample, the correlation of
    Hubble residuals with host-galaxy mass (left) and
    Hubble residuals with host-galaxy sSFR (right).
    Assuming a step function to describe the sample,
    red-dashed lines show the maximum likelihood average
    Hubble residual for each side of the step, which is
    indicated by a vertical grey-dashed line.  The solid-black
    line represents the best-fitting linear function.  In the left panel,
    histograms show the Hubble residuals of SNe in low-mass
    (dashed line) and high-mass (filled) host galaxies.
    In the right panel, histograms show the Hubble residuals
    of SNe in low-sSFR (dashed) and high-sSFR (filled)
    host galaxies.  We use the BBC method to
    generate the Hubble residuals shown here.}
  \label{fig:masscorr}
\end{figure*}

The relationship between SN\,Ia and their host galaxies is subject
to significant uncertainty and could bias cosmological parameters (e.g., \citealp{Childress14,Rigault18}).
To mitigate this uncertainty, we explored several methods of estimating the
potential bias to $w$ from the uncertain relation between SNe\,Ia
and their host galaxies.  First, in \S\ref{sec:results} we examined
the effect of allowing the host-galaxy mass step to evolve with redshift,
as predicted by \citet{Childress14} and observed with marginal significance
by \citetalias{Scolnic18} (although not detected by \citetalias{Jones18}).  In this work we have a limited redshift range over which the mass
step can evolve, and although we do not find significant evidence for
mass-step evolution, the uncertainties are much higher than in the Pantheon
analysis of \citetalias{Scolnic18}.  Allowing an evolving mass step
does not shift the measurement of $w$ when the BBC
analysis method is used (Table \ref{table:allw}).
The PSBEAMS measurement is shifted due to
observational biases as discussed in \S\ref{sec:nuisanceresults}.
Given that the
spectroscopically classified data do not prefer an evolving mass step,
we expect that the large $\Delta_M$ systematic uncertainty when using the PSBEAMS method is not a physical
effect.

Second, a number of recent papers have suggested alternative
relationships between SN\,Ia distance measurements and their
host-galaxy masses.
For example, some evidence has also shown that
metallicity, a function of both mass and SFR, may correlate
with Hubble residuals better than mass alone \citep{Hayden13}, a possible systematic
uncertainty that warrants investigation with additional data.
For the Foundation sample, the correlation of Hubble residuals
with host galaxy mass and sSFR is shown in Figure \ref{fig:masscorr}.

Recent work has also explored the
relationship between SN\,Ia Hubble residuals and host-galaxy properties
near the SN location.  Within the last year, \citet{Rigault18} found a strong
relationship between the SN distance measurement and the sSFR near
the SN location using SNFactory data, \citet{Roman18} found a strong relationship
between the SN distance and the local (and global) rest-frame $U-V$ color
of the host, and \citet{Jones18b} found a 3-$\sigma$ relationship between
the SN distance and the local host-galaxy mass \textit{after} global
mass correction.  In the \citet{Jones18b} data, the relationship between the SN distance
and local host-galaxy mass was larger than either the sSFR step or a
local color step, and the local
mass step was 60\% larger than the alternate steps
after global mass correction.
As our knowledge of the relationship between
SN\,Ia distances and host galaxies becomes increasingly fine-grained and
complex, it is important to test for these effects in cosmological analyses.

Although we cannot measure host-galaxy properties within $\sim$1--2 kpc of the SN location
for the $z > 0.1$ SN sample due to seeing limitations, we estimate the
bias due to a possible local mass or sSFR step in two ways.  First,
\citet{Jones18b} found that a step using the sSFR measured from the
global properties of the host galaxy was slightly more significant
than a local sSFR step, and so we measure $w$ by replacing the global
mass step with a global sSFR step.  The sSFR values are estimated
using the method discussed in \S\ref{sec:host},
but we reduce the sample slightly by using only
SN\,Ia with SDSS $u$-band data to ensure robust star-formation rates.
For the higher-$z$ SNe, this restriction is unimportant, but the majority of our sample
is not at high enough redshift to provide rest-frame $u$.  Including SNe for which robust sSFR
measurements cannot be computed, this restriction reduces
our sample by $\sim$300 SNe.  We then use both the BBC and PSBEAMS likelihoods
to estimate the distances after using a sSFR step instead of a mass step.

For the BBC method, we find an insignificant sSFR step of $\Delta_{\mathrm{sSFR}} = 0.018 \pm 0.011$~mag
for the G10 scatter model and a similarly small step for the C11 model.
Fitting with an undetected sSFR step instead of the detected mass step therefore shifts $w$ in a
way that is not justified by the data (a shift of $+$0.049).
For the PSBEAMS method, we find a larger sSFR step of $\Delta_{\mathrm{sSFR}} = 0.038 \pm 0.013$,
but fitting for a sSFR step instead of a mass step gives a statistics-only
measurement of $w$ that is shifted by just $-0.014$.  \citet{Jones18b} found a
somewhat larger sSFR step from low-$z$ data alone
of $0.054 \pm 0.020$; it may be that the higher-$z$ measurements with more limited
wavelength coverage have less accurate sSFR measurements, but the values of these two steps are
statistically consistent.  Both are also consistent with the sSFR step found by
\citet{Brout18} of $0.037 \pm 0.025$~mag.  We therefore do not expect that use of the mass step
instead of a sSFR step is biasing our measurements of $w$.

Second, we can compute $w$ using only SNe in locally massive regions of their
host galaxies.  From the \citet{Jones18b} public data, 83\% of SNe in globally
massive host galaxies that occur at host $R < 2$ (near their host center;
$R$ is discussed in \S\ref{sec:host})
are in locally massive regions.  This subset cuts the sample size by $\sim$40\%,
but if a strong change in $w$ is observed with this subset, it may mean that measurements
of dark energy are sensitive to the relationship between SN\,Ia and their local
host-galaxy environments.  Using the BBC method, $w$ is shifted by $-0.029$,
a shift equal to $\sim$90\% of the statistical uncertainty for the full
sample.  For the PSBEAMS method, however, $w$
is changed by less than 0.001.

These three tests are summarized in Table \ref{table:altsteps}.  They
show that $w$ is not sensitive to a mis-characterization
of the relationship between SN\,Ia and their host galaxies, although the statistically
insignificant, $\sim$3\% change in $w$ when controlling for a local mass step may
warrant investigation in future work.

\section{Conclusions}
\label{sec:conclusions}

We combine 1164 SNe\,Ia from the Pan-STARRS1 medium deep
survey with 174 SNe\,Ia from the Foundation Supernova Survey DR1 to
measure cosmological parameters from a single telescope
and photometric system.  The well-calibrated PS1 system makes this an excellent unified
sample for precision cosmology.  Future Foundation data releases
will build the SN\,Ia sample size up to 800 SNe\,Ia at $z < 0.1$.

Similar to the high-$z$ sample, Foundation is nearly an
untargeted, magnitude-limited sample, similar to the higher-$z$
data.  \citet{Foley18} notes that 86\% of the DR1 sample was
independently discovered by the untargeted ASAS-SN and PSST
surveys (though the untargeted
surveys were not the first surveys to discover 25\% of those SNe).
Including discoveries from other untargeted surveys brings the total to 94\%.
In sample selection, photometric system, and
host-galaxy demographics (Figure \ref{fig:massvz}), Foundation
is therefore a more homogeneous sample across the redshift range
than previous data sets.  One caveat is that the photometrically
classified MDS sample is biased towards more luminous host
galaxies due to the necessity of obtaining a spectroscopic
redshift.  A future analysis with photometric redshifts
could ameliorate this concern.

Foundation distances are \lowzdistoff~mag fainter than
distances from the previous low-$z$ data (\lowzdistsighyphen\ significance).
After combining our SN\,Ia distances with Planck priors,
we find $w = $ \w, an uncertainty \wreducpercent\ lower than
the previous \citetalias{Jones18} analysis that used only the CfA and CSP low-$z$
compilations.  Allowing $w$ to evolve with redshift using the
parametric form given by \citet{Linder03} gives $w_0 = $ \wo\ and
$w_a = $ \wa.

The systematic uncertainty attributed to SN selection biases
has decreased by \biascorsyspercent.  This decrease is due to the well-understood
selection criteria of the Foundation Supernova Survey and the similarity
of these criteria to previous high-$z$ analyses.  The photometric
calibration uncertainty has increased relative to the
previous analysis by 42\% as fewer independent photometric systems
are used, but adding additional high-$z$ or low-$z$ data to this
analysis, or using a SALT2 model re-trained with redder rest-frame data,
will reduce this source of uncertainty.

All Foundation SNe also have observations
in $i$ and $z$ filters, wavelengths at which SNe\,Ia are
better standard candles.  Currently, the SALT2 model
extends redward only to 7000 \AA\ and therefore SALT2
cannot fit these data.  Given that the Foundation
data extend to rest-frame $z$ band, they could be used
to re-train the SALT2 model on redder data and maximize the utility
of this data set.

In many ways the Foundation data allow us to remove
substantial uncertainty regarding SN\,Ia physics due to
the similarity of the PS1-observed SN data across
the redshift range.  The \lowzdistsighyphen\ discrepancy between
Foundation data and previous low-$z$ data may be a statistical
fluctuation but could also indicate
that unforeseen systematic uncertainties, related either to photometric
calibration or SN physics, are affecting the data.

Many additional SN\,Ia survey telescopes
are currently collecting data or will be in the near
future (DES, \citealp{DES18}; LSST, \citealp{LSST18};
{\it WFIRST}, \citealp{Hounsell18}).  Most or all
of these missions will have large samples of photometrically
classified SNe, and improving some of the methods discussed here to marginalize
over CC\,SN contamination
will be vital to the accuracy of future cosmological constraints.
However, these surveys will not observe
large samples of low-$z$ SNe.  With every subsequent addition to the Foundation
sample, the Hubble diagram anchor will improve in statistical
precision and the properties of dark energy will be
better understood.

\acknowledgements
We would like to thank D.\ Brout for useful discussions and assistance
with the BAO likelihoods and the anonymous referee for many
helpful comments.
D.O.J.\ is supported by a Gordon and Betty Moore Foundation
postdoctoral fellowship at the University of California, Santa Cruz.
This manuscript is based upon work supported by the National Aeronautics
and Space Administration under Contracts No.\ NNG16PJ34C and
NNG17PX03C issued through the {\it WFIRST} Science Investigation Teams Program.
R.J.F.\ and D.S.\ were supported in part by NASA grant 14-WPS14-0048.
The UCSC team is supported in part by NASA grants NNG16PJ34G and
NNG17PX03C; NSF grants AST-1518052 and AST-1815935; the Gordon \&
Betty Moore Foundation; the Heising-Simons Foundation;
and by fellowships from the Alfred P. Sloan Foundation
and the David and Lucile Packard Foundation to R.J.F.
This work was supported in part
by the Kavli Institute for Cosmological Physics at the University of
Chicago through grant NSF PHY-1125897 and an endowment from the Kavli
Foundation and its founder Fred Kavli.  D.S. is supported by NASA through
Hubble Fellowship grant HST-HF2-51383.001 awarded by the Space Telescope
Science Institute, which is operated by the Association of Universities
for Research in Astronomy, Inc., for NASA, under contract NAS 5-26555.
Supernova research at the Harvard-Smithsonian Center for Astrophysics
is supported in part by NSF grant AST-1516854.

The computations in this paper were aided by the
University of Chicago Research Computing Center
and the Odyssey cluster at Harvard University.  The
Odyssey cluster is supported by the FAS Division of Science, Research
Computing Group at Harvard University.

\appendix

\section{The BEAMS Likelihood Model}
In its simplest formulation, BEAMS models SN\,Ia and CC\,SN distances
with Gaussian likelihoods where the mean of the Gaussians
$-$ the distance, or average distance within a redshift bin $-$ are free
parameters.  From \citet{Kunz07} and using the notation in \citet{Kessler17} and \citetalias{Jones18},
the probablility of the model given the data, D, is:

\begin{equation}
  \begin{split}
        P(\theta|D) &\propto P(\theta) \times \prod_{i=1}^N(\mathcal{L}_i^{\mathrm{Ia}} + \mathcal{L}_i^{\mathrm{CC}}),\\
        \mathcal{L}_i^{\mathrm{Ia}} &= \mathrm{\tilde{P}}_i\mathrm{(Ia)D_{Ia}}(z_i,\mu_i,\mu_{\mathrm{model},i}),\\
        \mathrm{D_{Ia}} &= \frac{1}{\sigma_{\mu}\sqrt{2\pi}}\exp[-\chi^2_{\mathrm{HD}}/2].
  \end{split}
  \label{eqn:beamslike}
\end{equation}

\noindent P($\theta|$D) is the posterior probability of the model given the data, D, which is proportional
to the priors on free parameters $\theta$ and the product of the likelihoods for the N SNe in the sample.
For the $i$th SN, $\mathcal{L}_i^{\mathrm{Ia}}$ and $\mathcal{L}_i^{\mathrm{CC}}$ are the likelihoods of
the SN\,Ia and CC\,SN distribution, respectively.
The CC\,SN likelihood, $\mathcal{L}_i^{\mathrm{CC}}$, is identical to
the form of the Ia likelihood in Equation \ref{eqn:beamslike} except that the $\mu_{\mathrm{model},i}$ term is now the
related to the brightness of CC\,SNe, shape- and color-corrected as though they
were SNe\,Ia, rather than the cosmological distance modulus derived from
SN\,Ia standardization.
$\mathrm{\tilde{P}}_i\mathrm{(Ia)}$ is the adjusted prior probability that the $i$th SN is of Type Ia
(see Equation \ref{eqn:norm} below), and $\chi^2_{\mathrm{HD}}$
is the $\chi^2$ of the model distance compared to the measured distance.

In the methods outlined below,
$\mu_{\mathrm{model},i}$ is allowed to depend on the redshift, and distances are computed by
including the Tripp formula, Equation \ref{eqn:salt2}, in the likelihood.  The Tripp formula,
in turn, depends on the global free parameters
$\alpha$, $\beta$, and $\Delta_M$.
The PSBEAMS and BBC implementations of BEAMS are discussed
in much greater detail in \citetalias{Jones18} and \citet{Kessler17},
respectively.  However we give a broad outline of both methods,
and the differences between them, below.

\subsection{The PSBEAMS Method}

In the PSBEAMS likelihood, $\mu_{\mathrm{model}}$,
is linearly interpolated between a series of log($z$)-spaced control
points $\vec{z_b}$ across the redshift range (0.01,0.7).
\citet{Betoule14} find that the difference between this
approximation and $\Lambda$CDM is always smaller than 1~mmag across
the $0.01 < z < 1.3$ redshift range (see also \citealp{Marriner11}, who first
used redshift-binned distances that were independent of cosmological parameters).
PSBEAMS allows the dispersion
of SNe\,Ia, $\Sigma_{\mathrm{Ia}}$, to be a free parameter,
such that in Equation \ref{eqn:beamslike} we have $\sigma_{\mu}^2 = \sigma_i^2 + \Sigma_{\mathrm{Ia}}^2$.
The standard deviation of the SN\,Ia likelihood $-$ approximately equivalent to the intrinsic dispersion
of the SN\,Ia population $-$ is held constant as a function of
redshift.  The CC\,SN population, however, has a mean and dispersion that
evolve with redshift according to the same linear interpolation
method as $\mu_{\mathrm{model}}$.
The SN\,Ia and CC\,SN components of the likelihood
are multiplied by prior probabilities (discussed in the previous section) that
a given SN is a CC\,SN or a Type Ia.  As this is an abbreviated discussion
of the PSBEAMS likelihood model, we direct the reader to the full
mathematical formalism, which is given in \citetalias{Jones18}, \S3.1.

The relation between SN\,Ia distance and host-galaxy mass is
modeled by using two Gaussian distributions to model the
SN\,Ia likelihood, one for SNe\,Ia in high-mass host galaxies and
one for those in low-mass host galaxies (with a default high/low-mass
boundary at log(M$_{\ast}$/M$_{\odot}$) $=$ 10).
$\mathcal{L}_i^{\mathrm{Ia}}$ then becomes
$\mathcal{L}_i^{Ia,M<10} + \mathcal{L}_i^{Ia,M>10}$.  Each of these two likelihoods include
the prior probability that a given host galaxy
is high- or low-mass, and those probabilities are given by
the (assumed Gaussian) uncertainties on host mass from the method
discussed in \S\ref{sec:host}.

In total, the baseline likelihood model has 41 free parameters.
First, the
mean of the SN\,Ia distribution at 25 control points, $\mu_{\mathrm{model}}(\vec{z_b})$, which is
equivalent to the SN\,Ia distance measurement. Second, the mean
and standard deviation of the CC\,SN distribution at 5
control points (10 parameters).  Additionally, the Tripp parameters $\alpha$
and $\beta$, the host-galaxy mass step $\Delta_M$,
and the standard deviation of the SN\,Ia
distribution $\Sigma_{Ia}$ (not a function of redshift) are all free parameters.
Finally, we allow the prior probabilities that a SN is Type Ia
to be linearly shifted and re-normalized (2 parameters).  The relationship
between the normalization factor, $A$, the shift parameter, $S$, and
the adjusted probabilities for the $i$th SN are given by:

\begin{equation}
\begin{split}
\tilde{\mathrm{P}}_i(\mathrm{Ia}) = \frac{\mathrm{A} \times (\mathrm{P}_i\mathrm{(Ia) + S})}{1 - (\mathrm{P}_i(\mathrm{Ia}) + \mathrm{S}) + \mathrm{A} \times (\mathrm{P}_i(\mathrm{Ia}) + \mathrm{S})}\\
0 < \tilde{\mathrm{P}}_i\mathrm{(Ia)} < 1.
\end{split}
\label{eqn:norm}
\end{equation}

We apply loose priors on all parameters excepting the mean of
the SN\,Ia distribution $\mu_{\mathrm{model}}(\vec{z_b})$, so as not to impose priors on the SN\,Ia
distances, and sample the posterior with a Markov
Chain Monte Carlo (MCMC) algorithm (the \texttt{emcee}
Parallel-tempered Ensemble sampler; \citealp{Foreman13}).
Additional details about this procedure are given in
\citet{Jones17} and \citetalias{Jones18}, including the specific likelihood
equations and the values for the priors on each parameter.
The code is publicly available at \url{https://github.com/djones1040/BEAMS}.

\subsection{The BEAMS with Bias Corrections Method}
\label{sec:bbc}

The BBC likelihood has the same form as Equation \ref{eqn:beamslike} for SNe\,Ia,
but uses Monte Carlo SNANA simulations of the expected
CC\,SN population in the MDS to put a non-Gaussian
prior on the $z$-dependent Hubble residuals expected from CC\,SN contamination.
Alternatively, a more flexible, parameterized CC\,SN
model from \citet{Hlozek12} can be specified within BBC.
In contrast to the PSBEAMS method, which also uses a parameterized CC\,SN model,
\citet{Hlozek12} treats the CC\,SN distances and dispersion
as a polynomial function of the redshift.

Using the BBC method with a simulated CC\,SN
prior could be sensitive to inaccurate CC modeling in the simulations
the simulations.  However, we use both the simulation-based and \citet{Hlozek12}
CC\,SN likelihood models to ameliorate this concern.  We also
estimate biases in this method by simulating several variants of the
CC\,SN population (\S\ref{sec:sys}), each of which differ from the CC\,SN simulation used to
generate the simulation-based prior.

The BBC method has a parameter to re-normalize P(Ia) values, effectively the $A$ parameter
in Equation \ref{eqn:norm}, but unlike PSBEAMS does not have a parameter to
linearly shift P(Ia) values.  We found in previous work \citep{Jones17}
that PSNID is an excellent classifier, but classifies
a significant fraction of CC\,SNe as being of Type Ia
with 100\% confidence.  For this reason,
allowing a free parameter that linearly shifts probabilities up/down
(but restricting to $0 < \mathrm{P(Ia)} < 1$) is a necessary
parameter when using PSNID probabilities \citep{Jones17}.
Therefore, we use the NN
classifier as our baseline BBC classifier but include the use of
PSNID probabilities in the systematic uncertainty budget.

We test the BBC method to ensure that it is as reliable as our
nominal method (which was validated in \citealp{Jones17}).
We simulate photometrically classified SN samples
and investigate the change in measured SN\,Ia distances when CC\,SNe are included
in the sample versus when they are excluded.  This comparison gives
the expected bias in final SN\,Ia distances that could be
caused by inaccurate prior probabilities or imperfections in the likelihood model.
To ensure that our test sample is as close as possible to the real
data, we combine the set of real, spectroscopically classified SNe\,Ia used in this analysis
with 25 simulated samples of photometrically classified SNe.
Each of the 25 simulated samples contains the same number of photometrically
classified SNe as our real data; however, unlike the real photometrically
classified data, we know the true types of every SN.
For each of the 25 samples, we estimate the bias in measured distance by
comparing the BBC results from a sample \textit{with} simulated CC\,SNe
to that same sample after excluding simulated CC\,SNe.
The bias is $\lesssim$1~mmag across the entire redshift range
(Figure \ref{fig:ccmodbias}; ``default'' simulation).  However, we caution that
the method of estimating the probabilities is not entirely independent
of the BBC method; each SN\,Ia uses the same SALT2 simulation parameters as the Monte Carlo
simulation used to generate the data.  For this reason, we
test additional classification methods and simulations (\S\ref{sec:priors}),
which have increased systematic shifts.

\bibliographystyle{apj}

\begin{thebibliography}{108}
\expandafter\ifx\csname natexlab\endcsname\relax\def\natexlab#1{#1}\fi

\bibitem[{{Alam} {et~al.}(2017){Alam}, {Ata}, {Bailey}, {Beutler}, {Bizyaev},
  {Blazek}, {Bolton}, {Brownstein}, {Burden}, {Chuang}, {Comparat}, {Cuesta},
  {Dawson}, {Eisenstein}, {Escoffier}, {Gil-Mar{\'{\i}}n}, {Grieb}, {Hand},
  {Ho}, {Kinemuchi}, {Kirkby}, {Kitaura}, {Malanushenko}, {Malanushenko},
  {Maraston}, {McBride}, {Nichol}, {Olmstead}, {Oravetz}, {Padmanabhan},
  {Palanque-Delabrouille}, {Pan}, {Pellejero-Ibanez}, {Percival}, {Petitjean},
  {Prada}, {Price-Whelan}, {Reid}, {Rodr{\'{\i}}guez-Torres}, {Roe}, {Ross},
  {Ross}, {Rossi}, {Rubi{\~n}o-Mart{\'{\i}}n}, {Saito}, {Salazar-Albornoz},
  {Samushia}, {S{\'a}nchez}, {Satpathy}, {Schlegel}, {Schneider},
  {Sc{\'o}ccola}, {Seo}, {Sheldon}, {Simmons}, {Slosar}, {Strauss}, {Swanson},
  {Thomas}, {Tinker}, {Tojeiro}, {Maga{\~n}a}, {Vazquez}, {Verde}, {Wake},
  {Wang}, {Weinberg}, {White}, {Wood-Vasey}, {Y{\`e}che}, {Zehavi}, {Zhai}, \&
  {Zhao}}]{Alam17}
{Alam}, S., {Ata}, M., {Bailey}, S., {et~al.} 2017, \mnras, 470, 2617

\bibitem[{{Amanullah} {et~al.}(2015){Amanullah}, {Johansson}, {Goobar},
  {Ferretti}, {Papadogiannakis}, {Petrushevska}, {Brown}, {Cao}, {Contreras},
  {Dahle}, {Elias-Rosa}, {Fynbo}, {Gorosabel}, {Guaita}, {Hangard}, {Howell},
  {Hsiao}, {Kankare}, {Kasliwal}, {Leloudas}, {Lundqvist}, {Mattila}, {Nugent},
  {Phillips}, {Sandberg}, {Stanishev}, {Sullivan}, {Taddia}, {{\"O}stlin},
  {Asadi}, {Herrero-Illana}, {Jensen}, {Karhunen}, {Lazarevic}, {Varenius},
  {Santos}, {Sridhar}, {Wallstr{\"o}m}, \& {Wiegert}}]{Amanullah15}
{Amanullah}, R., {Johansson}, J., {Goobar}, A., {et~al.} 2015, \mnras, 453,
  3300

\bibitem[{{Anderson} {et~al.}(2014){Anderson}, {Aubourg}, {Bailey}, {Beutler},
  {Bhardwaj}, {Blanton}, {Bolton}, {Brinkmann}, {Brownstein}, {Burden},
  {Chuang}, {Cuesta}, {Dawson}, {Eisenstein}, {Escoffier}, {Gunn}, {Guo}, {Ho},
  {Honscheid}, {Howlett}, {Kirkby}, {Lupton}, {Manera}, {Maraston}, {McBride},
  {Mena}, {Montesano}, {Nichol}, {Nuza}, {Olmstead}, {Padmanabhan},
  {Palanque-Delabrouille}, {Parejko}, {Percival}, {Petitjean}, {Prada},
  {Price-Whelan}, {Reid}, {Roe}, {Ross}, {Ross}, {Sabiu}, {Saito}, {Samushia},
  {S{\'a}nchez}, {Schlegel}, {Schneider}, {Scoccola}, {Seo}, {Skibba},
  {Strauss}, {Swanson}, {Thomas}, {Tinker}, {Tojeiro}, {Maga{\~n}a}, {Verde},
  {Wake}, {Weaver}, {Weinberg}, {White}, {Xu}, {Y{\`e}che}, {Zehavi}, \&
  {Zhao}}]{Anderson14}
{Anderson}, L., {Aubourg}, {\'E}., {Bailey}, S., {et~al.} 2014, \mnras, 441, 24

\bibitem[{{Arnouts} \& {Ilbert}(2011)}]{Arnouts11}
{Arnouts}, S., \& {Ilbert}, O. 2011, {LePHARE: Photometric Analysis for
  Redshift Estimate}, Astrophysics Source Code Library

\bibitem[{{Astier} {et~al.}(2006){Astier}, {Guy}, {Regnault}, {Pain},
  {Aubourg}, {Balam}, {Basa}, {Carlberg}, {Fabbro}, {Fouchez}, {Hook},
  {Howell}, {Lafoux}, {Neill}, {Palanque-Delabrouille}, {Perrett}, {Pritchet},
  {Rich}, {Sullivan}, {Taillet}, {Aldering}, {Antilogus}, {Arsenijevic},
  {Balland}, {Baumont}, {Bronder}, {Courtois}, {Ellis}, {Filiol}, {Gon{\c
  c}alves}, {Goobar}, {Guide}, {Hardin}, {Lusset}, {Lidman}, {McMahon},
  {Mouchet}, {Mourao}, {Perlmutter}, {Ripoche}, {Tao}, \& {Walton}}]{Astier06}
{Astier}, P., {Guy}, J., {Regnault}, N., {et~al.} 2006, \aap, 447, 31

\bibitem[{{Becker}(2015)}]{Becker15}
{Becker}, A. 2015, {HOTPANTS: High Order Transform of PSF ANd Template
  Subtraction}, Astrophysics Source Code Library

\bibitem[{{Bertin} \& {Arnouts}(1996)}]{Bertin96}
{Bertin}, E., \& {Arnouts}, S. 1996, \aaps, 117, 393

\bibitem[{{Betoule} {et~al.}(2014){Betoule}, {Kessler}, {Guy}, {Mosher},
  {Hardin}, {Biswas}, {Astier}, {El-Hage}, {Konig}, {Kuhlmann}, {Marriner},
  {Pain}, {Regnault}, {Balland}, {Bassett}, {Brown}, {Campbell}, {Carlberg},
  {Cellier-Holzem}, {Cinabro}, {Conley}, {D'Andrea}, {DePoy}, {Doi}, {Ellis},
  {Fabbro}, {Filippenko}, {Foley}, {Frieman}, {Fouchez}, {Galbany}, {Goobar},
  {Gupta}, {Hill}, {Hlozek}, {Hogan}, {Hook}, {Howell}, {Jha}, {Le Guillou},
  {Leloudas}, {Lidman}, {Marshall}, {M{\"o}ller}, {Mour{\~a}o}, {Neveu},
  {Nichol}, {Olmstead}, {Palanque-Delabrouille}, {Perlmutter}, {Prieto},
  {Pritchet}, {Richmond}, {Riess}, {Ruhlmann-Kleider}, {Sako}, {Schahmaneche},
  {Schneider}, {Smith}, {Sollerman}, {Sullivan}, {Walton}, \&
  {Wheeler}}]{Betoule14}
{Betoule}, M., {Kessler}, R., {Guy}, J., {et~al.} 2014, \aap, 568, A22

\bibitem[{{Bohlin}(2014)}]{Bolin14}
{Bohlin}, R.~C. 2014, ArXiv e-prints

\bibitem[{{Brout} {et~al.}(2018){Brout}, {Scolnic}, {Kessler}, {D'Andrea},
  {Davis}, {Gupta}, {Hinton}, {Kim}, {Lasker}, {Lidman}, {Macaulay},
  {M{\"o}ller}, {Nichol}, {Sako}, {Smith}, {Sullivan}, {Zhang}, {Andersen},
  {Asorey}, {Avelino}, {Bassett}, {Brown}, {Calcino}, {Carollo}, {Challis},
  {Childress}, {Clocchiatti}, {Filippenko}, {Foley}, {Galbany}, {Glazebrook},
  {Hoormann}, {Kasai}, {Kirshner}, {Kuehn}, {Kuhlmann}, {Lewis}, {Mandel},
  {March}, {Miranda}, {Morganson}, {Muthukrishna}, {Nugent}, {Palmese}, {Pan},
  {Sharp}, {Sommer}, {Swann}, {Thomas}, {Tucker}, {Uddin}, {Wester}, {Abbott},
  {Allam}, {Annis}, {Avila}, {Bechtol}, {Bernstein}, {Bertin}, {Brooks},
  {Burke}, {Carnero Rosell}, {Carrasco Kind}, {Carretero}, {Castander},
  {Cunha}, {da Costa}, {Davis}, {De Vicente}, {DePoy}, {Desai}, {Diehl},
  {Doel}, {Drlica-Wagner}, {Eifler}, {Estrada}, {Fernandez}, {Flaugher},
  {Fosalba}, {Frieman}, {Garc{\'{\i}}a-Bellido}, {Gruen}, {Gruendl},
  {Gutierrez}, {Hartley}, {Hollowood}, {Honscheid}, {Hoyle}, {James}, {Jarvis},
  {Jeltema}, {Krause}, {Lahav}, {Li}, {Lima}, {Maia}, {Marriner}, {Marshall},
  {Martini}, {Menanteau}, {Miller}, {Miquel}, {Ogando}, {Plazas}, {Romer},
  {Roodman}, {Rykoff}, {Sanchez}, {Santiago}, {Scarpine}, {Schubnell},
  {Serrano}, {Sevilla-Noarbe}, {Smith}, {Soares-Santos}, {Sobreira}, {Suchyta},
  {Swanson}, {Tarle}, {Thomas}, {Troxel}, {Tucker}, {Vikram}, {Walker}, \&
  {Zhang}}]{Brout18}
{Brout}, D., {Scolnic}, D., {Kessler}, R., {et~al.} 2018, ArXiv e-prints

\bibitem[{{Brown} {et~al.}(2018){Brown}, {Stanek}, {Holoien}, {Kochanek},
  {Shappee}, {Prieto}, {Dong}, {Chen}, {Thompson}, {Beacom}, {Stritzinger},
  {Bersier}, \& {Brimacombe}}]{Brown18}
{Brown}, J.~S., {Stanek}, K.~Z., {Holoien}, T.~W.-S., {et~al.} 2018, ArXiv
  e-prints

\bibitem[{{Bruzual} \& {Charlot}(2003)}]{Bruzual03}
{Bruzual}, G., \& {Charlot}, S. 2003, \mnras, 344, 1000

\bibitem[{{Bulla} {et~al.}(2018){Bulla}, {Goobar}, \& {Dhawan}}]{Bulla18}
{Bulla}, M., {Goobar}, A., \& {Dhawan}, S. 2018, \mnras, 479, 3663

\bibitem[{{Burns} {et~al.}(2014){Burns}, {Stritzinger}, {Phillips}, {Hsiao},
  {Contreras}, {Persson}, {Folatelli}, {Boldt}, {Campillay}, {Castell{\'o}n},
  {Freedman}, {Madore}, {Morrell}, {Salgado}, \& {Suntzeff}}]{Burns14}
{Burns}, C.~R., {Stritzinger}, M., {Phillips}, M.~M., {et~al.} 2014, \apj, 789,
  32

\bibitem[{{Burns} {et~al.}(2018){Burns}, {Parent}, {Phillips}, {Stritzinger},
  {Krisciunas}, {Suntzeff}, {Hsiao}, {Contreras}, {Anais}, {Boldt}, {Busta},
  {Campillay}, {Castell{\'o}n}, {Folatelli}, {Freedman}, {Gonz{\'a}lez},
  {Hamuy}, {Heoflich}, {Krzeminski}, {Madore}, {Morrell}, {Persson}, {Roth},
  {Salgado}, {Ser{\'o}n}, \& {Torres}}]{Burns18}
{Burns}, C.~R., {Parent}, E., {Phillips}, M.~M., {et~al.} 2018, \apj, 869, 56

\bibitem[{{Chambers} {et~al.}(2016){Chambers}, {Magnier}, {Metcalfe},
  {Flewelling}, {Huber}, {Waters}, {Denneau}, {Draper}, {Farrow}, {Finkbeiner},
  {Holmberg}, {Koppenhoefer}, {Price}, {Saglia}, {Schlafly}, {Smartt},
  {Sweeney}, {Wainscoat}, {Burgett}, {Grav}, {Heasley}, {Hodapp}, {Jedicke},
  {Kaiser}, {Kudritzki}, {Luppino}, {Lupton}, {Monet}, {Morgan}, {Onaka},
  {Stubbs}, {Tonry}, {Banados}, {Bell}, {Bender}, {Bernard}, {Botticella},
  {Casertano}, {Chastel}, {Chen}, {Chen}, {Cole}, {Deacon}, {Frenk},
  {Fitzsimmons}, {Gezari}, {Goessl}, {Goggia}, {Goldman}, {Grebel}, {Hambly},
  {Hasinger}, {Heavens}, {Heckman}, {Henderson}, {Henning}, {Holman}, {Hopp},
  {Ip}, {Isani}, {Keyes}, {Koekemoer}, {Kotak}, {Long}, {Lucey}, {Liu},
  {Martin}, {McLean}, {Morganson}, {Murphy}, {Nieto-Santisteban}, {Norberg},
  {Peacock}, {Pier}, {Postman}, {Primak}, {Rae}, {Rest}, {Riess}, {Riffeser},
  {Rix}, {Roser}, {Schilbach}, {Schultz}, {Scolnic}, {Szalay}, {Seitz},
  {Shiao}, {Small}, {Smith}, {Soderblom}, {Taylor}, {Thakar}, {Thiel},
  {Thilker}, {Urata}, {Valenti}, {Walter}, {Watters}, {Werner}, {White},
  {Wood-Vasey}, \& {Wyse}}]{Chambers16}
{Chambers}, K.~C., {Magnier}, E.~A., {Metcalfe}, N., {et~al.} 2016, ArXiv
  e-prints

\bibitem[{{Chevallier} \& {Polarski}(2001)}]{Chevallier01}
{Chevallier}, M., \& {Polarski}, D. 2001, International Journal of Modern
  Physics D, 10, 213

\bibitem[{{Childress} {et~al.}(2014){Childress}, {Wolf}, \&
  {Zahid}}]{Childress14}
{Childress}, M.~J., {Wolf}, C., \& {Zahid}, H.~J. 2014, \mnras, 445, 1898

\bibitem[{{Chotard} {et~al.}(2011){Chotard}, {Gangler}, {Aldering},
  {Antilogus}, {Aragon}, {Bailey}, {Baltay}, {Bongard}, {Buton}, {Canto},
  {Childress}, {Copin}, {Fakhouri}, {Hsiao}, {Kerschhaggl}, {Kowalski},
  {Loken}, {Nugent}, {Paech}, {Pain}, {Pecontal}, {Pereira}, {Perlmutter},
  {Rabinowitz}, {Runge}, {Scalzo}, {Smadja}, {Tao}, {Thomas}, {Weaver}, {Wu},
  \& {Nearby Supernova Factory}}]{Chotard11}
{Chotard}, N., {Gangler}, E., {Aldering}, G., {et~al.} 2011, \aap, 529, L4

\bibitem[{{Conley} {et~al.}(2011){Conley}, {Guy}, {Sullivan}, {Regnault},
  {Astier}, {Balland}, {Basa}, {Carlberg}, {Fouchez}, {Hardin}, {Hook},
  {Howell}, {Pain}, {Palanque-Delabrouille}, {Perrett}, {Pritchet}, {Rich},
  {Ruhlmann-Kleider}, {Balam}, {Baumont}, {Ellis}, {Fabbro}, {Fakhouri},
  {Fourmanoit}, {Gonz{\'a}lez-Gait{\'a}n}, {Graham}, {Hudson}, {Hsiao},
  {Kronborg}, {Lidman}, {Mourao}, {Neill}, {Perlmutter}, {Ripoche}, {Suzuki},
  \& {Walker}}]{Conley11}
{Conley}, A., {Guy}, J., {Sullivan}, M., {et~al.} 2011, \apjs, 192, 1

\bibitem[{{Contreras} {et~al.}(2010){Contreras}, {Hamuy}, {Phillips},
  {Folatelli}, {Suntzeff}, {Persson}, {Stritzinger}, {Boldt}, {Gonz{\'a}lez},
  {Krzeminski}, {Morrell}, {Roth}, {Salgado}, {Jos{\'e} Maureira}, {Burns},
  {Freedman}, {Madore}, {Murphy}, {Wyatt}, {Li}, \& {Filippenko}}]{Contreras10}
{Contreras}, C., {Hamuy}, M., {Phillips}, M.~M., {et~al.} 2010, \aj, 139, 519

\bibitem[{{Dai} \& {Wang}(2016)}]{Dai16}
{Dai}, M., \& {Wang}, Y. 2016, \mnras, 459, 1819

\bibitem[{{DES Collaboration} {et~al.}(2018){DES Collaboration}, {Abbott},
  {Allam}, {Andersen}, {Angus}, {Asorey}, {Avelino}, {Avila}, {Bassett},
  {Bechtol}, {Bernstein}, {Bertin}, {Brooks}, {Brout}, {Brown}, {Burke},
  {Calcino}, {Carnero Rosell}, {Carollo}, {Carrasco Kind}, {Carretero},
  {Casas}, {Castander}, {Cawthon}, {Challis}, {Childress}, {Clocchiatti},
  {Cunha}, {D'Andrea}, {da Costa}, {Davis}, {Davis}, {De Vicente}, {DePoy},
  {Desai}, {Diehl}, {Doel}, {Drlica-Wagner}, {Eifler}, {Evrard}, {Fernandez},
  {Filippenko}, {Finley}, {Flaugher}, {Foley}, {Fosalba}, {Frieman}, {Galbany},
  {Garcia-Bellido}, {Gaztanaga}, {Giannantonio}, {Glazebrook}, {Goldstein},
  {Gonzalez-Gaitan}, {Gruen}, {Gruendl}, {Gschwend}, {Gupta}, {Gutierrez},
  {Hartley}, {Hinton}, {Hollowood}, {Honscheid}, {Hoormann}, {Hoyle}, {James},
  {Jeltema}, {Johnson}, {Johnson}, {Kasai}, {Kent}, {Kessler}, {Kim},
  {Kirshner}, {Kovacs}, {Krause}, {Kron}, {Kuehn}, {Kuhlmann}, {Kuropatkin},
  {Lahav}, {Lasker}, {Lewis}, {Li}, {Lidman}, {Lima}, {Lin}, {Macaulay},
  {Maia}, {Mandel}, {March}, {Marriner}, {Marshall}, {Martini}, {Menanteau},
  {Miller}, {Miquel}, {Miranda}, {Mohr}, {Morganson}, {Muthukrishna},
  {M{\"o}ller}, {Neilsen}, {Nichol}, {Nord}, {Nugent}, {Ogando}, {Palmese},
  {Pan}, {Plazas}, {Pursiainen}, {Romer}, {Roodman}, {Rozo}, {Rykoff}, {Sako},
  {Sanchez}, {Scarpine}, {Schindler}, {Schubnell}, {Scolnic}, {Serrano},
  {Sevilla-Noarbe}, {Sharp}, {Smith}, {Soares-Santos}, {Sobreira}, {Sommer},
  {Spinka}, {Suchyta}, {Sullivan}, {Swann}, {Tarle}, {Thomas}, {Thomas},
  {Troxel}, {Tucker}, {Uddin}, {Walker}, {Wiseman}, {Wolf}, {Yanny}, {Zhang},
  \& {Zhang}}]{DES18}
{DES Collaboration}, {Abbott}, T.~M.~C., {Allam}, S., {et~al.} 2018, ArXiv
  e-prints

\bibitem[{{Filippenko} {et~al.}(2001){Filippenko}, {Li}, {Treffers}, \&
  {Modjaz}}]{Filippenko01}
{Filippenko}, A.~V., {Li}, W.~D., {Treffers}, R.~R., \& {Modjaz}, M. 2001, in
  Astronomical Society of the Pacific Conference Series, Vol. 246, IAU Colloq.
  183: Small Telescope Astronomy on Global Scales, ed. B.~{Paczynski}, W.-P.
  {Chen}, \& C.~{Lemme}, 121

\bibitem[{{Flewelling} {et~al.}(2016){Flewelling}, {Magnier}, {Chambers},
  {Heasley}, {Holmberg}, {Huber}, {Sweeney}, {Waters}, {Chen}, {Farrow},
  {Hasinger}, {Henderson}, {Long}, {Metcalfe}, {Nieto-Santisteban}, {Norberg},
  {Saglia}, {Szalay}, {Rest}, {Thakar}, {Tonry}, {Valenti}, {Werner}, {White},
  {Denneau}, {Draper}, {Hodapp}, {Jedicke}, {Kaiser}, {Kudritzki}, {Price},
  {Wainscoat}, {Chastel}, {McClean}, {Postman}, \& {Shiao}}]{Flewelling16}
{Flewelling}, H.~A., {Magnier}, E.~A., {Chambers}, K.~C., {et~al.} 2016, ArXiv
  e-prints

\bibitem[{{Folatelli} {et~al.}(2010){Folatelli}, {Phillips}, {Burns},
  {Contreras}, {Hamuy}, {Freedman}, {Persson}, {Stritzinger}, {Suntzeff},
  {Krisciunas}, {Boldt}, {Gonz{\'a}lez}, {Krzeminski}, {Morrell}, {Roth},
  {Salgado}, {Madore}, {Murphy}, {Wyatt}, {Li}, {Filippenko}, \&
  {Miller}}]{Folatelli10}
{Folatelli}, G., {Phillips}, M.~M., {Burns}, C.~R., {et~al.} 2010, \aj, 139,
  120

\bibitem[{{Foley} \& {Mandel}(2013)}]{Foley13}
{Foley}, R.~J., \& {Mandel}, K. 2013, \apj, 778, 167

\bibitem[{{Foley} {et~al.}(2018){Foley}, {Scolnic}, {Rest}, {Jha}, {Pan},
  {Riess}, {Challis}, {Chambers}, {Coulter}, {Dettman}, {Foley}, {Fox},
  {Huber}, {Jones}, {Kilpatrick}, {Kirshner}, {Schultz}, {Siebert},
  {Flewelling}, {Gibson}, {Magnier}, {Miller}, {Primak}, {Smartt}, {Smith},
  {Wainscoat}, {Waters}, \& {Willman}}]{Foley18}
{Foley}, R.~J., {Scolnic}, D., {Rest}, A., {et~al.} 2018, \mnras, 475, 193

\bibitem[{{Foreman-Mackey} {et~al.}(2013){Foreman-Mackey}, {Hogg}, {Lang}, \&
  {Goodman}}]{Foreman13}
{Foreman-Mackey}, D., {Hogg}, D.~W., {Lang}, D., \& {Goodman}, J. 2013, \pasp,
  125, 306

\bibitem[{{Gaia Collaboration} {et~al.}(2016){Gaia Collaboration}, {Prusti},
  {de Bruijne}, {Brown}, {Vallenari}, {Babusiaux}, {Bailer-Jones}, {Bastian},
  {Biermann}, {Evans}, \& et~al.}]{Prusti16}
{Gaia Collaboration}, {Prusti}, T., {de Bruijne}, J.~H.~J., {et~al.} 2016,
  \aap, 595, A1

\bibitem[{{Garnavich} {et~al.}(1998){Garnavich}, {Jha}, {Challis},
  {Clocchiatti}, {Diercks}, {Filippenko}, {Gilliland}, {Hogan}, {Kirshner},
  {Leibundgut}, {Phillips}, {Reiss}, {Riess}, {Schmidt}, {Schommer}, {Smith},
  {Spyromilio}, {Stubbs}, {Suntzeff}, {Tonry}, \& {Carroll}}]{Garnavich98}
{Garnavich}, P.~M., {Jha}, S., {Challis}, P., {et~al.} 1998, \apj, 509, 74

\bibitem[{{Goobar} {et~al.}(2018){Goobar}, {Dhawan}, \& {Scolnic}}]{Goobar18}
{Goobar}, A., {Dhawan}, S., \& {Scolnic}, D. 2018, \mnras, 477, L75

\bibitem[{{Guillochon} {et~al.}(2017){Guillochon}, {Parrent}, {Kelley}, \&
  {Margutti}}]{Guillochon17}
{Guillochon}, J., {Parrent}, J., {Kelley}, L.~Z., \& {Margutti}, R. 2017, \apj,
  835, 64

\bibitem[{{Gupta} {et~al.}(2016){Gupta}, {Kuhlmann}, {Kovacs}, {Spinka},
  {Kessler}, {Goldstein}, {Liotine}, {Pomian}, {D'Andrea}, {Sullivan},
  {Carretero}, {Castander}, {Nichol}, {Finley}, {Fischer}, {Foley}, {Kim},
  {Papadopoulos}, {Sako}, {Scolnic}, {Smith}, {Tucker}, {Uddin}, {Wolf},
  {Yuan}, {Abbott}, {Abdalla}, {Benoit-L{\'e}vy}, {Bertin}, {Brooks}, {Carnero
  Rosell}, {Carrasco Kind}, {Cunha}, {da Costa}, {Desai}, {Doel}, {Eifler},
  {Evrard}, {Flaugher}, {Fosalba}, {Gazta{\~n}aga}, {Gruen}, {Gruendl},
  {James}, {Kuehn}, {Kuropatkin}, {Maia}, {Marshall}, {Miquel}, {Plazas},
  {Romer}, {S{\'a}nchez}, {Schubnell}, {Sevilla-Noarbe}, {Sobreira}, {Suchyta},
  {Swanson}, {Tarle}, {Walker}, \& {Wester}}]{Gupta16}
{Gupta}, R.~R., {Kuhlmann}, S., {Kovacs}, E., {et~al.} 2016, \aj, 152, 154

\bibitem[{{Guy} {et~al.}(2007){Guy}, {Astier}, {Baumont}, {Hardin}, {Pain},
  {Regnault}, {Basa}, {Carlberg}, {Conley}, {Fabbro}, {Fouchez}, {Hook},
  {Howell}, {Perrett}, {Pritchet}, {Rich}, {Sullivan}, {Antilogus}, {Aubourg},
  {Bazin}, {Bronder}, {Filiol}, {Palanque-Delabrouille}, {Ripoche}, \&
  {Ruhlmann-Kleider}}]{Guy07}
{Guy}, J., {Astier}, P., {Baumont}, S., {et~al.} 2007, \aap, 466, 11

\bibitem[{{Guy} {et~al.}(2010){Guy}, {Sullivan}, {Conley}, {Regnault},
  {Astier}, {Balland}, {Basa}, {Carlberg}, {Fouchez}, {Hardin}, {Hook},
  {Howell}, {Pain}, {Palanque-Delabrouille}, {Perrett}, {Pritchet}, {Rich},
  {Ruhlmann-Kleider}, {Balam}, {Baumont}, {Ellis}, {Fabbro}, {Fakhouri},
  {Fourmanoit}, {Gonz{\'a}lez-Gait{\'a}n}, {Graham}, {Hsiao}, {Kronborg},
  {Lidman}, {Mourao}, {Perlmutter}, {Ripoche}, {Suzuki}, \& {Walker}}]{Guy10}
{Guy}, J., {Sullivan}, M., {Conley}, A., {et~al.} 2010, \aap, 523, A7

\bibitem[{{Hayden} {et~al.}(2013){Hayden}, {Gupta}, {Garnavich}, {Mannucci},
  {Nichol}, \& {Sako}}]{Hayden13}
{Hayden}, B.~T., {Gupta}, R.~R., {Garnavich}, P.~M., {et~al.} 2013, \apj, 764,
  191

\bibitem[{{Hicken} {et~al.}(2009{\natexlab{a}}){Hicken}, {Wood-Vasey},
  {Blondin}, {Challis}, {Jha}, {Kelly}, {Rest}, \& {Kirshner}}]{Hicken09b}
{Hicken}, M., {Wood-Vasey}, W.~M., {Blondin}, S., {et~al.} 2009{\natexlab{a}},
  \apj, 700, 1097

\bibitem[{{Hicken} {et~al.}(2009{\natexlab{b}}){Hicken}, {Challis}, {Jha},
  {Kirshner}, {Matheson}, {Modjaz}, {Rest}, {Wood-Vasey}, {Bakos}, {Barton},
  {Berlind}, {Bragg}, {Brice{\~n}o}, {Brown}, {Caldwell}, {Calkins}, {Cho},
  {Ciupik}, {Contreras}, {Dendy}, {Dosaj}, {Durham}, {Eriksen}, {Esquerdo},
  {Everett}, {Falco}, {Fernandez}, {Gaba}, {Garnavich}, {Graves}, {Green},
  {Groner}, {Hergenrother}, {Holman}, {Hradecky}, {Huchra}, {Hutchison},
  {Jerius}, {Jordan}, {Kilgard}, {Krauss}, {Luhman}, {Macri}, {Marrone},
  {McDowell}, {McIntosh}, {McNamara}, {Megeath}, {Mochejska}, {Munoz},
  {Muzerolle}, {Naranjo}, {Narayan}, {Pahre}, {Peters}, {Peterson}, {Rines},
  {Ripman}, {Roussanova}, {Schild}, {Sicilia-Aguilar}, {Sokoloski}, {Smalley},
  {Smith}, {Spahr}, {Stanek}, {Barmby}, {Blondin}, {Stubbs}, {Szentgyorgyi},
  {Torres}, {Vaz}, {Vikhlinin}, {Wang}, {Westover}, {Woods}, \&
  {Zhao}}]{Hicken09a}
{Hicken}, M., {Challis}, P., {Jha}, S., {et~al.} 2009{\natexlab{b}}, \apj, 700,
  331

\bibitem[{{Hicken} {et~al.}(2012){Hicken}, {Challis}, {Kirshner}, {Rest},
  {Cramer}, {Wood-Vasey}, {Bakos}, {Berlind}, {Brown}, {Caldwell}, {Calkins},
  {Currie}, {de Kleer}, {Esquerdo}, {Everett}, {Falco}, {Fernandez},
  {Friedman}, {Groner}, {Hartman}, {Holman}, {Hutchins}, {Keys}, {Kipping},
  {Latham}, {Marion}, {Narayan}, {Pahre}, {Pal}, {Peters}, {Perumpilly},
  {Ripman}, {Sipocz}, {Szentgyorgyi}, {Tang}, {Torres}, {Vaz}, {Wolk}, \&
  {Zezas}}]{Hicken12}
{Hicken}, M., {Challis}, P., {Kirshner}, R.~P., {et~al.} 2012, \apjs, 200, 12

\bibitem[{{Hill} {et~al.}(2018){Hill}, {Shariff}, {Trotta}, {Ali-Khan}, {Jiao},
  {Liu}, {Moon}, {Parker}, {Paulus}, {van Dyk}, \& {Lucy}}]{Hill18}
{Hill}, R., {Shariff}, H., {Trotta}, R., {et~al.} 2018, \mnras, 481, 2766

\bibitem[{{Hlozek} {et~al.}(2012){Hlozek}, {Kunz}, {Bassett}, {Smith},
  {Newling}, {Varughese}, {Kessler}, {Bernstein}, {Campbell}, {Dilday},
  {Falck}, {Frieman}, {Kuhlmann}, {Lampeitl}, {Marriner}, {Nichol}, {Riess},
  {Sako}, \& {Schneider}}]{Hlozek12}
{Hlozek}, R., {Kunz}, M., {Bassett}, B., {et~al.} 2012, \apj, 752, 79

\bibitem[{{Holoien} {et~al.}(2017){Holoien}, {Brown}, {Stanek}, {Kochanek},
  {Shappee}, {Prieto}, {Dong}, {Brimacombe}, {Bishop}, {Bose}, {Beacom},
  {Bersier}, {Chen}, {Chomiuk}, {Falco}, {Godoy-Rivera}, {Morrell},
  {Pojmanski}, {Shields}, {Strader}, {Stritzinger}, {Thompson}, {Wo{\'z}niak},
  {Bock}, {Cacella}, {Conseil}, {Cruz}, {Fernandez}, {Kiyota}, {Koff},
  {Krannich}, {Marples}, {Masi}, {Monard}, {Nicholls}, {Nicolas}, {Post},
  {Stone}, \& {Wiethoff}}]{Holoien17}
{Holoien}, T.~W.-S., {Brown}, J.~S., {Stanek}, K.~Z., {et~al.} 2017, \mnras,
  471, 4966

\bibitem[{{Hounsell} {et~al.}(2018){Hounsell}, {Scolnic}, {Foley}, {Kessler},
  {Miranda}, {Avelino}, {Bohlin}, {Filippenko}, {Frieman}, {Jha}, {Kelly},
  {Kirshner}, {Mandel}, {Rest}, {Riess}, {Rodney}, \& {Strolger}}]{Hounsell18}
{Hounsell}, R., {Scolnic}, D., {Foley}, R.~J., {et~al.} 2018, \apj, 867, 23

\bibitem[{{Howell}(2001)}]{Howell01}
{Howell}, D.~A. 2001, \apjl, 554, L193

\bibitem[{{Huber} {et~al.}(2015){Huber}, {Chambers}, {Flewelling}, {Willman},
  {Primak}, {Schultz}, {Gibson}, {Magnier}, {Waters}, {Tonry}, {Wainscoat},
  {Smith}, {Wright}, {Smartt}, {Foley}, {Jha}, {Rest}, \& {Scolnic}}]{Huber15}
{Huber}, M., {Chambers}, K.~C., {Flewelling}, H., {et~al.} 2015, The
  Astronomer's Telegram, 7153

\bibitem[{{Jha} {et~al.}(2006){Jha}, {Kirshner}, {Challis}, {Garnavich},
  {Matheson}, {Soderberg}, {Graves}, {Hicken}, {Alves}, {Arce}, {Balog},
  {Barmby}, {Barton}, {Berlind}, {Bragg}, {Brice{\~n}o}, {Brown}, {Buckley},
  {Caldwell}, {Calkins}, {Carter}, {Concannon}, {Donnelly}, {Eriksen},
  {Fabricant}, {Falco}, {Fiore}, {Garcia}, {G{\'o}mez}, {Grogin}, {Groner},
  {Groot}, {Haisch}, {Hartmann}, {Hergenrother}, {Holman}, {Huchra},
  {Jayawardhana}, {Jerius}, {Kannappan}, {Kim}, {Kleyna}, {Kochanek},
  {Koranyi}, {Krockenberger}, {Lada}, {Luhman}, {Luu}, {Macri}, {Mader},
  {Mahdavi}, {Marengo}, {Marsden}, {McLeod}, {McNamara}, {Megeath}, {Moraru},
  {Mossman}, {Muench}, {Mu{\~n}oz}, {Muzerolle}, {Naranjo}, {Nelson-Patel},
  {Pahre}, {Patten}, {Peters}, {Peters}, {Raymond}, {Rines}, {Schild},
  {Sobczak}, {Spahr}, {Stauffer}, {Stefanik}, {Szentgyorgyi}, {Tollestrup},
  {V{\"a}is{\"a}nen}, {Vikhlinin}, {Wang}, {Willner}, {Wolk}, {Zajac}, {Zhao},
  \& {Stanek}}]{Jha06}
{Jha}, S., {Kirshner}, R.~P., {Challis}, P., {et~al.} 2006, \aj, 131, 527

\bibitem[{{Jones} {et~al.}(2015){Jones}, {Scolnic}, \& {Rodney}}]{Jones15}
{Jones}, D.~O., {Scolnic}, D.~M., \& {Rodney}, S.~A. 2015, {PythonPhot: Simple
  DAOPHOT-type photometry in Python}, Astrophysics Source Code Library

\bibitem[{{Jones} {et~al.}(2017){Jones}, {Scolnic}, {Riess}, {Kessler}, {Rest},
  {Kirshner}, {Berger}, {Ortega}, {Foley}, {Chornock}, {Challis}, {Burgett},
  {Chambers}, {Draper}, {Flewelling}, {Huber}, {Kaiser}, {Kudritzki},
  {Metcalfe}, {Wainscoat}, \& {Waters}}]{Jones17}
{Jones}, D.~O., {Scolnic}, D.~M., {Riess}, A.~G., {et~al.} 2017, \apj, 843, 6

\bibitem[{{Jones} {et~al.}(2018{\natexlab{a}}){Jones}, {Scolnic}, {Riess},
  {Rest}, {Kirshner}, {Berger}, {Kessler}, {Pan}, {Foley}, {Chornock},
  {Ortega}, {Challis}, {Burgett}, {Chambers}, {Draper}, {Flewelling}, {Huber},
  {Kaiser}, {Kudritzki}, {Metcalfe}, {Tonry}, {Wainscoat}, {Waters}, {Gall},
  {Kotak}, {McCrum}, {Smartt}, \& {Smith}}]{Jones18}
---. 2018{\natexlab{a}}, \apj, 857, 51

\bibitem[{{Jones} {et~al.}(2018{\natexlab{b}}){Jones}, {Riess}, {Scolnic},
  {Pan}, {Johnson}, {Coulter}, {Dettman}, {Foley}, {Foley}, {Huber}, {Jha},
  {Kilpatrick}, {Kirshner}, {Rest}, {Schultz}, \& {Siebert}}]{Jones18b}
{Jones}, D.~O., {Riess}, A.~G., {Scolnic}, D.~M., {et~al.} 2018{\natexlab{b}},
  \apj, 867, 108

\bibitem[{{J{\"o}nsson} {et~al.}(2010){J{\"o}nsson}, {Sullivan}, {Hook},
  {Basa}, {Carlberg}, {Conley}, {Fouchez}, {Howell}, {Perrett}, \&
  {Pritchet}}]{Jonsson10}
{J{\"o}nsson}, J., {Sullivan}, M., {Hook}, I., {et~al.} 2010, \mnras, 405, 535

\bibitem[{{Kelly} {et~al.}(2010){Kelly}, {Hicken}, {Burke}, {Mandel}, \&
  {Kirshner}}]{Kelly10}
{Kelly}, P.~L., {Hicken}, M., {Burke}, D.~L., {Mandel}, K.~S., \& {Kirshner},
  R.~P. 2010, \apj, 715, 743

\bibitem[{{Kessler} \& {Scolnic}(2017)}]{Kessler17}
{Kessler}, R., \& {Scolnic}, D. 2017, \apj, 836, 56

\bibitem[{{Kessler} {et~al.}(2009){Kessler}, {Becker}, {Cinabro}, {Vanderplas},
  {Frieman}, {Marriner}, {Davis}, {Dilday}, {Holtzman}, {Jha}, {Lampeitl},
  {Sako}, {Smith}, {Zheng}, {Nichol}, {Bassett}, {Bender}, {Depoy}, {Doi},
  {Elson}, {Filippenko}, {Foley}, {Garnavich}, {Hopp}, {Ihara}, {Ketzeback},
  {Kollatschny}, {Konishi}, {Marshall}, {McMillan}, {Miknaitis}, {Morokuma},
  {M{\"o}rtsell}, {Pan}, {Prieto}, {Richmond}, {Riess}, {Romani}, {Schneider},
  {Sollerman}, {Takanashi}, {Tokita}, {van der Heyden}, {Wheeler}, {Yasuda}, \&
  {York}}]{Kessler09}
{Kessler}, R., {Becker}, A.~C., {Cinabro}, D., {et~al.} 2009, \apjs, 185, 32

\bibitem[{{Kessler} {et~al.}(2013){Kessler}, {Guy}, {Marriner}, {Betoule},
  {Brinkmann}, {Cinabro}, {El-Hage}, {Frieman}, {Jha}, {Mosher}, \&
  {Schneider}}]{Kessler13}
{Kessler}, R., {Guy}, J., {Marriner}, J., {et~al.} 2013, \apj, 764, 48

\bibitem[{{Kessler} {et~al.}(2015){Kessler}, {Marriner}, {Childress},
  {Covarrubias}, {D'Andrea}, {Finley}, {Fischer}, {Foley}, {Goldstein},
  {Gupta}, {Kuehn}, {Marcha}, {Nichol}, {Papadopoulos}, {Sako}, {Scolnic},
  {Smith}, {Sullivan}, {Wester}, {Yuan}, {Abbott}, {Abdalla}, {Allam},
  {Benoit-L{\'e}vy}, {Bernstein}, {Bertin}, {Brooks}, {Carnero Rosell},
  {Carrasco Kind}, {Castander}, {Crocce}, {da Costa}, {Desai}, {Diehl},
  {Eifler}, {Fausti Neto}, {Flaugher}, {Frieman}, {Gerdes}, {Gruen}, {Gruendl},
  {Honscheid}, {James}, {Kuropatkin}, {Li}, {Maia}, {Marshall}, {Martini},
  {Miller}, {Miquel}, {Nord}, {Ogando}, {Plazas}, {Reil}, {Romer}, {Roodman},
  {Sanchez}, {Sevilla-Noarbe}, {Smith}, {Soares-Santos}, {Sobreira}, {Tarle},
  {Thaler}, {Thomas}, {Tucker}, {Walker}, \& {DES Collaboration}}]{Kessler15}
{Kessler}, R., {Marriner}, J., {Childress}, M., {et~al.} 2015, \aj, 150, 172

\bibitem[{{Kessler} {et~al.}(2018){Kessler}, {Brout}, {D'Andrea}, {Davis},
  {Hinton}, {Kim}, {Lasker}, {Lidman}, {Macaulay}, {M{\"o}ller}, {Sako},
  {Scolnic}, {Smith}, {Sullivan}, {Zhang}, {Andersen}, {Asorey}, {Avelino},
  {Calcino}, {Carollo}, {Challis}, {Childress}, {Clocchiatti}, {Crawford},
  {Filippenko}, {Foley}, {Glazebrook}, {Hoormann}, {Kasai}, {Kirshner},
  {Lewis}, {Mandel}, {March}, {Morganson}, {Muthukrishna}, {Nugent}, {Pan},
  {Sommer}, {Swann}, {Thomas}, {Tucker}, {Uddin}, {Abbott}, {Allam}, {Annis},
  {Avila}, {Banerji}, {Bechtol}, {Bertin}, {Brooks}, {Buckley-Geer}, {Burke},
  {Carnero Rosell}, {Carrasco Kind}, {Carretero}, {Castander}, {Crocce}, {da
  Costa}, {Davis}, {De Vicente}, {Desai}, {Diehl}, {Doel}, {Eifler},
  {Flaugher}, {Fosalba}, {Frieman}, {Garcia-Bellido}, {Gaztanaga}, {Gerdes},
  {Gruen}, {Gruendl}, {Gutierrez}, {Hartley}, {Hollowood}, {Honscheid},
  {James}, {Johnson}, {Johnson}, {Krause}, {Kuehn}, {Kuropatkin}, {Lahav},
  {Li}, {Lima}, {Marshall}, {Martini}, {Menanteau}, {Miller}, {Miquel}, {Nord},
  {Plazas}, {Roodman}, {Sanchez}, {Scarpine}, {Schindler}, {Schubnell},
  {Serrano}, {Sevilla-Noarbe}, {Soares-Santos}, {Sobreira}, {Suchyta}, {Tarle},
  {Thomas}, {Walker}, \& {Zhang}}]{Kessler18}
{Kessler}, R., {Brout}, D., {D'Andrea}, C.~B., {et~al.} 2018, ArXiv e-prints

\bibitem[{{Kim} {et~al.}(2018){Kim}, {Smith}, {Sullivan}, \& {Lee}}]{Kim18}
{Kim}, Y.-L., {Smith}, M., {Sullivan}, M., \& {Lee}, Y.-W. 2018, \apj, 854, 24

\bibitem[{{Knights} {et~al.}(2013){Knights}, {Bassett}, {Varughese}, {Hlozek},
  {Kunz}, {Smith}, \& {Newling}}]{Knights13}
{Knights}, M., {Bassett}, B.~A., {Varughese}, M., {et~al.} 2013, \jcap, 1, 039

\bibitem[{{Knop} {et~al.}(2003){Knop}, {Aldering}, {Amanullah}, {Astier},
  {Blanc}, {Burns}, {Conley}, {Deustua}, {Doi}, {Ellis}, {Fabbro}, {Folatelli},
  {Fruchter}, {Garavini}, {Garmond}, {Garton}, {Gibbons}, {Goldhaber},
  {Goobar}, {Groom}, {Hardin}, {Hook}, {Howell}, {Kim}, {Lee}, {Lidman},
  {Mendez}, {Nobili}, {Nugent}, {Pain}, {Panagia}, {Pennypacker}, {Perlmutter},
  {Quimby}, {Raux}, {Regnault}, {Ruiz-Lapuente}, {Sainton}, {Schaefer},
  {Schahmaneche}, {Smith}, {Spadafora}, {Stanishev}, {Sullivan}, {Walton},
  {Wang}, {Wood-Vasey}, \& {Yasuda}}]{Knop03}
{Knop}, R.~A., {Aldering}, G., {Amanullah}, R., {et~al.} 2003, \apj, 598, 102

\bibitem[{{Kowalski} {et~al.}(2008){Kowalski}, {Rubin}, {Aldering},
  {Agostinho}, {Amadon}, {Amanullah}, {Balland}, {Barbary}, {Blanc}, {Challis},
  {Conley}, {Connolly}, {Covarrubias}, {Dawson}, {Deustua}, {Ellis}, {Fabbro},
  {Fadeyev}, {Fan}, {Farris}, {Folatelli}, {Frye}, {Garavini}, {Gates},
  {Germany}, {Goldhaber}, {Goldman}, {Goobar}, {Groom}, {Haissinski}, {Hardin},
  {Hook}, {Kent}, {Kim}, {Knop}, {Lidman}, {Linder}, {Mendez}, {Meyers},
  {Miller}, {Moniez}, {Mour{\~a}o}, {Newberg}, {Nobili}, {Nugent}, {Pain},
  {Perdereau}, {Perlmutter}, {Phillips}, {Prasad}, {Quimby}, {Regnault},
  {Rich}, {Rubenstein}, {Ruiz-Lapuente}, {Santos}, {Schaefer}, {Schommer},
  {Smith}, {Soderberg}, {Spadafora}, {Strolger}, {Strovink}, {Suntzeff},
  {Suzuki}, {Thomas}, {Walton}, {Wang}, {Wood-Vasey}, \& {Yun}}]{Kowalski08}
{Kowalski}, M., {Rubin}, D., {Aldering}, G., {et~al.} 2008, \apj, 686, 749

\bibitem[{{Kunz} {et~al.}(2007){Kunz}, {Bassett}, \& {Hlozek}}]{Kunz07}
{Kunz}, M., {Bassett}, B.~A., \& {Hlozek}, R.~A. 2007, \prd, 75, 103508

\bibitem[{{Lampeitl} {et~al.}(2010){Lampeitl}, {Smith}, {Nichol}, {Bassett},
  {Cinabro}, {Dilday}, {Foley}, {Frieman}, {Garnavich}, {Goobar}, {Im}, {Jha},
  {Marriner}, {Miquel}, {Nordin}, {{\"O}stman}, {Riess}, {Sako}, {Schneider},
  {Sollerman}, \& {Stritzinger}}]{Lampeitl10}
{Lampeitl}, H., {Smith}, M., {Nichol}, R.~C., {et~al.} 2010, \apj, 722, 566

\bibitem[{{Lewis} \& {Bridle}(2002)}]{Lewis02}
{Lewis}, A., \& {Bridle}, S. 2002, \prd, 66, 103511

\bibitem[{{Li} {et~al.}(2011){Li}, {Leaman}, {Chornock}, {Filippenko},
  {Poznanski}, {Ganeshalingam}, {Wang}, {Modjaz}, {Jha}, {Foley}, \&
  {Smith}}]{Li11}
{Li}, W., {Leaman}, J., {Chornock}, R., {et~al.} 2011, \mnras, 412, 1441

\bibitem[{{Linder}(2003)}]{Linder03}
{Linder}, E.~V. 2003, Physical Review Letters, 90, 091301

\bibitem[{{Mandel} {et~al.}(2011){Mandel}, {Narayan}, \& {Kirshner}}]{Mandel11}
{Mandel}, K.~S., {Narayan}, G., \& {Kirshner}, R.~P. 2011, \apj, 731, 120

\bibitem[{{Marriner} {et~al.}(2011){Marriner}, {Bernstein}, {Kessler},
  {Lampeitl}, {Miquel}, {Mosher}, {Nichol}, {Sako}, {Schneider}, \&
  {Smith}}]{Marriner11}
{Marriner}, J., {Bernstein}, J.~P., {Kessler}, R., {et~al.} 2011, \apj, 740, 72

\bibitem[{{M{\'e}nard} {et~al.}(2010){M{\'e}nard}, {Scranton}, {Fukugita}, \&
  {Richards}}]{Menard10}
{M{\'e}nard}, B., {Scranton}, R., {Fukugita}, M., \& {Richards}, G. 2010,
  \mnras, 405, 1025

\bibitem[{{Perlmutter} {et~al.}(1999){Perlmutter}, {Aldering}, {Goldhaber},
  {Knop}, {Nugent}, {Castro}, {Deustua}, {Fabbro}, {Goobar}, {Groom}, {Hook},
  {Kim}, {Kim}, {Lee}, {Nunes}, {Pain}, {Pennypacker}, {Quimby}, {Lidman},
  {Ellis}, {Irwin}, {McMahon}, {Ruiz-Lapuente}, {Walton}, {Schaefer}, {Boyle},
  {Filippenko}, {Matheson}, {Fruchter}, {Panagia}, {Newberg}, {Couch}, \&
  {Project}}]{Perlmutter99}
{Perlmutter}, S., {Aldering}, G., {Goldhaber}, G., {et~al.} 1999, \apj, 517,
  565

\bibitem[{{Pierel} {et~al.}(2018){Pierel}, {Rodney}, {Avelino}, {Bianco},
  {Filippenko}, {Foley}, {Friedman}, {Hicken}, {Hounsell}, {Jha}, {Kessler},
  {Kirshner}, {Mandel}, {Narayan}, {Scolnic}, \& {Strolger}}]{Pierel18}
{Pierel}, J.~D.~R., {Rodney}, S., {Avelino}, A., {et~al.} 2018, \pasp, 130,
  114504

\bibitem[{{Planck Collaboration} {et~al.}(2016){Planck Collaboration}, {Ade},
  {Aghanim}, {Arnaud}, {Ashdown}, {Aumont}, {Baccigalupi}, {Banday},
  {Barreiro}, {Bartlett}, \& et~al.}]{Planck16}
{Planck Collaboration}, {Ade}, P.~A.~R., {Aghanim}, N., {et~al.} 2016, \aap,
  594, A13

\bibitem[{{Planck Collaboration} {et~al.}(2018){Planck Collaboration},
  {Aghanim}, {Akrami}, {Ashdown}, {Aumont}, {Baccigalupi}, {Ballardini},
  {Banday}, {Barreiro}, {Bartolo}, {Basak}, {Battye}, {Benabed}, {Bernard},
  {Bersanelli}, {Bielewicz}, {Bock}, {Bond}, {Borrill}, {Bouchet}, {Boulanger},
  {Bucher}, {Burigana}, {Butler}, {Calabrese}, {Cardoso}, {Carron},
  {Challinor}, {Chiang}, {Chluba}, {Colombo}, {Combet}, {Contreras}, {Crill},
  {Cuttaia}, {de Bernardis}, {de Zotti}, {Delabrouille}, {Delouis}, {Di
  Valentino}, {Diego}, {Dor{\'e}}, {Douspis}, {Ducout}, {Dupac}, {Dusini},
  {Efstathiou}, {Elsner}, {En{\ss}lin}, {Eriksen}, {Fantaye}, {Farhang},
  {Fergusson}, {Fernandez-Cobos}, {Finelli}, {Forastieri}, {Frailis},
  {Franceschi}, {Frolov}, {Galeotta}, {Galli}, {Ganga}, {G{\'e}nova-Santos},
  {Gerbino}, {Ghosh}, {Gonz{\'a}lez-Nuevo}, {G{\'o}rski}, {Gratton},
  {Gruppuso}, {Gudmundsson}, {Hamann}, {Handley}, {Herranz}, {Hivon}, {Huang},
  {Jaffe}, {Jones}, {Karakci}, {Keih{\"a}nen}, {Keskitalo}, {Kiiveri}, {Kim},
  {Kisner}, {Knox}, {Krachmalnicoff}, {Kunz}, {Kurki-Suonio}, {Lagache},
  {Lamarre}, {Lasenby}, {Lattanzi}, {Lawrence}, {Le Jeune}, {Lemos},
  {Lesgourgues}, {Levrier}, {Lewis}, {Liguori}, {Lilje}, {Lilley}, {Lindholm},
  {L{\'o}pez-Caniego}, {Lubin}, {Ma}, {Mac{\'{\i}}as-P{\'e}rez}, {Maggio},
  {Maino}, {Mandolesi}, {Mangilli}, {Marcos-Caballero}, {Maris}, {Martin},
  {Martinelli}, {Mart{\'{\i}}nez-Gonz{\'a}lez}, {Matarrese}, {Mauri}, {McEwen},
  {Meinhold}, {Melchiorri}, {Mennella}, {Migliaccio}, {Millea}, {Mitra},
  {Miville-Desch{\^e}nes}, {Molinari}, {Montier}, {Morgante}, {Moss}, {Natoli},
  {N{\o}rgaard-Nielsen}, {Pagano}, {Paoletti}, {Partridge}, {Patanchon},
  {Peiris}, {Perrotta}, {Pettorino}, {Piacentini}, {Polastri}, {Polenta},
  {Puget}, {Rachen}, {Reinecke}, {Remazeilles}, {Renzi}, {Rocha}, {Rosset},
  {Roudier}, {Rubi{\~n}o-Mart{\'{\i}}n}, {Ruiz-Granados}, {Salvati}, {Sandri},
  {Savelainen}, {Scott}, {Shellard}, {Sirignano}, {Sirri}, {Spencer},
  {Sunyaev}, {Suur-Uski}, {Tauber}, {Tavagnacco}, {Tenti}, {Toffolatti},
  {Tomasi}, {Trombetti}, {Valenziano}, {Valiviita}, {Van Tent}, {Vibert},
  {Vielva}, {Villa}, {Vittorio}, {Wandelt}, {Wehus}, {White}, {White},
  {Zacchei}, \& {Zonca}}]{Planck18}
{Planck Collaboration}, {Aghanim}, N., {Akrami}, Y., {et~al.} 2018, ArXiv
  e-prints

\bibitem[{{Rest} {et~al.}(2005){Rest}, {Stubbs}, {Becker}, {Miknaitis},
  {Miceli}, {Covarrubias}, {Hawley}, {Smith}, {Suntzeff}, {Olsen}, {Prieto},
  {Hiriart}, {Welch}, {Cook}, {Nikolaev}, {Huber}, {Prochtor}, {Clocchiatti},
  {Minniti}, {Garg}, {Challis}, {Keller}, \& {Schmidt}}]{Rest05}
{Rest}, A., {Stubbs}, C., {Becker}, A.~C., {et~al.} 2005, \apj, 634, 1103

\bibitem[{{Rest} {et~al.}(2014){Rest}, {Scolnic}, {Foley}, {Huber}, {Chornock},
  {Narayan}, {Tonry}, {Berger}, {Soderberg}, {Stubbs}, {Riess}, {Kirshner},
  {Smartt}, {Schlafly}, {Rodney}, {Botticella}, {Brout}, {Challis}, {Czekala},
  {Drout}, {Hudson}, {Kotak}, {Leibler}, {Lunnan}, {Marion}, {McCrum},
  {Milisavljevic}, {Pastorello}, {Sanders}, {Smith}, {Stafford}, {Thilker},
  {Valenti}, {Wood-Vasey}, {Zheng}, {Burgett}, {Chambers}, {Denneau}, {Draper},
  {Flewelling}, {Hodapp}, {Kaiser}, {Kudritzki}, {Magnier}, {Metcalfe},
  {Price}, {Sweeney}, {Wainscoat}, \& {Waters}}]{Rest14}
{Rest}, A., {Scolnic}, D., {Foley}, R.~J., {et~al.} 2014, \apj, 795, 44

\bibitem[{{Riess} {et~al.}(1998){Riess}, {Filippenko}, {Challis},
  {Clocchiatti}, {Diercks}, {Garnavich}, {Gilliland}, {Hogan}, {Jha},
  {Kirshner}, {Leibundgut}, {Phillips}, {Reiss}, {Schmidt}, {Schommer},
  {Smith}, {Spyromilio}, {Stubbs}, {Suntzeff}, \& {Tonry}}]{Riess98}
{Riess}, A.~G., {Filippenko}, A.~V., {Challis}, P., {et~al.} 1998, \aj, 116,
  1009

\bibitem[{{Riess} {et~al.}(1999){Riess}, {Kirshner}, {Schmidt}, {Jha},
  {Challis}, {Garnavich}, {Esin}, {Carpenter}, {Grashius}, {Schild}, {Berlind},
  {Huchra}, {Prosser}, {Falco}, {Benson}, {Brice{\~n}o}, {Brown}, {Caldwell},
  {dell'Antonio}, {Filippenko}, {Goodman}, {Grogin}, {Groner}, {Hughes},
  {Green}, {Jansen}, {Kleyna}, {Luu}, {Macri}, {McLeod}, {McLeod}, {McNamara},
  {McLean}, {Milone}, {Mohr}, {Moraru}, {Peng}, {Peters}, {Prestwich},
  {Stanek}, {Szentgyorgyi}, \& {Zhao}}]{Riess99}
{Riess}, A.~G., {Kirshner}, R.~P., {Schmidt}, B.~P., {et~al.} 1999, \aj, 117,
  707

\bibitem[{{Riess} {et~al.}(2004){Riess}, {Strolger}, {Tonry}, {Casertano},
  {Ferguson}, {Mobasher}, {Challis}, {Filippenko}, {Jha}, {Li}, {Chornock},
  {Kirshner}, {Leibundgut}, {Dickinson}, {Livio}, {Giavalisco}, {Steidel},
  {Ben{\'{\i}}tez}, \& {Tsvetanov}}]{Riess04}
{Riess}, A.~G., {Strolger}, L.-G., {Tonry}, J., {et~al.} 2004, \apj, 607, 665

\bibitem[{{Riess} {et~al.}(2007){Riess}, {Strolger}, {Casertano}, {Ferguson},
  {Mobasher}, {Gold}, {Challis}, {Filippenko}, {Jha}, {Li}, {Tonry}, {Foley},
  {Kirshner}, {Dickinson}, {MacDonald}, {Eisenstein}, {Livio}, {Younger}, {Xu},
  {Dahl{\'e}n}, \& {Stern}}]{Riess07}
{Riess}, A.~G., {Strolger}, L.-G., {Casertano}, S., {et~al.} 2007, \apj, 659,
  98

\bibitem[{{Riess} {et~al.}(2018){Riess}, {Casertano}, {Yuan}, {Macri},
  {Anderson}, {MacKenty}, {Bowers}, {Clubb}, {Filippenko}, {Jones}, \&
  {Tucker}}]{Riess18}
{Riess}, A.~G., {Casertano}, S., {Yuan}, W., {et~al.} 2018, \apj, 855, 136

\bibitem[{{Rigault} {et~al.}(2013){Rigault}, {Copin}, {Aldering}, {Antilogus},
  {Aragon}, {Bailey}, {Baltay}, {Bongard}, {Buton}, {Canto}, {Cellier-Holzem},
  {Childress}, {Chotard}, {Fakhouri}, {Feindt}, {Fleury}, {Gangler},
  {Greskovic}, {Guy}, {Kim}, {Kowalski}, {Lombardo}, {Nordin}, {Nugent},
  {Pain}, {P{\'e}contal}, {Pereira}, {Perlmutter}, {Rabinowitz}, {Runge},
  {Saunders}, {Scalzo}, {Smadja}, {Tao}, {Thomas}, \& {Weaver}}]{Rigault13}
{Rigault}, M., {Copin}, Y., {Aldering}, G., {et~al.} 2013, \aap, 560, A66

\bibitem[{{Rigault} {et~al.}(2015){Rigault}, {Aldering}, {Kowalski}, {Copin},
  {Antilogus}, {Aragon}, {Bailey}, {Baltay}, {Baugh}, {Bongard}, {Boone},
  {Buton}, {Chen}, {Chotard}, {Fakhouri}, {Feindt}, {Fagrelius}, {Fleury},
  {Fouchez}, {Gangler}, {Hayden}, {Kim}, {Leget}, {Lombardo}, {Nordin}, {Pain},
  {Pecontal}, {Pereira}, {Perlmutter}, {Rabinowitz}, {Runge}, {Rubin},
  {Saunders}, {Smadja}, {Sofiatti}, {Suzuki}, {Tao}, \& {Weaver}}]{Rigault15}
{Rigault}, M., {Aldering}, G., {Kowalski}, M., {et~al.} 2015, \apj, 802, 20

\bibitem[{{Rigault} {et~al.}(2018){Rigault}, {Brinnel}, {Aldering},
  {Antilogus}, {Aragon}, {Bailey}, {Baltay}, {Barbary}, {Bongard}, {Boone},
  {Buton}, {Childress}, {Chotard}, {Copin}, {Dixon}, {Fagrelius}, {Feindt},
  {Fouchez}, {Gangler}, {Hayden}, {Hillebrandt}, {Howell}, {Kim}, {Kowalski},
  {Kuesters}, {Leget}, {Lombardo}, {Lin}, {Nordin}, {Pain}, {Pecontal},
  {Pereira}, {Perlmutter}, {Rabinowitz}, {Runge}, {Rubin}, {Saunders},
  {Smadja}, {Sofiatti}, {Suzuki}, {Taubenberger}, {Tao}, \&
  {Thomas}}]{Rigault18}
{Rigault}, M., {Brinnel}, V., {Aldering}, G., {et~al.} 2018, ArXiv e-prints

\bibitem[{{Roberts} {et~al.}(2017){Roberts}, {Lochner}, {Fonseca}, {Bassett},
  {Lablanche}, \& {Agarwal}}]{Roberts17}
{Roberts}, E., {Lochner}, M., {Fonseca}, J., {et~al.} 2017, Journal of
  Cosmology and Astro-Particle Physics, 2017, 036

\bibitem[{{Roman} {et~al.}(2018){Roman}, {Hardin}, {Betoule}, {Astier},
  {Balland}, {Ellis}, {Fabbro}, {Guy}, {Hook}, {Howell}, {Lidman}, {Mitra},
  {M{\"o}ller}, {Mour{\~a}o}, {Neveu}, {Palanque-Delabrouille}, {Pritchet},
  {Regnault}, {Ruhlmann-Kleider}, {Saunders}, \& {Sullivan}}]{Roman18}
{Roman}, M., {Hardin}, D., {Betoule}, M., {et~al.} 2018, \aap, 615, A68

\bibitem[{{Ross} {et~al.}(2015){Ross}, {Samushia}, {Howlett}, {Percival},
  {Burden}, \& {Manera}}]{Ross15}
{Ross}, A.~J., {Samushia}, L., {Howlett}, C., {et~al.} 2015, \mnras, 449, 835

\bibitem[{{Sako} {et~al.}(2011){Sako}, {Bassett}, {Connolly}, {Dilday},
  {Cambell}, {Frieman}, {Gladney}, {Kessler}, {Lampeitl}, {Marriner}, {Miquel},
  {Nichol}, {Schneider}, {Smith}, \& {Sollerman}}]{Sako11}
{Sako}, M., {Bassett}, B., {Connolly}, B., {et~al.} 2011, \apj, 738, 162

\bibitem[{{Sako} {et~al.}(2018){Sako}, {Bassett}, {Becker}, {Brown},
  {Campbell}, {Wolf}, {Cinabro}, {D'Andrea}, {Dawson}, {DeJongh}, {Depoy},
  {Dilday}, {Doi}, {Filippenko}, {Fischer}, {Foley}, {Frieman}, {Galbany},
  {Garnavich}, {Goobar}, {Gupta}, {Hill}, {Hayden}, {Hlozek}, {Holtzman},
  {Hopp}, {Jha}, {Kessler}, {Kollatschny}, {Leloudas}, {Marriner}, {Marshall},
  {Miquel}, {Morokuma}, {Mosher}, {Nichol}, {Nordin}, {Olmstead}, {{\"O}stman},
  {Prieto}, {Richmond}, {Romani}, {Sollerman}, {Stritzinger}, {Schneider},
  {Smith}, {Wheeler}, {Yasuda}, \& {Zheng}}]{Sako18}
{Sako}, M., {Bassett}, B., {Becker}, A.~C., {et~al.} 2018, \pasp, 130, 064002

\bibitem[{{Schechter} {et~al.}(1993){Schechter}, {Mateo}, \&
  {Saha}}]{Schechter93}
{Schechter}, P.~L., {Mateo}, M., \& {Saha}, A. 1993, \pasp, 105, 1342

\bibitem[{{Schlafly} {et~al.}(2012){Schlafly}, {Finkbeiner}, {Juri{\'c}},
  {Magnier}, {Burgett}, {Chambers}, {Grav}, {Hodapp}, {Kaiser}, {Kudritzki},
  {Martin}, {Morgan}, {Price}, {Rix}, {Stubbs}, {Tonry}, \&
  {Wainscoat}}]{Schlafly12}
{Schlafly}, E.~F., {Finkbeiner}, D.~P., {Juri{\'c}}, M., {et~al.} 2012, \apj,
  756, 158

\bibitem[{{Scolnic} \& {Kessler}(2016)}]{Scolnic16}
{Scolnic}, D., \& {Kessler}, R. 2016, \apjl, 822, L35

\bibitem[{{Scolnic} {et~al.}(2014{\natexlab{a}}){Scolnic}, {Rest}, {Riess},
  {Huber}, {Foley}, {Brout}, {Chornock}, {Narayan}, {Tonry}, {Berger},
  {Soderberg}, {Stubbs}, {Kirshner}, {Rodney}, {Smartt}, {Schlafly},
  {Botticella}, {Challis}, {Czekala}, {Drout}, {Hudson}, {Kotak}, {Leibler},
  {Lunnan}, {Marion}, {McCrum}, {Milisavljevic}, {Pastorello}, {Sanders},
  {Smith}, {Stafford}, {Thilker}, {Valenti}, {Wood-Vasey}, {Zheng}, {Burgett},
  {Chambers}, {Denneau}, {Draper}, {Flewelling}, {Hodapp}, {Kaiser},
  {Kudritzki}, {Magnier}, {Metcalfe}, {Price}, {Sweeney}, {Wainscoat}, \&
  {Waters}}]{Scolnic14b}
{Scolnic}, D., {Rest}, A., {Riess}, A., {et~al.} 2014{\natexlab{a}}, \apj, 795,
  45

\bibitem[{{Scolnic} {et~al.}(2015){Scolnic}, {Casertano}, {Riess}, {Rest},
  {Schlafly}, {Foley}, {Finkbeiner}, {Tang}, {Burgett}, {Chambers}, {Draper},
  {Flewelling}, {Hodapp}, {Huber}, {Kaiser}, {Kudritzki}, {Magnier},
  {Metcalfe}, \& {Stubbs}}]{Scolnic15}
{Scolnic}, D., {Casertano}, S., {Riess}, A., {et~al.} 2015, \apj, 815, 117

\bibitem[{{Scolnic} {et~al.}(2014{\natexlab{b}}){Scolnic}, {Riess}, {Foley},
  {Rest}, {Rodney}, {Brout}, \& {Jones}}]{Scolnic14}
{Scolnic}, D.~M., {Riess}, A.~G., {Foley}, R.~J., {et~al.} 2014{\natexlab{b}},
  \apj, 780, 37

\bibitem[{{Scolnic} {et~al.}(2018){Scolnic}, {Jones}, {Rest}, {Pan},
  {Chornock}, {Foley}, {Huber}, {Kessler}, {Narayan}, {Riess}, {Rodney},
  {Berger}, {Brout}, {Challis}, {Drout}, {Finkbeiner}, {Lunnan}, {Kirshner},
  {Sanders}, {Schlafly}, {Smartt}, {Stubbs}, {Tonry}, {Wood-Vasey}, {Foley},
  {Hand}, {Johnson}, {Burgett}, {Chambers}, {Draper}, {Hodapp}, {Kaiser},
  {Kudritzki}, {Magnier}, {Metcalfe}, {Bresolin}, {Gall}, {Kotak}, {McCrum}, \&
  {Smith}}]{Scolnic18}
{Scolnic}, D.~M., {Jones}, D.~O., {Rest}, A., {et~al.} 2018, \apj, 859, 101

\bibitem[{{Spergel} {et~al.}(2015){Spergel}, {Gehrels}, {Baltay}, {Bennett},
  {Breckinridge}, {Donahue}, {Dressler}, {Gaudi}, {Greene}, {Guyon}, {Hirata},
  {Kalirai}, {Kasdin}, {Macintosh}, {Moos}, {Perlmutter}, {Postman},
  {Rauscher}, {Rhodes}, {Wang}, {Weinberg}, {Benford}, {Hudson}, {Jeong},
  {Mellier}, {Traub}, {Yamada}, {Capak}, {Colbert}, {Masters}, {Penny},
  {Savransky}, {Stern}, {Zimmerman}, {Barry}, {Bartusek}, {Carpenter}, {Cheng},
  {Content}, {Dekens}, {Demers}, {Grady}, {Jackson}, {Kuan}, {Kruk}, {Melton},
  {Nemati}, {Parvin}, {Poberezhskiy}, {Peddie}, {Ruffa}, {Wallace}, {Whipple},
  {Wollack}, \& {Zhao}}]{Spergel15}
{Spergel}, D., {Gehrels}, N., {Baltay}, C., {et~al.} 2015, ArXiv e-prints

\bibitem[{{Stetson}(1987)}]{Stetson87}
{Stetson}, P.~B. 1987, \pasp, 99, 191

\bibitem[{{Stritzinger} {et~al.}(2011){Stritzinger}, {Phillips}, {Boldt},
  {Burns}, {Campillay}, {Contreras}, {Gonzalez}, {Folatelli}, {Morrell},
  {Krzeminski}, {Roth}, {Salgado}, {DePoy}, {Hamuy}, {Freedman}, {Madore},
  {Marshall}, {Persson}, {Rheault}, {Suntzeff}, {Villanueva}, {Li}, \&
  {Filippenko}}]{Stritzinger11}
{Stritzinger}, M.~D., {Phillips}, M.~M., {Boldt}, L.~N., {et~al.} 2011, \aj,
  142, 156

\bibitem[{{Sullivan} {et~al.}(2006){Sullivan}, {Le Borgne}, {Pritchet},
  {Hodsman}, {Neill}, {Howell}, {Carlberg}, {Astier}, {Aubourg}, {Balam},
  {Basa}, {Conley}, {Fabbro}, {Fouchez}, {Guy}, {Hook}, {Pain},
  {Palanque-Delabrouille}, {Perrett}, {Regnault}, {Rich}, {Taillet}, {Baumont},
  {Bronder}, {Ellis}, {Filiol}, {Lusset}, {Perlmutter}, {Ripoche}, \&
  {Tao}}]{Sullivan06}
{Sullivan}, M., {Le Borgne}, D., {Pritchet}, C.~J., {et~al.} 2006, \apj, 648,
  868

\bibitem[{{Sullivan} {et~al.}(2010){Sullivan}, {Conley}, {Howell}, {Neill},
  {Astier}, {Balland}, {Basa}, {Carlberg}, {Fouchez}, {Guy}, {Hardin}, {Hook},
  {Pain}, {Palanque-Delabrouille}, {Perrett}, {Pritchet}, {Regnault}, {Rich},
  {Ruhlmann-Kleider}, {Baumont}, {Hsiao}, {Kronborg}, {Lidman}, {Perlmutter},
  \& {Walker}}]{Sullivan10}
{Sullivan}, M., {Conley}, A., {Howell}, D.~A., {et~al.} 2010, \mnras, 406, 782

\bibitem[{{Sullivan} {et~al.}(2011){Sullivan}, {Guy}, {Conley}, {Regnault},
  {Astier}, {Balland}, {Basa}, {Carlberg}, {Fouchez}, {Hardin}, {Hook},
  {Howell}, {Pain}, {Palanque-Delabrouille}, {Perrett}, {Pritchet}, {Rich},
  {Ruhlmann-Kleider}, {Balam}, {Baumont}, {Ellis}, {Fabbro}, {Fakhouri},
  {Fourmanoit}, {Gonz{\'a}lez-Gait{\'a}n}, {Graham}, {Hudson}, {Hsiao},
  {Kronborg}, {Lidman}, {Mourao}, {Neill}, {Perlmutter}, {Ripoche}, {Suzuki},
  \& {Walker}}]{Sullivan11}
{Sullivan}, M., {Guy}, J., {Conley}, A., {et~al.} 2011, \apj, 737, 102

\bibitem[{{Tartaglia} {et~al.}(2018){Tartaglia}, {Sand}, {Valenti}, {Wyatt},
  {Anderson}, {Arcavi}, {Ashall}, {Botticella}, {Cartier}, {Chen}, {Cikota},
  {Coulter}, {Della Valle}, {Foley}, {Gal-Yam}, {Galbany}, {Gall}, {Haislip},
  {Harmanen}, {Hosseinzadeh}, {Howell}, {Hsiao}, {Inserra}, {Jha}, {Kankare},
  {Kilpatrick}, {Kouprianov}, {Kuncarayakti}, {Maccarone}, {Maguire},
  {Mattila}, {Mazzali}, {McCully}, {Melandri}, {Morrell}, {Phillips},
  {Pignata}, {Piro}, {Prentice}, {Reichart}, {Rojas-Bravo}, {Smartt}, {Smith},
  {Sollerman}, {Stritzinger}, {Sullivan}, {Taddia}, \& {Young}}]{Tartaglia18}
{Tartaglia}, L., {Sand}, D.~J., {Valenti}, S., {et~al.} 2018, \apj, 853, 62

\bibitem[{{The LSST Dark Energy Science Collaboration} {et~al.}(2018){The LSST
  Dark Energy Science Collaboration}, {Mandelbaum}, {Eifler}, {Hlo{\v z}ek},
  {Collett}, {Gawiser}, {Scolnic}, {Alonso}, {Awan}, {Biswas}, {Blazek},
  {Burchat}, {Chisari}, {Dell'Antonio}, {Digel}, {Frieman}, {Goldstein},
  {Hook}, {Ivezi{\'c}}, {Kahn}, {Kamath}, {Kirkby}, {Kitching}, {Krause},
  {Leget}, {Marshall}, {Meyers}, {Miyatake}, {Newman}, {Nichol}, {Rykoff},
  {Sanchez}, {Slosar}, {Sullivan}, \& {Troxel}}]{LSST18}
{The LSST Dark Energy Science Collaboration}, {Mandelbaum}, R., {Eifler}, T.,
  {et~al.} 2018, ArXiv e-prints

\bibitem[{{Tonry} {et~al.}(2003){Tonry}, {Schmidt}, {Barris}, {Candia},
  {Challis}, {Clocchiatti}, {Coil}, {Filippenko}, {Garnavich}, {Hogan},
  {Holland}, {Jha}, {Kirshner}, {Krisciunas}, {Leibundgut}, {Li}, {Matheson},
  {Phillips}, {Riess}, {Schommer}, {Smith}, {Sollerman}, {Spyromilio},
  {Stubbs}, \& {Suntzeff}}]{Tonry03}
{Tonry}, J.~L., {Schmidt}, B.~P., {Barris}, B., {et~al.} 2003, \apj, 594, 1

\bibitem[{{Tonry} {et~al.}(2018){Tonry}, {Denneau}, {Heinze}, {Stalder},
  {Smith}, {Smartt}, {Stubbs}, {Weiland}, \& {Rest}}]{Tonry18}
{Tonry}, J.~L., {Denneau}, L., {Heinze}, A.~N., {et~al.} 2018, ArXiv e-prints

\bibitem[{{Tripp}(1998)}]{Tripp98}
{Tripp}, R. 1998, \aap, 331, 815

\bibitem[{{Wood-Vasey} {et~al.}(2007){Wood-Vasey}, {Miknaitis}, {Stubbs},
  {Jha}, {Riess}, {Garnavich}, {Kirshner}, {Aguilera}, {Becker}, {Blackman},
  {Blondin}, {Challis}, {Clocchiatti}, {Conley}, {Covarrubias}, {Davis},
  {Filippenko}, {Foley}, {Garg}, {Hicken}, {Krisciunas}, {Leibundgut}, {Li},
  {Matheson}, {Miceli}, {Narayan}, {Pignata}, {Prieto}, {Rest}, {Salvo},
  {Schmidt}, {Smith}, {Sollerman}, {Spyromilio}, {Tonry}, {Suntzeff}, \&
  {Zenteno}}]{WoodVasey07}
{Wood-Vasey}, W.~M., {Miknaitis}, G., {Stubbs}, C.~W., {et~al.} 2007, \apj,
  666, 694

\end{thebibliography}

\end{document}